\documentclass{declare}
\usepackage[absolute,overlay]{textpos}
\usepackage{tikz}
\usepackage{hyperref}
\usepackage{url}
\usepackage{amsthm}
\usepackage[table]{xcolor}
\usepackage{hyperref}

\usepackage{fontawesome5}
\usepackage[smartEllipses]{markdown}
\usepackage{pifont}
\usepackage{xcolor}
\usepackage{bbm}
\usepackage[table]{xcolor} 
\usepackage{multirow}   
\usepackage{subcaption}
\usepackage{wrapfig}
\usepackage{xspace}

\usepackage{wrapfig}
\usepackage{multirow}
\usepackage{graphicx,xcolor,float}
\usepackage{threeparttable}
\usepackage[ruled,vlined]{algorithm2e}
\usepackage{colortbl}
\usepackage{color}
\usepackage{multirow}
\usepackage{tabularx}
\usepackage{float}
\usepackage{amsmath, amssymb}
\usepackage{tikz}
\usetikzlibrary{shapes.geometric, arrows.meta, positioning, calc, fit, backgrounds}
\usepackage{graphicx}
\usepackage{booktabs}
\usepackage{arydshln}
\usepackage{enumitem}
\usepackage{wrapfig}
\usepackage{caption}
\usepackage{graphicx}
\usepackage{makecell}
\usepackage{float} 
\usepackage{soul}
\usepackage{mathtools}
\usepackage{bbding}
\usepackage{makecell}
\usepackage{tabularx}
\usepackage{amssymb,mathrsfs,amsmath}
\usepackage{pifont}
\usepackage{xcolor}
\usepackage{bbm}
\usepackage[table]{xcolor} 
\usepackage{multirow}   

\usepackage{xspace}
\usepackage{seqsplit}

\usepackage{fontawesome5}

\usepackage{listings}

\tcbuselibrary{skins}

\definecolor{gtgray}{HTML}{F1F3F5}
\definecolor{posgreen}{HTML}{E6F4EA}
\definecolor{negred}{HTML}{FCE8E6}
\definecolor{bestgreen}{HTML}{C8E6C9}

\newcommand{\pos}[1]{\cellcolor{posgreen}{#1}}
\newcommand{\negp}[1]{\cellcolor{negred}{#1}}
\newcommand{\best}[1]{\cellcolor{bestgreen}\textbf{#1}}

\definecolor{slate}{HTML}{536171}
\definecolor{arrowgray}{HTML}{52606D}
\definecolor{plateedge}{HTML}{A9B2BC}
\definecolor{backedge}{HTML}{C6CED6}
\definecolor{backfill}{HTML}{F5F7F8}
\definecolor{paperfill}{HTML}{DDE7EF}
\definecolor{gapfill}{HTML}{E3EFEA}
\definecolor{tealedge}{HTML}{6F9186}
\definecolor{questionfill}{HTML}{E8E4EF}
\definecolor{questionedge}{HTML}{7D718F}
\definecolor{thetafill}{HTML}{F2EBDD}
\definecolor{thetaedge}{HTML}{A79672}
\definecolor{panelfill}{HTML}{FAFAF7}
\definecolor{panelborder}{HTML}{D5D9DD}
\definecolor{notetext}{HTML}{39434D}

\usepackage{xcolor}
\usepackage{enumitem}

\definecolor{promptbg}{HTML}{F8FAFC}
\definecolor{promptframe}{HTML}{CBD5E1}

\newtcolorbox{promptbox}[1]{
  enhanced,
  breakable,
  colback=promptbg,
  colframe=promptframe,
  boxrule=0.45pt,
  arc=2mm,
  left=1.5mm,
  right=1.5mm,
  top=1mm,
  bottom=1mm,
  before skip=0.8em,
  after skip=0.8em,
  title=\textbf{#1},
  fonttitle=\small,
  coltitle=black
}

\newcommand{\promptfield}[1]{\vspace{0.35em}\noindent\textbf{#1}\par}

\newcommand{\dataset}{RQ-Bench}

\usepackage{xcolor}

\usepackage{enumitem}

\usepackage{tikz}

\usetikzlibrary{positioning, arrows.meta}

\definecolor{rqblue}{HTML}{EAF2FF}
\definecolor{rqblueframe}{HTML}{2B5FAB}
\definecolor{gapred}{HTML}{FFF3F0}
\definecolor{gapredframe}{HTML}{B6472F}
\definecolor{ideagreen}{HTML}{EFFAF3}
\definecolor{ideagreenframe}{HTML}{2E7D4F}
\definecolor{softgray}{HTML}{F8FAFC}

\newtcolorbox{examplecard}[2]{
  enhanced,
  breakable,
  colback=#1,
  colframe=#2,
  boxrule=0.45pt,
  arc=2mm,
  left=1.5mm,
  right=1.5mm,
  top=1.2mm,
  bottom=1.2mm,
  before skip=0.5em,
  after skip=0.5em
}

\usepackage{xcolor}
\usepackage{enumitem}
\usepackage{tabularx}
\usepackage{ragged2e}
\usepackage{tikz}
\usetikzlibrary{arrows.meta}

\definecolor{gapbg}{HTML}{FFF8F5}
\definecolor{gapframe}{HTML}{C56A4A}
\definecolor{rqbg}{HTML}{F4F8FF}
\definecolor{rqframe}{HTML}{4A78B8}
\definecolor{ideabg}{HTML}{F5FBF6}
\definecolor{ideaframe}{HTML}{4A8F61}

\newtcolorbox{dapobox}[3]{
  enhanced,
  colback=#1,
  colframe=#2,
  boxrule=0.45pt,
  arc=2mm,
  left=2mm,
  right=2mm,
  top=1.5mm,
  bottom=1.5mm,
  height=#3,
  valign=top
}

\runningtitle{Sinhahajari et al. (2026). On the Limits of LLM-as-Judge for Scientific Novelty Assessment.}

\runningstatus{Under review}
\title{On the Limits of LLM-as-Judge for Scientific Novelty Assessment}

\author[1]{Soumitra Sinhahajari}
\author[1]{Navonil Majumder}
\author[1]{Soujanya Poria}
\affiliation[1]{\small{DeCLaRe Lab, Nanyang Technological University}}
\hfdataset{rq-bench}
\correspondence{Soujanya Poria (\email{soujanya.poria@ntu.edu.sg})}
\definecolor{forestgreen}{RGB}{34,139,34}

\newcommand{\grpo}[1]{\texttt{GRPO}}
\abstract{
LLMs are increasingly used to generate and judge scientific ideas. This makes novelty evaluation a central problem. Full idea evaluation is difficult because it often requires judging a method, its feasibility, and its empirical promise. We therefore study a cleaner upstream object: the research question (RQ). RQ generation is a prerequisite for scientific ideation, and RQs can be compared against questions pursued in real papers. We introduce \dataset{}, a benchmark built from recent arXiv papers. For each paper, we reconstruct author-anchored RQs from its cited background, gaps, and contributions. These RQs are not the only valid questions for the same background. They are author-anchored reference points for testing novelty judgments. We evaluate model-generated RQs with standalone LLM judging, comparative LLM judging, and human expert evaluation. 
LLM judges consistently rate model-generated RQs as highly novel, producing a \emph{novelty mirage}; in comparative evaluations, this preference becomes even stronger. Domain experts, however, reach the opposite conclusion and prefer the author-anchored reference questions. We further find that many generated RQs are narrow or source-bound, a dimension that LLM judges often miss unless explicitly tested. Overall, the contradictory novelty evaluations between LLM judges and human experts raise a serious concern about the reliability of using LLMs to assess the scientific novelty of research questions.
}

\date{\today}

\usepackage{hyperref}

\usepackage{booktabs}
\usepackage{soul}
\usepackage[table]{xcolor}

\crefformat{section}{\S#2#1#3} 
\crefformat{subsection}{\S#2#1#3}
\crefformat{subsubsection}{\S#2#1#3}

\let\realcite\cite
\renewcommand{\cite}[1]{\ifx.#1.\hl{[?]}\else\realcite{#1}\fi}

\begin{document}

\maketitle

\section{Introduction}

LLMs are now used not only to generate scientific ideas, but also to judge them. This led us to ask: what are the limits of LLM-as-judge for evaluating scientific novelty? If LLMs generate a scientific artifact and other LLMs judge its novelty, the whole pipeline depends on whether LLM-as-judge is reliable. This is especially important for ``AI scientist'' systems, where LLM agents retrieve papers, propose ideas, design experiments, and draft manuscripts~\cite{Luetal2024, Yamadaetal2025, Baeketal2024}.

Most existing work evaluates the final research idea. This is hard. A full idea may sound novel, but judging it often requires checking the method, feasibility, experiments, baselines, and likely impact. This is a long-tailed evaluation problem. We therefore study a simpler but still central object: the research question (RQ). A good RQ comes before a good idea. It asks what should be studied before a method is proposed.

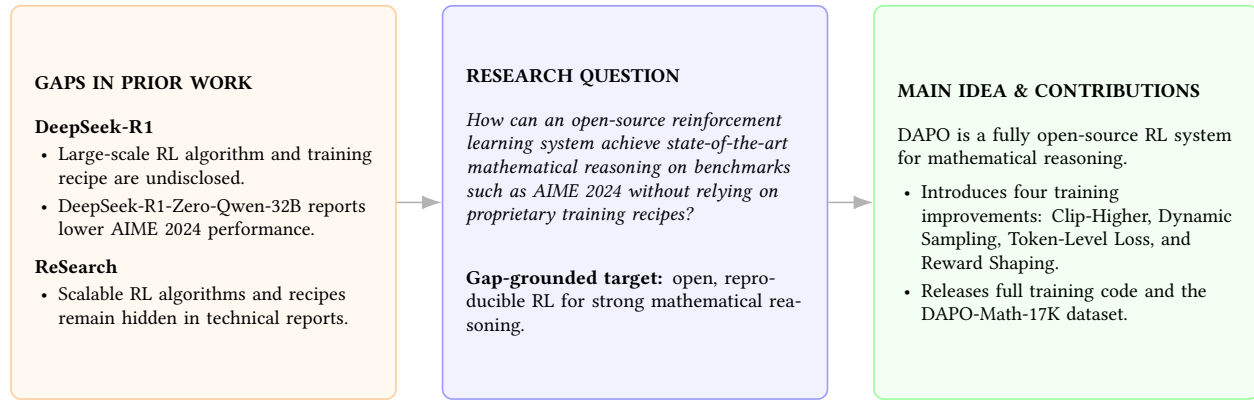
\begin{figure*}[t]
\centering
\resizebox{\textwidth}{!}{
\begin{tikzpicture}[
  node distance=8mm, 
  box/.style={
    draw,
    thick,
    rounded corners=5pt,
    text width=6cm, 
    align=left,
    inner sep=12pt,
    font=\small,
    minimum height=7cm 
  },
  arrow/.style={
    -{Latex[length=3.5mm, width=2.5mm]},
    thick,
    gray!60
  }
]

\node[box, fill=orange!5, draw=orange!40] (gaps) {
  \textbf{GAPS IN PRIOR WORK}\\[1em]
  \textbf{DeepSeek-R1}
  \begin{itemize}[leftmargin=1.2em, itemsep=0.2em, topsep=0.2em, parsep=0pt]
    \item Large-scale RL algorithm and training recipe are undisclosed.
    \item DeepSeek-R1-Zero-Qwen-32B reports lower AIME 2024 performance.
  \end{itemize}\vspace{0.5em}
  
  \textbf{ReSearch}
  \begin{itemize}[leftmargin=1.2em, itemsep=0.2em, topsep=0.2em, parsep=0pt]
    \item Scalable RL algorithms and recipes remain hidden in technical reports.
  \end{itemize}
};

\node[box, fill=blue!5, draw=blue!40, right=of gaps] (rq) {
  \textbf{RESEARCH QUESTION}\\[1em]
  \textit{How can an open-source reinforcement learning system achieve state-of-the-art mathematical reasoning on benchmarks such as AIME 2024 without relying on proprietary training recipes?}\\[2em]
  
  \textbf{Gap-grounded target:} open, reproducible RL for strong mathematical reasoning.
};

\node[box, fill=green!5, draw=green!40, right=of rq] (idea) {
  \textbf{MAIN IDEA \& CONTRIBUTIONS}\\[1em]
  DAPO is a fully open-source RL system for mathematical reasoning.\\[0.5em]
  
  \begin{itemize}[leftmargin=1.2em, itemsep=0.2em, topsep=0.2em, parsep=0pt]
    \item Introduces four training improvements: Clip-Higher, Dynamic Sampling, Token-Level Loss, and Reward Shaping.
    \item Releases full training code and the DAPO-Math-17K dataset.
  \end{itemize}
};

\draw[arrow] (gaps) -- (rq);
\draw[arrow] (rq) -- (idea);

\end{tikzpicture}%
}
\caption{Example of gap-grounded RQ formulation. Prior-work gaps motivate a research question, which is then addressed by the paper's concrete idea and contributions.}
\label{fig:dapo-example}
\end{figure*}

Classical accounts of research also place question formulation before the final argument. \citet{Boothetal2008} describe research as a movement from \emph{topic} to \emph{question} to \emph{problem} to \emph{argument}. In practice, however, gaps, questions, and ideas are often mixed together. Figure~\ref{fig:dapo-example} shows the difference:
\begin{itemize}
    \item A \emph{gap} is a missing piece in the literature, such as an unanswered question, an unresolved disagreement, or an untested assumption.
    \item A \emph{research question} is the question that opens a space for investigation. It does not yet commit to a method or an answer.
    \item An \emph{idea} is a proposed method, framework, or approach for answering one or more RQs.
\end{itemize}

Existing benchmarks for LLM-driven scientific ideation usually start from a human-provided topic, scope, or question. IdeaBench~\cite{Guoetal2024} uses curated background papers. ResearchBench~\cite{Liuetal2025} and \citet{Sietal2024} evaluate later research stages, but they still depend on a predefined question or topic. These settings are useful, but they bypass an earlier step: can a model infer a meaningful RQ from prior literature, and can LLM judges reliably assess its novelty?

We introduce \dataset{} to study this question. For each target paper, we collect influential cited works and local citation contexts. We then extract the gaps used by the authors and reconstruct the RQs addressed by the paper. We call these \emph{author-anchored RQs}. They are used as reference RQs. They are not the only valid RQs that could be asked from the same background. A background can support many good questions. The author-anchored RQ is useful because it gives a human reference point: it was pursued by real authors and led to a real paper.

Using this setup, we test the reliability of LLM novelty judgments. We compare model-generated RQs against author-anchored RQs under three views: standalone LLM scoring, comparative LLM scoring, and human expert evaluation. We also study the scope of generated RQs. This matters because an RQ can be polished and gap-shaped, but still be too narrow, source-bound, or diagnostic to open a strong research direction.

Our study asks the following questions:
\begin{itemize}
    \item \textbf{Novelty judging:} Do LLM judges give stable novelty judgments for scientific RQs?
    \item \textbf{Reference comparison:} Do model-generated RQs appear more novel than author-anchored RQs under standalone and comparative evaluation?
    \item \textbf{Human alignment:} Do LLM judges agree with human experts when they compare the same RQs?
    \item \textbf{Scope:} Are LLM-generated RQs broad enough, or are they narrow and source-bound?
    \item \textbf{Author-anchor overlap:} Do models reproduce the author-pursued direction, or do they mainly generate alternatives?
\end{itemize}

Our findings are as follows. First, LLM judges assigned high novelty scores to model-generated RQs in the standalone setting and preferred them over author-anchored RQs. Second, comparative judging amplified this, as shown by the sharp increase in win rates. Third, human experts preferred author-anchored RQs far more often than LLM judges did. Fourth, experts found that model generated RQs are narrow or source-bound. When we evaluated narrowness explicitly, the results aligned with experts' non-obviousness judgments. This suggests that narrowness is an important dimension missing from current novelty evaluations by LLMs. Finally, low author-anchor overlap should not be read as failure by itself. It mainly shows that models often generate alternatives rather than the direction pursued by the paper. The larger problem is whether those alternatives are judged reliably.

\section{Related Work}

\paragraph{Autonomous Scientific Discovery.} Automated scientific discovery has evolved from early rule-based systems \cite{Langley1987, Buchanan1984, Lenat1983} and domain-specific machine learning \cite{King2009, Schmidt2009, Jumper2021, Merchant2023} to modern foundation models. Recent systems frame LLMs as \emph{AI scientists} that can propose ideas, write code, run experiments, and draft papers~\cite{Luetal2024, Yamadaetal2025, Beeletal2025}. Other systems add multiple agents, web search, or critique loops~\cite{Baeketal2024, DeepResearchBench2025, GoogleERA2025}. These systems usually begin at ideation. They often assume that the research question or topic is already given.

\paragraph{LLM Ideation and Research Benchmarks.} Existing benchmarks mainly evaluate downstream scientific artifacts. IdeaBench~\cite{Guoetal2024}, AI Idea Bench~\cite{Qiuetal2025}, and LiveIdeaBench~\cite{Ruanetal2024} evaluate generated ideas from curated papers or topics. ResearchBench~\cite{Liuetal2025} and DiscoveryBench~\cite{Majumderetal2024} evaluate hypothesis composition and data-driven reasoning. These tasks are important, but full idea evaluation is hard because it can require judging feasibility, method quality, experiments, and impact. RQ generation is a cleaner upstream testbed. It still requires novelty judgment, but it does not require judging a full method.

\paragraph{LLM-as-Judge for Scientific Novelty.} LLM-as-judge is widely used because open-ended generations are hard to score automatically. But scientific novelty is different from many standard evaluation tasks. It is high-level, partly subjective, and sensitive to framing. Even human peer review can show disagreement, so a single novelty score should not be treated as absolute. Still, a useful judge should show stable trends across evaluation formats and should not sharply disagree with expert humans. Our work studies this reliability question in the RQ setting.

\paragraph{Research Question Formulation.} Question formulation is a central part of research. A research question is not the same as a topic or a method. It turns a gap in prior work into an object of inquiry. We use this distinction to build \dataset{} and to test whether LLMs and LLM-judges can handle the pre-ideation stage of scientific work.

\section{Formulation}
Let \(P=\{p_1,\ldots,p_m\}\) be the background papers given to the model. Let \(q\) be a research question. We view RQ generation as a two-step process. First, the model infers grounded gaps from the papers. Then it writes an RQ from those gaps.

\[
P \rightarrow \mathcal{G} \rightarrow q,
\]
where \(\mathcal{G}=\{g_1,\ldots,g_k\}\) is the set of inferred gaps. This simple view is useful because an RQ should not be a free-form idea. It should be grounded in prior work. At the same time, the model may use its broader scientific knowledge when it connects the papers and turns gaps into questions.

\section{Benchmark Dataset Creation}

Let \(M\) denote a target paper. We first extract a set of influential cited papers \(\mathcal{C}=\{c_1,\ldots,c_n\}\) and the target paper's main idea and contributions \(I\). For each influential citation \(c_i\), we collect its local citation context \(x_i\), consisting of the citation sentence in \(M\) and its surrounding sentences. Let \(\mathcal{X}=\{x_1,\ldots,x_n\}\). We then infer grounded gaps \(\mathcal{G}\) that are both linked to the cited works and addressed by the target paper's idea and contributions. Finally, we derive the research question \(q\) from these grounded gaps:

\[
\mathcal{C} \sim p_{\phi}(\mathcal{C}\mid M), \qquad
I \sim p_{\phi}(I\mid M),
\]
\[
\mathcal{X}=f_{\mathrm{ctx}}(M,\mathcal{C}),
\]
\[
\mathcal{G} \sim p_{\phi}(\mathcal{G}\mid \mathcal{C},\mathcal{X},I,M), \qquad q \sim p_{\phi}(q\mid \mathcal{G},I,M).
\]

\begin{figure*}
    \centering
    \includegraphics[width=0.9\linewidth]{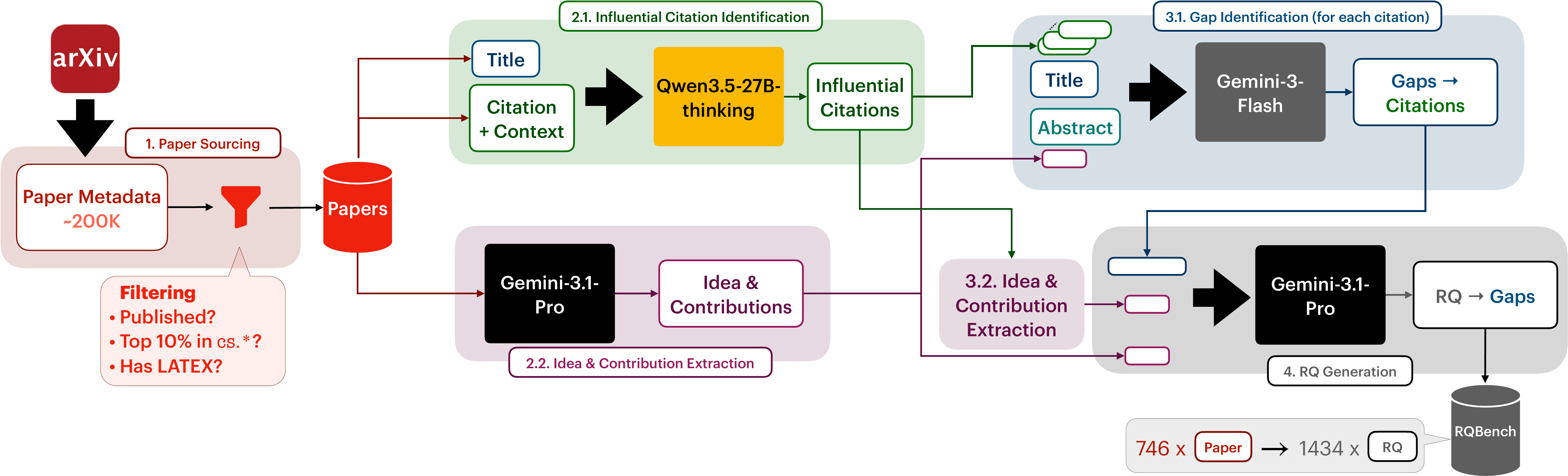}
    \caption{A high-level overview of the benchmark dataset-creation pipeline.}
    \label{fig:pipeline}
\end{figure*}

Here, \(\phi\) denotes the LLM-assisted extraction and construction pipeline as shown in \Cref{fig:pipeline}. The variable \(\mathcal{G}\) ensures that the resulting research question is not an unconstrained idea, but is grounded in the way the target paper uses influential prior work to motivate its own contributions.

The following sections detail our LLM-assisted pipeline that curates recent arXiv computer science papers to extract influential citations, core contributions, and grounded research gaps. These elements are ultimately synthesized into a comprehensive benchmark dataset comprising 1,434 research questions derived from 746 source papers.

\paragraph{Paper Sourcing.} The dataset is built from recent computer science papers (\texttt{cs.AI, cs.CL,} $\dots$) on \url{https://arxiv.org}. In our current version, we consider about $\sim$200K papers published between \texttt{01-01-2025} and \texttt{10-4-2026}. Restricting the benchmark to papers that are certainly unseen by all LLMs would reduce scale and coverage. We therefore use recent papers and report author-anchor overlap as a diagnostic. If a model memorizes a paper or its RQ, this score should be high. If the score is low, the model may be generating a different valid question, or it may simply not recover the author-pursued direction.

\paragraph{Filtering.} The metadata of the above papers are retrieved with arxiv API, and we keep only the top 10\% papers with available \LaTeX{} source from each subfield of computer science (\texttt{cs.*}) based on their citation counts as a proxy for paper quality. Among these papers, we subsequently keep only those published in a conference or a journal, as indexed by Semantic Scholar\footnote{\url{https://www.semanticscholar.org}}; papers not indexed by Semantic Scholar are excluded.

\paragraph{Influential Citation Identification.}

Good research papers are typically well grounded in existing literature, which shapes the research questions posed, the ideas proposed, and the contributions presented. To identify the most influential cited works within a paper, we prompt \texttt{qwen3.5-27b-thinking}~\cite{qwen3.5} with the following context: 1) the cited paper's \emph{title}, 2) citation frequency of the cited work in the paper, 3) each \emph{section name} in which the citation appears---helping to discount loosely related references (e.g., in the introduction or related work), 3) each citation of the work with its \emph{local context}, consisting of the parent sentence and three preceding and following sentences, to capture how central the cited work is to the paper.

The model then classifies each citation as an \emph{inspiration}, a \emph{competitor/baseline}, or as irrelevant, minimally influential, or a canonical reference (e.g., the original Transformer paper~\cite{vaswani2017attention}). Citations categorized as \emph{inspiration} or \emph{competitor} are subsequently treated as influential.

\paragraph{Idea and Contribution Extraction.}

On the other hand, the key ideas and contributions of a paper are intended to address its underlying research questions, whether stated explicitly or implicitly. Accordingly, we extract these ideas and contributions from the paper content using \texttt{gemini-3.1-pro}\footnote{\url{https://ai.google.dev/gemini-api/docs/models/gemini-3.1-pro-preview}} to inform and guide the subsequent extraction of research questions. We employ a frontier LLM for this step because accurately capturing these attributes is central, and its outputs are consistently higher quality than those of open-source alternatives such as \texttt{qwen3.5}.

\paragraph{Gap Identification.}

New research ideas are typically proposed to address some gaps within a field. These gaps may not necessarily originate from the prior works in the domain. We are particularly interested in contributions that are explicitly motivated by previous research, as such the contributions yield gaps that are grounded in the literature and directly raise the research questions addressed by the paper. Thus to inform and guide the subsequent research question extraction, we retrieve the gaps that are both identified and addressed by the authors, by prompting \texttt{gemini-3-flash}\footnote{\url{https://ai.google.dev/gemini-api/docs/models/gemini-3-flash-preview}} with the pre-extracted ideas and main contributions of the paper along with the influential citations with their in-paper \emph{local contexts}. Notably in this step, the use of a lighter LLM is motivated by the smaller context size and the relatively simpler and extractive nature of the task, as the gaps are identified for each influential citation through independent prompts. This is aimed to isolate the individual role of the cited works in the ideas and contributions of the paper.

\paragraph{Research Question Generation.}

Finally, research question generation is conceived as a bridge between the gaps in the literature, which are informed by the influential works cited in the paper, and the ideas and contributions of the paper that often address the research questions and, in turn, the gaps. Accordingly, \texttt{gemini-3.1-pro} is prompted with the gaps and the ideas and contributions of the influential citations to obtain the author-anchored research questions.

\begin{table}[t]
\centering
\small
\setlength{\tabcolsep}{4pt}
\renewcommand{\arraystretch}{1.05}
\begin{tabular}{lr}
\toprule
\textbf{Statistic} & \textbf{Value} \\
\midrule
\rowcolor{gray!15} \multicolumn{2}{c}{\textit{Scale}} \\
Research questions (RQs)              & 1{,}434 \\
Source papers                         & 746 \\
Unique cited papers                   & 1{,}375 \\
RQ--citation links                    & 2{,}464 \\
Source--predecessor links             & 2{,}098 \\
Grounded gap statements               & 3{,}151 \\
arXiv CS sub-fields                   & 13 \\
\rowcolor{gray!15} \multicolumn{2}{c}{\textit{Per source paper}} \\
RQs: Mean\,/\,Median\,/\,Max          & 1.92\,/\,2\,/\,4 \\
Papers with 1\,/\,2\,/\,3\,/\,4 RQs   & 229\,/\,361\,/\,141\,/\,15 \\
Influential citations: & \\
\quad Mean\,/\,Median\,/\,Max         & 2.81\,/\,3\,/\,9 \\
\rowcolor{gray!15} \multicolumn{2}{c}{\textit{Cited refs per RQ}} \\
Mean\,/\,Median\,/\,Max               & 1.72\,/\,1\,/\,7 \\
\% grounded in 1\,/\,2\,/\,$\geq$3 ref(s) & 52.4\,/\,30.6\,/\,16.9 \\
\rowcolor{gray!15} \multicolumn{2}{c}{\textit{Gaps}} \\
Per RQ: Mean\,/\,Median\,/\,Max       & 2.20\,/\,2\,/\,11 \\
Per Influential citation:   & \\
\quad Mean\,/\,Median\,/\,Max         & 1.50\,/\,1\,/\,8 \\
\rowcolor{gray!15} \multicolumn{2}{c}{\textit{Cited-paper reuse}} \\
RQs per cited paper (mean\,/\,max)    & 1.79\,/\,87 \\
\% used by exactly one RQ             & 67.3 \\
\rowcolor{gray!15} \multicolumn{2}{c}{\textit{Question text}} \\
Length in words (mean\,/\,max)        & 24.7\,/\,50 \\
\rowcolor{gray!15} \multicolumn{2}{c}{\textit{Source-paper metadata}} \\
Contributions (mean\,/\,max)          & 3.66\,/\,6 \\
Methodological\,/\,Appl.\,/\,Comb.    & 731\,/\,13\,/\,2 \\
\bottomrule
\end{tabular}
\caption{Summary statistics of \dataset{}.}
\label{tab:dataset-stats}
\end{table}

\paragraph{Final Dataset Details.}
The final dataset \dataset{} comprises 1{,}434 research questions grounded in 746 arXiv Computer Science papers spanning 13 sub-fields, with 2{,}464 RQ-citation links to 1{,}375 unique cited papers and 3{,}151 distinct gap statements in total. Each record contains the RQ text, the source paper's arXiv ID, sub-field, novelty type, problem statement, and main-idea headline with its stated contributions; for every cited paper referenced by an RQ, we additionally release the section-wise extracted body text. RQs whose cited papers lacked retrievable section text were filtered out (27 RQs dropped). Per-statistic summaries are given in \Cref{tab:dataset-stats}; the sub-field breakdown is in \Cref{tab:dataset-subfields}.

\paragraph{Author-Anchored RQs as Reference Points.}
We use the reconstructed RQs as author-anchored reference points. They are not the only valid questions for the same background. A set of papers can support many plausible RQs. The value of the author-anchored RQ is that it gives a concrete human reference: it was pursued by real authors and led to a published paper. This lets us test whether LLM judges evaluate model-generated alternatives consistently against the same reference.

\paragraph{Verification.}
With two checks, we verified the plausibility of the author-anchored LLM-extracted RQs being addressed by the source papers. Firstly, we asked \texttt{gemini-3.1-pro} to judge each RQ against the full text of its source paper. Secondly, human judges performed the same check on 50 randomly sampled RQs. \texttt{gemini-3.1-pro} approved 98\% of the RQs, while human judges approved 100\% of the sampled RQs. These results suggest that the extracted RQs are faithful to the source papers and support the quality of \dataset{}.

\section{Evaluation Setup}
\label{sec:evaluation-setup}
\subsection{Novelty Scores}

Let \(Q_i^m=\{q_{i,1}^m,\ldots,q_{i,5}^m\}\) be the five RQs generated by model \(m\) for instance \(i\). Let \(q_i^\text{GT}\) be the author-anchored reference RQ.

We score each RQ on three metrics:
\[
\mathcal{K}=\{\text{Orig},\text{Gap},\text{NonObs}\}.
\]
These denote originality, gap addressing, and non-obviousness. Each metric receives a score from 0 to 3. The rubric is given in \Cref{app:novelty}. We report results for each metric separately. We do not define novelty as the sum of the three scores.

\paragraph{Scoring settings.}
We use two scoring settings:

In \textbf{standalone scoring}, each RQ is scored independently, without being shown alongside the author-anchored reference RQ or other generated RQs. This setting measures the absolute score assigned to each RQ under the rubric.

In \textbf{comparative scoring}, the author-anchored reference RQ and the five model-generated RQs are scored in the same context. This setting measures whether the judge prefers at least one model-generated RQ over the reference RQ for a given metric.

\paragraph{Metric-level Pass@5 comparison.}
For both scoring settings, we compare the best model-generated RQ among the five candidates against the author-anchored reference RQ. For a scoring setting \(c\in\{\mathrm{stand},\mathrm{comp}\}\), instance \(i\), model \(m\), metric \(k\), and judge \(j\), let
\[
s_{i,k}^{(c,j)}(q)
\]
denote the score assigned to RQ \(q\). We define the model-reference score difference as
\[
d_{i,k}^{(c,j)}
=
\max_{q\in Q_i^m}s_{i,k}^{(c,j)}(q)
-
s_{i,k}^{(c,j)}(q_i^\text{GT}).
\]
Thus, \(d_{i,k}^{(c,j)}>0\) means that judge \(j\) assigned at least one generated RQ a higher score than the reference RQ for metric \(k\). This is a metric-level Pass@5 comparison.

\paragraph{Per-judge win, tie, and lose rates.}
For each judge separately, we assign the outcome
\[
r_{i,k}^{(c,j)}
=
\begin{cases}
\text{win}, & d_{i,k}^{(c,j)}>0,\\
\text{tie}, & d_{i,k}^{(c,j)}=0,\\
\text{lose}, & d_{i,k}^{(c,j)}<0.
\end{cases}
\]
The corresponding per-judge rates are
\[
\mathrm{WinRate}_{m,k}^{(c,j)}
=
\frac{1}{N}\sum_{i=1}^{N}
\mathbbm{1}[r_{i,k}^{(c,j)}=\text{win}],
\]
\[
\mathrm{TieRate}_{m,k}^{(c,j)}
=
\frac{1}{N}\sum_{i=1}^{N}
\mathbbm{1}[r_{i,k}^{(c,j)}=\text{tie}],
\]
\[
\mathrm{LoseRate}_{m,k}^{(c,j)}
=
1-\mathrm{WinRate}_{m,k}^{(c,j)}
-\mathrm{TieRate}_{m,k}^{(c,j)}.
\]

\paragraph{Combined-judge win, tie, and lose rates.}
For analyses that require agreement across judges, we also define a combined-judge outcome. With two judges \(j\in\{1,2\}\), the combined outcome is
\[
r_{i,k}^{(c)}
=
\begin{cases}
\text{win}, & d_{i,k}^{(c,1)}>0 \ \text{and}\ d_{i,k}^{(c,2)}>0,\\
\text{tie}, & d_{i,k}^{(c,1)}\geq 0,\ d_{i,k}^{(c,2)}\geq 0, \\
& \quad \text{and not both are positive},\\
\text{lose}, & \text{otherwise}.
\end{cases}
\]
Thus, a combined win requires both judges to score the best model-generated RQ above the reference RQ. A combined tie means that neither judge scores the model below the reference, but the condition for a win is not met. All remaining cases are counted as losses.

The combined-judge rates are
\[
\mathrm{WinRate}_{m,k}^{(c)}
=
\frac{1}{N}\sum_{i=1}^{N}
\mathbbm{1}[r_{i,k}^{(c)}=\text{win}],
\]
\[
\mathrm{TieRate}_{m,k}^{(c)}
=
\frac{1}{N}\sum_{i=1}^{N}
\mathbbm{1}[r_{i,k}^{(c)}=\text{tie}],
\]
\[
\mathrm{LoseRate}_{m,k}^{(c)}
=
1-\mathrm{WinRate}_{m,k}^{(c)}
-\mathrm{TieRate}_{m,k}^{(c)}.
\]

For comparative scoring, we report only the win, tie, and lose rates defined above. We do not report average comparative scores, because comparative scoring is intended to measure relative preference between generated and reference RQs within the same judging context.

\paragraph{Standalone score summaries.}
For standalone scoring, we additionally report absolute score summaries. For judge \(j\), the mean standalone score of the author-anchored reference RQs is
\[
\mathrm{GT}_{k}^{(\mathrm{stand},j)}
=
\frac{1}{N}\sum_{i=1}^{N}
s_{i,k}^{(\mathrm{stand},j)}(q_i^\text{GT}).
\]
For model \(m\), the standalone Best@5 score is
\[
\mathrm{Best@5}_{m,k}^{(\mathrm{stand},j)}
=
\frac{1}{N}\sum_{i=1}^{N}
\max_{q\in Q_i^m}
s_{i,k}^{(\mathrm{stand},j)}(q).
\]
These score summaries are reported only for standalone scoring, alongside the standalone win, tie, and lose rates.

\subsection{Scope and Narrowness}
\label{subsec:scope-metrics}

Novelty is not only about whether a question mentions a gap. A question can be gap-shaped but still be narrow. We therefore study scope as a separate dimension. We focus on two signals. \emph{Source-boundedness} asks whether the RQ stays too close to one background paper, dataset, metric, or ablation. \emph{Diagnostic framing} asks whether the RQ mainly asks for an analysis or failure diagnosis instead of opening a broader research direction. Diagnostic RQs are not always bad. They become a problem when they are mistaken for strong novelty.

We score each RQ on two metrics:
\[
\mathcal{N}=\{\text{SrcBound},\text{DiagFrame}\}.
\]
These denote source-boundedness, and diagnostic framing. Each metric receives a score from 0 to 3; the lower the score is the broader the RQ. The rubric is given in \Cref{app:narrowness}. We report results for each metric separately. The standalone and comparative scoring setups remain identical to the novelty scoring setups.

\subsection{Author-Anchor Overlap}

Finally, we report author-anchor overlap. It measures whether the generated RQ follows the direction pursued by the source paper:
\[
\mathrm{Overlap@5}_m=\frac{1}{N}\sum_i \max_{q\in Q_i^m}\mathrm{Match}(q,q_i^\text{GT}).
\]
Here, \(\mathrm{Match}\in\{0,1,2,3,4\}\). A score of 1 means ``same broad topic, different question.'' A score of 2 means ``substantial overlap, but different scope.'' This metric is not an absolute quality score. Low overlap can mean that the model generated a different question from the same background.

\section{Experiments}

\begin{table*}[t]
\centering
\small
\setlength{\tabcolsep}{5pt}

\definecolor{rqband}{RGB}{222,235,247}   

\begin{tcolorbox}[
    enhanced,
    colback=white,
    colframe=gray!35,
    boxrule=0.4pt,
    arc=2mm,
    left=2mm, right=2mm, top=2mm, bottom=2mm,
    drop shadow=black!12,
    width=\textwidth
]
\centering
\begin{tabular}{@{} c @{\hspace{6mm}} !{\color{gray!45}\vrule width 0.5pt} @{\hspace{6mm}} l @{}}

{\renewcommand{\arraystretch}{1.45}%
\begin{tabular}[t]{lr}
\toprule
Model & $\mathrm{Overlap@5}_m$ \\
\midrule
\texttt{qwen3-30b}       & 1.00 \\
\texttt{gpt-oss-20b}     & 1.16 \\
\texttt{deepseek-v4-pro} & 1.19 \\
\texttt{gemma-4-31b-it}  & 1.05 \\
\texttt{gemini-3.1-pro}  & 1.07 \\
\texttt{gpt-5.5}         & \best{1.33} \\
\bottomrule
\end{tabular}}

&

{\renewcommand{\arraystretch}{1.08}%
\begin{tabular}[t]{@{} c @{\hspace{5pt}} >{\raggedright\arraybackslash}p{0.46\textwidth} @{}}
\toprule
\multicolumn{2}{@{}l}{\textbf{Scoring rubric}\; (candidate RQ vs.\ GT)} \\
\midrule
\textbf{0} & \textbf{Unrelated.} No topic overlap. \\[3pt]
\rowcolor{rqband}
\textbf{1} & \textbf{Same topic, different question.} Differs in object of  inquiry, answer type, or scope. \\
\rowcolor{rqband}
\textbf{2} & \textbf{Substantial overlap.} Same object of inquiry, but the scope or angle differs noticeably. \\[3pt]
\textbf{3} & \textbf{Semantic equivalent.} A paraphrase of GT. \\[3pt]
\textbf{4} & \textbf{Subsumes GT.} Strictly more general; scope fully contains GT. \\
\bottomrule
\end{tabular}}

\end{tabular}

\vspace{1mm}
{\footnotesize\color{black!70}\raggedright
Shaded rows mark the \colorbox{rqband}{\,1--2\,} band that all evaluated models occupy.\par}
\end{tcolorbox}

\caption{Author-anchor overlap. $\mathrm{Overlap@5}_m$ measures how close the best
generated RQ is to the author-pursued direction, scored against the 0--4 rubric (right).
It is a diagnostic score, not an absolute quality score.}
\label{tab:overlap-results}
\end{table*}

\begin{table*}[p]
\centering
\renewcommand{\arraystretch}{1.15}
\begin{subtable}[t]{0.48\linewidth}
\vspace{0pt}
\centering
\setlength{\tabcolsep}{4pt}
\begin{tcolorbox}[enhanced,colback=white,colframe=gray!35,boxrule=0.4pt,arc=2mm,left=1mm,right=1mm,top=1mm,bottom=1mm,drop shadow=black!12]
\resizebox{\linewidth}{!}{
\begin{tabular}{l >{\centering\arraybackslash}p{24mm} >{\centering\arraybackslash}p{24mm} >{\centering\arraybackslash}p{24mm}}
\toprule
\multirow{2}{*}{\textbf{Model}} & \multicolumn{3}{c}{\shortstack{\textbf{Novelty Score}\\ \textbf{(mean\,$\pm$\,std)}}} \\
\cmidrule(lr){2-4}
 & Non-Obv. & Orig. & Gap-Addr. \\
\midrule
\texttt{qwen3-30b-a3b-thinking-2507} & \negp{1.34\,{\scriptsize$\pm$0.64}} & \negp{1.88\,{\scriptsize$\pm$0.63}} & \pos{2.72\,{\scriptsize$\pm$0.45}} \\
\texttt{gpt-oss-20b} & \pos{1.60\,{\scriptsize$\pm$0.57}} & \pos{2.16\,{\scriptsize$\pm$0.63}} & \pos{2.79\,{\scriptsize$\pm$0.41}} \\
\texttt{gemma-4-31b-it} & \pos{1.81\,{\scriptsize$\pm$0.53}} & \pos{2.37\,{\scriptsize$\pm$0.62}} & \pos{2.89\,{\scriptsize$\pm$0.32}} \\
\texttt{deepseek-v4-pro} & \pos{1.78\,{\scriptsize$\pm$0.51}} & \pos{2.26\,{\scriptsize$\pm$0.61}} & \pos{2.94\,{\scriptsize$\pm$0.23}} \\
\midrule
\texttt{gemini-3.1-pro} & \pos{1.81\,{\scriptsize$\pm$0.50}} & \pos{2.39\,{\scriptsize$\pm$0.65}} & \pos{2.96\,{\scriptsize$\pm$0.20}} \\
\texttt{gpt-5.5} & \best{1.98\,{\scriptsize$\pm$0.25}} & \best{2.67\,{\scriptsize$\pm$0.48}} & \best{2.98\,{\scriptsize$\pm$0.12}} \\
\midrule
\rowcolor{gtgray}
\textbf{Author-Anchored RQ} & \textbf{1.41\,{\scriptsize$\pm$0.87}} & \textbf{1.90\,{\scriptsize$\pm$1.00}} & \textbf{2.01\,{\scriptsize$\pm$0.97}} \\
\bottomrule
\end{tabular}}
\end{tcolorbox}
\caption{Judge: \texttt{gemini-3.1-pro} -- mean\,\texorpdfstring{$\pm$}{+/-}\,std of each model's best per-dimension RQ score.}
\label{tab:nov-mean-gemini-31-pro}
\end{subtable}%
\hfill%
\begin{subtable}[t]{0.48\linewidth}
\vspace{0pt}
\centering
\setlength{\tabcolsep}{4pt}
\begin{tcolorbox}[enhanced,colback=white,colframe=gray!35,boxrule=0.4pt,arc=2mm,left=1mm,right=1mm,top=1mm,bottom=1mm,drop shadow=black!12]
\resizebox{\linewidth}{!}{
\begin{tabular}{l >{\centering\arraybackslash}p{24mm} >{\centering\arraybackslash}p{24mm} >{\centering\arraybackslash}p{24mm}}
\toprule
\multirow{2}{*}{\textbf{Model}} & \multicolumn{3}{c}{\shortstack{\textbf{Win\,(Tie) Rate vs.}\\ \textbf{Author-Anchored RQ (\%)}}} \\
\cmidrule(lr){2-4}
 & Non-Obv. & Orig. & Gap-Addr. \\
\midrule
\texttt{qwen3-30b-a3b-thinking-2507} & \negp{27.5\,(37.0)} & \negp{30.7\,(32.1)} & \pos{51.3\,(42.4)} \\
\texttt{gpt-oss-20b} & \negp{33.7\,(42.5)} & \negp{38.2\,(36.0)} & \pos{54.5\,(40.4)} \\
\texttt{gemma-4-31b-it} & \negp{40.0\,(44.4)} & \negp{45.7\,(33.9)} & \pos{59.8\,(37.0)} \\
\texttt{deepseek-v4-pro} & \negp{39.4\,(44.4)} & \negp{43.0\,(32.1)} & \pos{62.4\,(36.5)} \\
\midrule
\texttt{gemini-3.1-pro} & \negp{39.4\,(45.9)} & \negp{46.6\,(34.1)} & \pos{63.5\,(35.7)} \\
\texttt{gpt-5.5} & \best{44.5\,(48.4)} & \best{56.1\,(33.8)} & \best{64.8\,(34.9)} \\
\midrule
\rowcolor{gtgray}
\textbf{Author-Anchored RQ} & \textbf{--} & \textbf{--} & \textbf{--} \\
\bottomrule
\end{tabular}}
\end{tcolorbox}
\caption{Judge: \texttt{gemini-3.1-pro} -- win\,(tie) rate vs.\ the author-anchored reference.}
\label{tab:nov-win-gemini-31-pro}
\end{subtable}
\par\medskip
\begin{subtable}[t]{0.48\linewidth}
\vspace{0pt}
\centering
\setlength{\tabcolsep}{4pt}
\begin{tcolorbox}[enhanced,colback=white,colframe=gray!35,boxrule=0.4pt,arc=2mm,left=1mm,right=1mm,top=1mm,bottom=1mm,drop shadow=black!12]
\resizebox{\linewidth}{!}{
\begin{tabular}{l >{\centering\arraybackslash}p{24mm} >{\centering\arraybackslash}p{24mm} >{\centering\arraybackslash}p{24mm}}
\toprule
\multirow{2}{*}{\textbf{Model}} & \multicolumn{3}{c}{\shortstack{\textbf{Novelty Score}\\ \textbf{(mean\,$\pm$\,std)}}} \\
\cmidrule(lr){2-4}
 & Non-Obv. & Orig. & Gap-Addr. \\
\midrule
\texttt{qwen3-30b-a3b-thinking-2507} & \pos{1.93\,{\scriptsize$\pm$0.43}} & \pos{2.49\,{\scriptsize$\pm$0.54}} & \pos{2.81\,{\scriptsize$\pm$0.39}} \\
\texttt{gpt-oss-20b} & \pos{2.03\,{\scriptsize$\pm$0.40}} & \pos{2.67\,{\scriptsize$\pm$0.48}} & \pos{2.87\,{\scriptsize$\pm$0.34}} \\
\texttt{gemma-4-31b-it} & \pos{2.02\,{\scriptsize$\pm$0.38}} & \pos{2.69\,{\scriptsize$\pm$0.48}} & \pos{2.90\,{\scriptsize$\pm$0.30}} \\
\texttt{deepseek-v4-pro} & \pos{2.06\,{\scriptsize$\pm$0.36}} & \pos{2.67\,{\scriptsize$\pm$0.47}} & \pos{2.92\,{\scriptsize$\pm$0.27}} \\
\midrule
\texttt{gemini-3.1-pro} & \pos{2.08\,{\scriptsize$\pm$0.39}} & \pos{2.72\,{\scriptsize$\pm$0.46}} & \pos{2.95\,{\scriptsize$\pm$0.22}} \\
\texttt{gpt-5.5} & \best{2.18\,{\scriptsize$\pm$0.38}} & \best{2.91\,{\scriptsize$\pm$0.28}} & \best{2.98\,{\scriptsize$\pm$0.13}} \\
\midrule
\rowcolor{gtgray}
\textbf{Author-Anchored RQ} & \textbf{1.68\,{\scriptsize$\pm$0.77}} & \textbf{2.17\,{\scriptsize$\pm$0.84}} & \textbf{1.96\,{\scriptsize$\pm$0.84}} \\
\bottomrule
\end{tabular}}
\end{tcolorbox}
\caption{Judge: \texttt{deepseek-v4-pro} -- mean\,\texorpdfstring{$\pm$}{+/-}\,std of each model's best per-dimension RQ score.}
\label{tab:nov-mean-deepseek-v4-pro}
\end{subtable}%
\hfill%
\begin{subtable}[t]{0.48\linewidth}
\vspace{0pt}
\centering
\setlength{\tabcolsep}{4pt}
\begin{tcolorbox}[enhanced,colback=white,colframe=gray!35,boxrule=0.4pt,arc=2mm,left=1mm,right=1mm,top=1mm,bottom=1mm,drop shadow=black!12]
\resizebox{\linewidth}{!}{
\begin{tabular}{l >{\centering\arraybackslash}p{24mm} >{\centering\arraybackslash}p{24mm} >{\centering\arraybackslash}p{24mm}}
\toprule
\multirow{2}{*}{\textbf{Model}} & \multicolumn{3}{c}{\shortstack{\textbf{Win\,(Tie) Rate vs.}\\ \textbf{Author-Anchored RQ (\%)}}} \\
\cmidrule(lr){2-4}
 & Non-Obv. & Orig. & Gap-Addr. \\
\midrule
\texttt{qwen3-30b-a3b-thinking-2507} & \negp{32.7\,(51.2)} & \negp{39.5\,(41.0)} & \pos{64.7\,(31.7)} \\
\texttt{gpt-oss-20b} & \negp{35.7\,(52.3)} & \negp{45.1\,(43.5)} & \pos{67.0\,(31.9)} \\
\texttt{gemma-4-31b-it} & \negp{35.9\,(51.6)} & \negp{48.1\,(39.3)} & \pos{70.6\,(27.7)} \\
\texttt{deepseek-v4-pro} & \negp{37.1\,(53.1)} & \negp{46.6\,(40.6)} & \pos{70.8\,(28.3)} \\
\midrule
\texttt{gemini-3.1-pro} & \negp{37.8\,(51.2)} & \negp{47.9\,(41.9)} & \pos{72.6\,(26.8)} \\
\texttt{gpt-5.5} & \best{43.3\,(49.1)} & \best{56.7\,(39.7)} & \best{74.5\,(25.3)} \\
\midrule
\rowcolor{gtgray}
\textbf{Author-Anchored RQ} & \textbf{--} & \textbf{--} & \textbf{--} \\
\bottomrule
\end{tabular}}
\end{tcolorbox}
\caption{Judge: \texttt{deepseek-v4-pro} -- win\,(tie) rate vs.\ the author-anchored reference.}
\label{tab:nov-win-deepseek-v4-pro}
\end{subtable}
\par\medskip
\begin{subtable}[t]{0.48\linewidth}
\vspace{0pt}
\centering
\setlength{\tabcolsep}{4pt}
\begin{tcolorbox}[enhanced,colback=white,colframe=gray!35,boxrule=0.4pt,arc=2mm,left=1mm,right=1mm,top=1mm,bottom=1mm,drop shadow=black!12]
\resizebox{\linewidth}{!}{
\begin{tabular}{l >{\centering\arraybackslash}p{24mm} >{\centering\arraybackslash}p{24mm} >{\centering\arraybackslash}p{24mm}}
\toprule
\multirow{2}{*}{\textbf{Model}} & \multicolumn{3}{c}{\shortstack{\textbf{Novelty Score}\\ \textbf{(mean\,$\pm$\,std)}}} \\
\cmidrule(lr){2-4}
 & Non-Obv. & Orig. & Gap-Addr. \\
\midrule
\texttt{qwen3-30b-a3b-thinking-2507} & \pos{1.64\,{\scriptsize$\pm$0.45}} & \pos{2.18\,{\scriptsize$\pm$0.50}} & \pos{2.77\,{\scriptsize$\pm$0.36}} \\
\texttt{gpt-oss-20b} & \pos{1.81\,{\scriptsize$\pm$0.40}} & \pos{2.41\,{\scriptsize$\pm$0.47}} & \pos{2.83\,{\scriptsize$\pm$0.30}} \\
\texttt{gemma-4-31b-it} & \pos{1.92\,{\scriptsize$\pm$0.38}} & \pos{2.53\,{\scriptsize$\pm$0.45}} & \pos{2.89\,{\scriptsize$\pm$0.25}} \\
\texttt{deepseek-v4-pro} & \pos{1.92\,{\scriptsize$\pm$0.35}} & \pos{2.47\,{\scriptsize$\pm$0.45}} & \pos{2.93\,{\scriptsize$\pm$0.19}} \\
\midrule
\texttt{gemini-3.1-pro} & \pos{1.95\,{\scriptsize$\pm$0.35}} & \pos{2.55\,{\scriptsize$\pm$0.46}} & \pos{2.95\,{\scriptsize$\pm$0.16}} \\
\texttt{gpt-5.5} & \best{2.08\,{\scriptsize$\pm$0.24}} & \best{2.79\,{\scriptsize$\pm$0.29}} & \best{2.98\,{\scriptsize$\pm$0.09}} \\
\midrule
\rowcolor{gtgray}
\textbf{Author-Anchored RQ} & \textbf{1.55\,{\scriptsize$\pm$0.73}} & \textbf{2.03\,{\scriptsize$\pm$0.84}} & \textbf{1.98\,{\scriptsize$\pm$0.79}} \\
\bottomrule
\end{tabular}}
\end{tcolorbox}
\caption{Two-Judge Combined -- mean\,\texorpdfstring{$\pm$}{+/-}\,std of each model's best per-dimension RQ score.}
\label{tab:nov-mean-two-judge-average}
\end{subtable}%
\hfill%
\begin{subtable}[t]{0.48\linewidth}
\vspace{0pt}
\centering
\setlength{\tabcolsep}{4pt}
\begin{tcolorbox}[enhanced,colback=white,colframe=gray!35,boxrule=0.4pt,arc=2mm,left=1mm,right=1mm,top=1mm,bottom=1mm,drop shadow=black!12]
\resizebox{\linewidth}{!}{
\begin{tabular}{l >{\centering\arraybackslash}p{24mm} >{\centering\arraybackslash}p{24mm} >{\centering\arraybackslash}p{24mm}}
\toprule
\multirow{2}{*}{\textbf{Model}} & \multicolumn{3}{c}{\shortstack{\textbf{Win\,(Tie) Rate vs.}\\ \textbf{Author-Anchored RQ (\%)}}} \\
\cmidrule(lr){2-4}
 & Non-Obv. & Orig. & Gap-Addr. \\
\midrule
\texttt{qwen3-30b-a3b-thinking-2507} & \negp{15.5\,(40.2)} & \negp{20.6\,(33.9)} & \negp{36.7\,(52.9)} \\
\texttt{gpt-oss-20b} & \negp{19.4\,(48.0)} & \negp{27.0\,(39.6)} & \negp{40.9\,(51.4)} \\
\texttt{gemma-4-31b-it} & \negp{22.3\,(52.9)} & \negp{29.3\,(41.0)} & \negp{46.1\,(48.1)} \\
\texttt{deepseek-v4-pro} & \negp{23.0\,(52.9)} & \negp{28.0\,(37.9)} & \negp{47.8\,(49.8)} \\
\midrule
\texttt{gemini-3.1-pro} & \negp{24.0\,(52.2)} & \negp{30.4\,(41.6)} & \pos{50.8\,(46.9)} \\
\texttt{gpt-5.5} & \best{27.2\,(59.1)} & \best{38.4\,(44.3)} & \best{53.2\,(45.7)} \\
\midrule
\rowcolor{gtgray}
\textbf{Author-Anchored RQ} & \textbf{--} & \textbf{--} & \textbf{--} \\
\bottomrule
\end{tabular}}
\end{tcolorbox}
\caption{Two-Judge Combined -- win\,(tie) rate vs.\ the author-anchored reference.}
\label{tab:nov-win-two-judge-average}
\end{subtable}

\caption{
\textbf{Standalone novelty assessment of generated research questions.}
Rows report scores from the \texttt{gemini-3.1-pro} judge, the \texttt{deepseek-v4-pro} judge, and the combined-judge evaluation. Scores are reported separately for non-obviousness, originality, and gap addressing. For each instance and metric, we compare the author-anchored reference RQ with the best-of-five model generated RQs for that metric, i.e., a metric-level Pass@5 comparison. \emph{Left}: mean\,\texorpdfstring{$\pm$}{+/-}\,std of the Best@5 standalone score. \emph{Right}: win (tie) rate against the reference. \pos{Green}/\negp{red} denotes scores above/below the reference on the left, and win rates above/below 50\% on the right; \best{bold} marks the best model in each column.
}
\label{tab:novelty-standalone-full}
\end{table*}

\begin{table*}[p]
\centering
\renewcommand{\arraystretch}{1.15}
\begin{subtable}[t]{0.48\linewidth}
\vspace{0pt}
\centering
\setlength{\tabcolsep}{4pt}
\begin{tcolorbox}[enhanced,colback=white,colframe=gray!35,boxrule=0.4pt,arc=2mm,left=1mm,right=1mm,top=1mm,bottom=1mm,drop shadow=black!12]
\resizebox{\linewidth}{!}{
\begin{tabular}{l >{\centering\arraybackslash}p{24mm} >{\centering\arraybackslash}p{24mm} >{\centering\arraybackslash}p{24mm}}
\toprule
\multirow{2}{*}{\textbf{Model}} & \multicolumn{3}{c}{\shortstack{\textbf{Win\,(Tie) Rate -- Standalone}\\ \textbf{vs.\ Author-Anchored RQ (\%)}}} \\
\cmidrule(lr){2-4}
 & Non-Obv. & Orig. & Gap-Addr. \\
\midrule
\texttt{qwen3-30b-a3b-thinking-2507} & \negp{27.5\,(37.0)} & \negp{30.7\,(32.1)} & \pos{51.3\,(42.4)} \\
\texttt{gpt-oss-20b} & \negp{33.7\,(42.5)} & \negp{38.2\,(36.0)} & \pos{54.5\,(40.4)} \\
\texttt{gemma-4-31b-it} & \negp{40.0\,(44.4)} & \negp{45.7\,(33.9)} & \pos{59.8\,(37.0)} \\
\texttt{deepseek-v4-pro} & \negp{39.4\,(44.4)} & \negp{43.0\,(32.1)} & \pos{62.4\,(36.5)} \\
\midrule
\texttt{gemini-3.1-pro} & \negp{39.4\,(45.9)} & \negp{46.6\,(34.1)} & \pos{63.5\,(35.7)} \\
\texttt{gpt-5.5} & \best{44.5\,(48.4)} & \best{56.1\,(33.8)} & \best{64.8\,(34.9)} \\
\midrule
\rowcolor{gtgray}
\textbf{Author-Anchored RQ} & \textbf{--} & \textbf{--} & \textbf{--} \\
\bottomrule
\end{tabular}}
\end{tcolorbox}
\caption{Judge: \texttt{gemini-3.1-pro} -- standalone win\,(tie) rate.}
\label{tab:win-std-gemini-31-pro}
\end{subtable}%
\hfill%
\begin{subtable}[t]{0.48\linewidth}
\vspace{0pt}
\centering
\setlength{\tabcolsep}{4pt}
\begin{tcolorbox}[enhanced,colback=white,colframe=gray!35,boxrule=0.4pt,arc=2mm,left=1mm,right=1mm,top=1mm,bottom=1mm,drop shadow=black!12]
\resizebox{\linewidth}{!}{
\begin{tabular}{l >{\centering\arraybackslash}p{24mm} >{\centering\arraybackslash}p{24mm} >{\centering\arraybackslash}p{24mm}}
\toprule
\multirow{2}{*}{\textbf{Model}} & \multicolumn{3}{c}{\shortstack{\textbf{Win\,(Tie) Rate -- Comparative}\\ \textbf{vs.\ Author-Anchored RQ (\%)}}} \\
\cmidrule(lr){2-4}
 & Non-Obv. & Orig. & Gap-Addr. \\
\midrule
\texttt{qwen3-30b-a3b-thinking-2507} & \negp{39.5\,(20.9)} & \negp{36.4\,(22.0)} & \pos{51.7\,(26.1)} \\
\texttt{gpt-oss-20b} & \pos{50.5\,(23.7)} & \negp{50.0\,(19.7)} & \pos{59.6\,(25.4)} \\
\texttt{gemma-4-31b-it} & \pos{69.9\,(17.9)} & \pos{63.9\,(15.9)} & \pos{74.9\,(17.6)} \\
\texttt{deepseek-v4-pro} & \pos{63.9\,(19.3)} & \pos{62.0\,(17.4)} & \pos{78.9\,(13.8)} \\
\midrule
\texttt{gemini-3.1-pro} & \pos{69.7\,(18.6)} & \pos{63.9\,(19.0)} & \pos{80.0\,(14.1)} \\
\texttt{gpt-5.5} & \best{79.3\,(12.4)} & \best{74.0\,(17.6)} & \best{88.9\,(9.1)} \\
\midrule
\rowcolor{gtgray}
\textbf{Author-Anchored RQ} & \textbf{--} & \textbf{--} & \textbf{--} \\
\bottomrule
\end{tabular}}
\end{tcolorbox}
\caption{Judge: \texttt{gemini-3.1-pro} -- comparative win\,(tie) rate.}
\label{tab:win-cmp-gemini-31-pro}
\end{subtable}
\par\medskip
\begin{subtable}[t]{0.48\linewidth}
\vspace{0pt}
\centering
\setlength{\tabcolsep}{4pt}
\begin{tcolorbox}[enhanced,colback=white,colframe=gray!35,boxrule=0.4pt,arc=2mm,left=1mm,right=1mm,top=1mm,bottom=1mm,drop shadow=black!12]
\resizebox{\linewidth}{!}{
\begin{tabular}{l >{\centering\arraybackslash}p{24mm} >{\centering\arraybackslash}p{24mm} >{\centering\arraybackslash}p{24mm}}
\toprule
\multirow{2}{*}{\textbf{Model}} & \multicolumn{3}{c}{\shortstack{\textbf{Win\,(Tie) Rate -- Standalone}\\ \textbf{vs.\ Author-Anchored RQ (\%)}}} \\
\cmidrule(lr){2-4}
 & Non-Obv. & Orig. & Gap-Addr. \\
\midrule
\texttt{qwen3-30b-a3b-thinking-2507} & \negp{32.7\,(51.2)} & \negp{39.5\,(41.0)} & \pos{64.7\,(31.7)} \\
\texttt{gpt-oss-20b} & \negp{35.7\,(52.3)} & \negp{45.1\,(43.5)} & \pos{67.0\,(31.9)} \\
\texttt{gemma-4-31b-it} & \negp{35.9\,(51.6)} & \negp{48.1\,(39.3)} & \pos{70.6\,(27.7)} \\
\texttt{deepseek-v4-pro} & \negp{37.1\,(53.1)} & \negp{46.6\,(40.6)} & \pos{70.8\,(28.3)} \\
\midrule
\texttt{gemini-3.1-pro} & \negp{37.8\,(51.2)} & \negp{47.9\,(41.9)} & \pos{72.6\,(26.8)} \\
\texttt{gpt-5.5} & \best{43.3\,(49.1)} & \best{56.7\,(39.7)} & \best{74.5\,(25.3)} \\
\midrule
\rowcolor{gtgray}
\textbf{Author-Anchored RQ} & \textbf{--} & \textbf{--} & \textbf{--} \\
\bottomrule
\end{tabular}}
\end{tcolorbox}
\caption{Judge: \texttt{deepseek-v4-pro} -- standalone win\,(tie) rate.}
\label{tab:win-std-deepseek-v4-pro}
\end{subtable}%
\hfill%
\begin{subtable}[t]{0.48\linewidth}
\vspace{0pt}
\centering
\setlength{\tabcolsep}{4pt}
\begin{tcolorbox}[enhanced,colback=white,colframe=gray!35,boxrule=0.4pt,arc=2mm,left=1mm,right=1mm,top=1mm,bottom=1mm,drop shadow=black!12]
\resizebox{\linewidth}{!}{
\begin{tabular}{l >{\centering\arraybackslash}p{24mm} >{\centering\arraybackslash}p{24mm} >{\centering\arraybackslash}p{24mm}}
\toprule
\multirow{2}{*}{\textbf{Model}} & \multicolumn{3}{c}{\shortstack{\textbf{Win\,(Tie) Rate -- Comparative}\\ \textbf{vs.\ Author-Anchored RQ (\%)}}} \\
\cmidrule(lr){2-4}
 & Non-Obv. & Orig. & Gap-Addr. \\
\midrule
\texttt{qwen3-30b-a3b-thinking-2507} & \negp{31.2\,(41.5)} & \negp{31.0\,(37.5)} & \pos{52.6\,(34.8)} \\
\texttt{gpt-oss-20b} & \negp{43.4\,(39.4)} & \negp{43.6\,(38.5)} & \pos{60.6\,(32.1)} \\
\texttt{gemma-4-31b-it} & \negp{48.3\,(38.0)} & \negp{45.5\,(39.9)} & \pos{66.0\,(28.2)} \\
\texttt{deepseek-v4-pro} & \negp{49.5\,(38.2)} & \negp{46.7\,(38.9)} & \pos{66.6\,(30.7)} \\
\midrule
\texttt{gemini-3.1-pro} & \pos{52.8\,(35.7)} & \negp{46.5\,(38.2)} & \pos{67.2\,(28.9)} \\
\texttt{gpt-5.5} & \best{53.3\,(39.2)} & \best{51.4\,(39.9)} & \best{71.1\,(27.0)} \\
\midrule
\rowcolor{gtgray}
\textbf{Author-Anchored RQ} & \textbf{--} & \textbf{--} & \textbf{--} \\
\bottomrule
\end{tabular}}
\end{tcolorbox}
\caption{Judge: \texttt{deepseek-v4-pro} -- comparative win\,(tie) rate.}
\label{tab:win-cmp-deepseek-v4-pro}
\end{subtable}
\par\medskip
\begin{subtable}[t]{0.48\linewidth}
\vspace{0pt}
\centering
\setlength{\tabcolsep}{4pt}
\begin{tcolorbox}[enhanced,colback=white,colframe=gray!35,boxrule=0.4pt,arc=2mm,left=1mm,right=1mm,top=1mm,bottom=1mm,drop shadow=black!12]
\resizebox{\linewidth}{!}{
\begin{tabular}{l >{\centering\arraybackslash}p{24mm} >{\centering\arraybackslash}p{24mm} >{\centering\arraybackslash}p{24mm}}
\toprule
\multirow{2}{*}{\textbf{Model}} & \multicolumn{3}{c}{\shortstack{\textbf{Win\,(Tie) Rate -- Standalone}\\ \textbf{vs.\ Author-Anchored RQ (\%)}}} \\
\cmidrule(lr){2-4}
 & Non-Obv. & Orig. & Gap-Addr. \\
\midrule
\texttt{qwen3-30b-a3b-thinking-2507} & \negp{15.5\,(40.2)} & \negp{20.6\,(33.9)} & \negp{36.7\,(52.9)} \\
\texttt{gpt-oss-20b} & \negp{19.4\,(48.0)} & \negp{27.0\,(39.6)} & \negp{40.9\,(51.4)} \\
\texttt{gemma-4-31b-it} & \negp{22.3\,(52.9)} & \negp{29.3\,(41.0)} & \negp{46.1\,(48.1)} \\
\texttt{deepseek-v4-pro} & \negp{23.0\,(52.9)} & \negp{28.0\,(37.9)} & \negp{47.8\,(49.8)} \\
\midrule
\texttt{gemini-3.1-pro} & \negp{24.0\,(52.2)} & \negp{30.4\,(41.6)} & \pos{50.8\,(46.9)} \\
\texttt{gpt-5.5} & \best{27.2\,(59.1)} & \best{38.4\,(44.3)} & \best{53.2\,(45.7)} \\
\midrule
\rowcolor{gtgray}
\textbf{Author-Anchored RQ} & \textbf{--} & \textbf{--} & \textbf{--} \\
\bottomrule
\end{tabular}}
\end{tcolorbox}
\caption{Two-Judge Combined -- standalone win\,(tie) rate.}
\label{tab:win-std-two-judge-average}
\end{subtable}%
\hfill%
\begin{subtable}[t]{0.48\linewidth}
\vspace{0pt}
\centering
\setlength{\tabcolsep}{4pt}
\begin{tcolorbox}[enhanced,colback=white,colframe=gray!35,boxrule=0.4pt,arc=2mm,left=1mm,right=1mm,top=1mm,bottom=1mm,drop shadow=black!12]
\resizebox{\linewidth}{!}{
\begin{tabular}{l >{\centering\arraybackslash}p{24mm} >{\centering\arraybackslash}p{24mm} >{\centering\arraybackslash}p{24mm}}
\toprule
\multirow{2}{*}{\textbf{Model}} & \multicolumn{3}{c}{\shortstack{\textbf{Win\,(Tie) Rate -- Comparative}\\ \textbf{vs.\ Author-Anchored RQ (\%)}}} \\
\cmidrule(lr){2-4}
 & Non-Obv. & Orig. & Gap-Addr. \\
\midrule
\texttt{qwen3-30b-a3b-thinking-2507} & \negp{20.4\,(27.7)} & \negp{21.6\,(26.3)} & \negp{33.3\,(37.1)} \\
\texttt{gpt-oss-20b} & \negp{31.9\,(33.4)} & \negp{31.7\,(32.6)} & \negp{41.1\,(38.3)} \\
\texttt{gemma-4-31b-it} & \negp{42.3\,(34.0)} & \negp{38.5\,(32.8)} & \pos{51.2\,(36.1)} \\
\texttt{deepseek-v4-pro} & \negp{39.9\,(34.5)} & \negp{38.5\,(34.3)} & \pos{54.4\,(35.7)} \\
\midrule
\texttt{gemini-3.1-pro} & \negp{42.5\,(37.6)} & \negp{38.5\,(35.9)} & \pos{56.6\,(33.6)} \\
\texttt{gpt-5.5} & \best{49.1\,(36.8)} & \best{46.0\,(37.8)} & \best{63.4\,(33.1)} \\
\midrule
\rowcolor{gtgray}
\textbf{Author-Anchored RQ} & \textbf{--} & \textbf{--} & \textbf{--} \\
\bottomrule
\end{tabular}}
\end{tcolorbox}
\caption{Two-Judge Combined -- comparative win\,(tie) rate.}
\label{tab:win-cmp-two-judge-average}
\end{subtable}

\caption{
\textbf{Win (tie) rates of generated research questions against the author-anchored reference RQ}
under \emph{standalone} scoring (\emph{left}) and \emph{comparative} scoring (\emph{right}).
Rows report results for the \texttt{gemini-3.1-pro} judge, the \texttt{deepseek-v4-pro} judge, and the combined-judge evaluation.
For each metric, we compare the reference RQ with the best-of-five model generated RQs, i.e., a metric-level Pass@5 comparison. Results are reported as win\% (tie\%). Models are sorted by non-obviousness, then originality, then gap addressing within open- and closed-weight groups. \pos{Green}/\negp{red} indicates win rates above/below 50\%; \best{bold} marks the best model in each column.
}
\label{tab:win-standalone-comparative}
\end{table*}

\subsection{Experimental Setup}

We evaluate the research question generation capabilities of frontier language models through a controlled, LLM-as-a-judge empirical framework. The following details the baseline models, the judge configurations, and the step-by-step task formulation used to extract and score novel research questions.

\paragraph{Models Under Evaluation.}

We evaluate a mix of open and proprietary high-thinking models:
\texttt{\seqsplit{qwen3-30b-a3b-thinking-2507}}, \texttt{deepseek-v4-pro},
\texttt{gemma-4-31b-it}, \texttt{gpt-oss-20b},
\texttt{gpt-5.5}, and \texttt{gemini-3.1-pro}. For models with explicit
reasoning controls, we use their high-thinking or native reasoning mode.
All models are systematically deployed with their respective thinking modes activated and configured with manufacturer-recommended default sampling hyperparameters.

\paragraph{Judge Configurations.} To ensure systematic semantic verification, we employ two advanced reasoning models as judges: \texttt{gemini-3.1-pro} and \texttt{deepseek-v4-pro}. Both judges are configured with maximum reasoning budgets (\texttt{thinking\_level: high}) and greedy decoding (\texttt{temperature = 0.0}) to enforce rigorous, reproducible evaluations.

\paragraph{Task Formulation.} 
Each author-anchored research question is rooted in one or more influential papers. To assess the models' capacity for research ideation, we provide them with these source texts and prompt them to extract critical gaps, such as methodological limitations or open challenges. Based directly on these identified gaps, the models must formulate exactly five specific, answerable research questions. For multi-paper scenarios, the models are instructed to cross-reference the literature and synthesize convergent limitations into cohesive, novel questions. The generated outputs are then scored using the evaluation framework detailed in Section~\ref{sec:evaluation-setup}.

\subsection{Results and Discussions}

\paragraph{Author-Anchor Overlap.}
We report $\mathrm{Overlap@5}_m$ to show whether models reproduce the author-pursued direction. As shown in Table~\ref{tab:overlap-results}, most models score close to 1.0. This means they often find the broad area, but ask a different question. The strongest model, \texttt{gpt-5.5}, reaches 1.33. Low overlap may indicate a different underlying phenomenon. A background can support many valid RQs. The result mainly shows that models often generate alternatives rather than the question pursued by the paper.

\paragraph{Standalone Scoring.}
Under standalone LLM judging, almost all model RQs receive higher best-of-five novelty scores than the author-anchored reference across all three metrics. The strongest model is consistently \texttt{gpt-5.5}, achieving the highest mean scores in non-obviousness, originality, and gap addressing. However, these elevated absolute scores do not translate into decisive wins. As shown in Table~\ref{tab:novelty-standalone-full}, while models frequently tie with the GT, their strict win rates remain notably low. Under the two-judge combined evaluation, even \texttt{gpt-5.5} secures clear win rates of only 27.2\% for non-obviousness and 38.4\% for originality. Among the three evaluated dimensions, gap addressing yields the highest win rates, whereas non-obviousness remains the most challenging dimension for models to demonstrably surpass the reference. This suggests that standalone judging assigns high novelty scores to both author-anchored and model-generated RQs. Since each RQ is evaluated independently, high scores do not necessarily translate into clear wins. As a result, standalone evaluation produces high tie rates despite the elevated absolute scores.

\paragraph{Comparative Scoring.}
The evaluation dynamics shift considerably when the judge assesses the reference RQ and model RQs side by side. As shown in Table~\ref{tab:win-standalone-comparative}, comparative scoring reduces the frequency of ties and significantly increases the strict win rates for model-generated RQs. Under the two-judge combined setting, the strict win rate for \texttt{gpt-5.5} on non-obviousness jumps from 27.2\% in standalone scoring to 49.1\% in comparative scoring, accompanied by a sharp drop in its tie rate from 59.1\% to 36.8\%. When evaluating multiple RQs in the same shared context, LLM judges exhibit a clear tendency to break ties in favor of the generated outputs, amplifying the perceived performance of the models across all three novelty metrics. 

Comparative evaluation forces judges to break ties, leading to a sharp increase in win rates for model-generated RQs and a corresponding drop in tie rates. This raises the question of whether the generated RQs are genuinely more novel than the author-anchored RQs, especially with respect to non-obviousness. To verify this, we proceed to human expert evaluation.

\begin{table*}[t]
\centering
\renewcommand{\arraystretch}{1.2}

\begin{subtable}[t]{0.55\linewidth}
\centering
\setlength{\tabcolsep}{10pt}
\begin{tcolorbox}[
    enhanced, colback=white, colframe=gray!35, boxrule=0.4pt,
    arc=2mm, left=1mm, right=1mm, top=1mm, bottom=1mm, drop shadow=black!12]
\resizebox{\linewidth}{!}{
\begin{tabular}{l cccc}
\toprule
\multirow{2}{*}{\textbf{Evaluator}}
& \multicolumn{4}{c}{\textbf{Win Rate Agreement: Non-Obviousness}} \\
\cmidrule(lr){2-5}
& Expert-1 & Expert-2 & \texttt{gemini-3.1-pro} & \texttt{deepseek-v4-pro} \\
\midrule
Expert-1
& \cellcolor{gray!15}-- & \cellcolor{blue!24}60\% & \cellcolor{blue!9}22\% & \cellcolor{blue!11}28\% \\

Expert-2
& \cellcolor{blue!24}60\% & \cellcolor{gray!15}-- & \cellcolor{blue!16}40\% & \cellcolor{blue!14}34\% \\

\texttt{gemini-3.1-pro}
& \cellcolor{blue!9}22\% & \cellcolor{blue!16}40\% & \cellcolor{gray!15}-- & \cellcolor{blue!21}52\% \\

\texttt{deepseek-v4-pro}
& \cellcolor{blue!11}28\% & \cellcolor{blue!14}34\% & \cellcolor{blue!21}52\% & \cellcolor{gray!15}-- \\
\bottomrule
\end{tabular}}
\end{tcolorbox}
\caption{Agreement (\%) on comparative non-obviousness win rates between human experts and LLM judges.}
\label{tab:agreement_non_obviousness}
\end{subtable}%
\hfill%
\begin{subtable}[t]{0.43\linewidth}
\centering
\setlength{\tabcolsep}{12pt}
\begin{tcolorbox}[
    enhanced, colback=white, colframe=gray!35, boxrule=0.4pt,
    arc=2mm, left=1mm, right=1mm, top=1mm, bottom=1mm, drop shadow=black!12]
\resizebox{\linewidth}{!}{
\begin{tabular}{l ccc}
\toprule
\multirow{2}{*}{\textbf{Evaluator}}
& \multicolumn{3}{c}{\textbf{Win Rate Distribution: \texttt{gpt-5.5} vs. GT}} \\
\cmidrule(lr){2-4}
& \texttt{gpt-5.5} Wins & GT Wins & Ties \\
\midrule
Expert-1
& \cellcolor{blue!6} 7 (14\%) & \cellcolor{blue!31} \textbf{39 (78\%)} & \cellcolor{gray!10} 4 (8\%) \\

Expert-2
& \cellcolor{blue!15} 19 (38\%) & \cellcolor{blue!22} \textbf{28 (56\%)} & \cellcolor{gray!10} 3 (6\%) \\

\midrule
\texttt{gemini-3.1-pro}
& \cellcolor{blue!33} \textbf{41 (82\%)} & \cellcolor{blue!3} 4 (8\%) & \cellcolor{gray!10} 5 (10\%) \\

\texttt{deepseek-v4-pro}
& \cellcolor{blue!21} \textbf{26 (52\%)} & \cellcolor{blue!7} 9 (18\%) & \cellcolor{gray!10} 15 (30\%) \\
\bottomrule
\end{tabular}}
\end{tcolorbox}
\caption{Win rate distribution for comparative non-obviousness scoring between GT and best-of-five RQs generated by \texttt{gpt-5.5}.}
\label{tab:winrate_non_obviousness}
\end{subtable}

\caption{Comparison of human expert and LLM judge evaluations on comparative \textbf{non-obviousness} scoring, over 50 randomly selected samples.}
\label{tab:expert_vs_llm_judge_comparative_non_obviousness}
\end{table*}

\paragraph{Reliability of LLM-as-a-Judge.}
Since LLM judges consistently assigned high novelty scores to model-generated RQs, we conducted a human expert evaluation to assess their reliability. We sampled 50 instances from \texttt{cs.CL} and \texttt{cs.LG}. For each instance, we compared the author-anchored reference RQ (GT) with the best-of-five RQ generated by \texttt{gpt-5.5}. Domain experts from the same fields performed blinded comparative evaluations. To rigorously test the models' ability to assess deeper scientific value, domain experts specifically evaluated the RQs based on the critical \textbf{non-obviousness} dimension, using the same rubric as the LLM judges.

Table~\ref{tab:agreement_non_obviousness} shows a strong disagreement between human and LLM evaluations regarding non-obviousness. Agreement within the same evaluator type was relatively high (60\% for the Expert-Expert pair and 52\% for the LLM-LLM pair), but agreement between humans and LLMs dropped to as low as 22\%. This mismatch also appears in the win rates shown in Table~\ref{tab:winrate_non_obviousness}. Human evaluators strongly preferred the author-anchored GT for non-obviousness, assigning it win rates of 78\% and 56\%. In contrast, LLM judges heavily favored the \texttt{gpt-5.5} outputs, awarding them non-obviousness win rates of 82\% (\texttt{gemini-3.1-pro}) and 52\% (\texttt{deepseek-v4-pro}).

Experts noted that many generated questions imitate the structure of research gaps, but remain less original and less non-obvious. They also reported that many questions are narrow and too tied to the background papers. This suggests that LLM judges may reward surface gap language while missing deeper aspects of novelty. Overall, the contradictory novelty evaluations between LLM judges and human experts question the reliability and trustworthiness of using LLMs to assess the scientific novelty of research questions.

\begin{tcolorbox}[
    colback=olive!2!white,
    colframe=olive!120!black,
    title=\textbf{Key Finding: Evaluation Reliability},
    fonttitle=\bfseries
]
\textbf{Question:} How reliable is the LLM-as-a-judge paradigm for RQ generation as compared to human judgments?

\vspace{0.5em}
\hrule
\vspace{0.5em}

\textbf{Answer:} To verify whether the high novelty scores assigned by LLM judges to model-generated RQs were reliable, we conducted a human expert evaluation. We then ran an expert study on 50 samples, specifically assessing the critical \textbf{non-obviousness} dimension. Table~\ref{tab:agreement_non_obviousness} and Table~\ref{tab:winrate_non_obviousness} show a strong gap between human experts and LLM judges. Human experts preferred the author-anchored GT RQs. LLM judges preferred the best-of-five \texttt{gpt-5.5} RQs. \textbf{Experts noted that many \texttt{gpt-5.5} RQs identify gaps, but still lack originality, non-obviousness, and broader scientific scope. This raises concerns about using current LLMs as judges for RQ novelty.}
\end{tcolorbox}

\section{Scope and Narrowness: What LLM Judges Miss in Novelty Scoring}
\label{sec:alternative-assessment-metric}

\begin{table*}[t]
\centering
\setlength{\tabcolsep}{6pt}
\renewcommand{\arraystretch}{1.18}

\begin{tcolorbox}[
    enhanced, colback=white, colframe=gray!35, boxrule=0.4pt,
    arc=2mm, left=1mm, right=1mm, top=1mm, bottom=1mm, drop shadow=black!12
]
\begin{tabular*}{\linewidth}{@{}l@{\extracolsep{\fill}}cccc@{}}
\toprule
\multirow{2}{*}{\textbf{Model}}
& \multicolumn{2}{c}{\textbf{Qwen3-Embedding-4B}}
& \multicolumn{2}{c}{\textbf{all-mpnet-base-v2}} \\
\cmidrule(lr){2-3}\cmidrule(lr){4-5}
& Single bg & Multi bg & Single bg & Multi bg \\
\midrule

\texttt{gemma-4-31b-it}
& \negp{0.550} & \negp{0.598} & \negp{0.404} & \negp{0.454} \\

\texttt{gemini-3.1-pro}
& \negp{0.585} & \negp{0.621} & \negp{0.482} & \negp{0.523} \\

\texttt{gpt-5.5}
& \negp{0.590} & \negp{0.619} & \negp{0.458} & \negp{0.489} \\

\texttt{gpt-oss-20b}
& \negp{0.591} & \best{0.631} & \best{0.483} & \negp{0.531} \\

\texttt{deepseek-v4-pro}
& \negp{0.596} & \negp{0.631} & \negp{0.482} & \best{0.538} \\

\texttt{qwen3-30b-a3b-thinking-2507}
& \best{0.602} & \negp{0.628} & \negp{0.459} & \negp{0.504} \\

\midrule
\rowcolor{gtgray}
\textbf{Author-Anchored RQ}
& \textbf{0.612} & \textbf{0.645} & \textbf{0.538} & \textbf{0.576} \\

\bottomrule
\end{tabular*}
\end{tcolorbox}
\caption{Cosine similarity between generated research questions and background paper contributions across two embedding spaces. \textit{Note: bg indicates background papers.}}
\label{tab:similarity-with-bg}
\end{table*}

\paragraph{Semantic Similarity with Background Papers.} During expert evaluation, the experts often described LLM-generated RQs as narrow---many questions looked like a small tweak to the ideas present in the background paper rather than a broader research direction. A natural proxy to evaluate this tendency is semantic similarity: are narrow RQs simply closer to the background papers? To test this, we compared each RQ with its background papers using \texttt{qwen3-embedding-4b} and \texttt{all-mpnet-base-v2}. For model RQs, we selected, per instance, the RQ with the highest mean standalone novelty score across the three metrics (non-obviousness, originality, and gap-addressing). For contexts with multiple background papers, we report the maximum similarity to any background paper.

However, the results in \cref{tab:similarity-with-bg} are counter-intuitive: across both embedding models and evaluation settings, GT research questions consistently show higher semantic similarity to the background papers than the LLM-generated questions. This suggests that semantic similarity does not capture the notion of \emph{narrowness} identified by human experts.

\begin{table*}[p]
\centering
\renewcommand{\arraystretch}{1.15}
\begin{subtable}[t]{0.48\linewidth}
\vspace{0pt}
\centering
\setlength{\tabcolsep}{4pt}
\begin{tcolorbox}[enhanced,colback=white,colframe=gray!35,boxrule=0.4pt,arc=2mm,left=1mm,right=1mm,top=1mm,bottom=1mm,drop shadow=black!12]
\resizebox{\linewidth}{!}{
\begin{tabular}{l >{\centering\arraybackslash}p{26mm} >{\centering\arraybackslash}p{26mm}}
\toprule
\multirow{2}{*}{\textbf{Model}} & \multicolumn{2}{c}{\shortstack{\textbf{Narrowness Score}\\ \textbf{(mean\,$\pm$\,std, lower better)}}} \\
\cmidrule(lr){2-3}
 & \shortstack{Source-\\Boundedness} & \shortstack{Diagnostic\\Framing} \\
\midrule
\texttt{gemma-4-31b-it} & \negp{1.55\,{\scriptsize$\pm$0.78}} & \pos{0.05\,{\scriptsize$\pm$0.26}} \\
\texttt{gpt-oss-20b} & \negp{1.15\,{\scriptsize$\pm$0.90}} & \pos{0.05\,{\scriptsize$\pm$0.27}} \\
\texttt{qwen3-30b-a3b-thinking-2507} & \negp{1.14\,{\scriptsize$\pm$1.06}} & \negp{0.28\,{\scriptsize$\pm$0.63}} \\
\texttt{deepseek-v4-pro} & \negp{1.08\,{\scriptsize$\pm$0.89}} & \pos{0.05\,{\scriptsize$\pm$0.27}} \\
\midrule
\texttt{gpt-5.5} & \negp{1.35\,{\scriptsize$\pm$0.84}} & \best{0.01\,{\scriptsize$\pm$0.11}} \\
\texttt{gemini-3.1-pro} & \best{1.00\,{\scriptsize$\pm$0.88}} & \pos{0.02\,{\scriptsize$\pm$0.18}} \\
\midrule
\rowcolor{gtgray}
\textbf{Author-Anchored RQ} & \textbf{0.47\,{\scriptsize$\pm$0.76}} & \textbf{0.07\,{\scriptsize$\pm$0.39}} \\
\bottomrule
\end{tabular}}
\end{tcolorbox}
\caption{Judge: \texttt{gemini-3.1-pro} -- mean\,\texorpdfstring{$\pm$}{+/-}\,std of each model's best per-dimension RQ score (lower is better).}
\label{tab:narr-mean-gemini-31-pro}
\end{subtable}%
\hfill%
\begin{subtable}[t]{0.48\linewidth}
\vspace{0pt}
\centering
\setlength{\tabcolsep}{4pt}
\begin{tcolorbox}[enhanced,colback=white,colframe=gray!35,boxrule=0.4pt,arc=2mm,left=1mm,right=1mm,top=1mm,bottom=1mm,drop shadow=black!12]
\resizebox{\linewidth}{!}{
\begin{tabular}{l >{\centering\arraybackslash}p{26mm} >{\centering\arraybackslash}p{26mm}}
\toprule
\multirow{2}{*}{\textbf{Model}} & \multicolumn{2}{c}{\shortstack{\textbf{Win\,(Lose) Rate vs.}\\ \textbf{Author-Anchored RQ (\%)}}} \\
\cmidrule(lr){2-3}
 & \shortstack{Source-\\Boundedness} & \shortstack{Diagnostic\\Framing} \\
\midrule
\texttt{gemma-4-31b-it} & \negp{6.4\,(68.9)} & \pos{3.2\,(3.0)} \\
\texttt{gpt-oss-20b} & \negp{11.0\,(51.7)} & \negp{3.3\,(4.2)} \\
\texttt{qwen3-30b-a3b-thinking-2507} & \negp{11.8\,(48.3)} & \negp{3.3\,(18.7)} \\
\texttt{deepseek-v4-pro} & \negp{11.7\,(49.0)} & \pos{3.4\,(3.3)} \\
\midrule
\texttt{gpt-5.5} & \negp{7.3\,(61.1)} & \best{3.4\,(0.5)} \\
\texttt{gemini-3.1-pro} & \best{11.8\,(46.4)} & \pos{3.4\,(0.9)} \\
\midrule
\rowcolor{gtgray}
\textbf{Author-Anchored RQ} & \textbf{--} & \textbf{--} \\
\bottomrule
\end{tabular}}
\end{tcolorbox}
\caption{Judge: \texttt{gemini-3.1-pro} -- win\,(lose) rate vs.\ the author-anchored reference.}
\label{tab:narr-win-gemini-31-pro}
\end{subtable}
\par\medskip
\begin{subtable}[t]{0.48\linewidth}
\vspace{0pt}
\centering
\setlength{\tabcolsep}{4pt}
\begin{tcolorbox}[enhanced,colback=white,colframe=gray!35,boxrule=0.4pt,arc=2mm,left=1mm,right=1mm,top=1mm,bottom=1mm,drop shadow=black!12]
\resizebox{\linewidth}{!}{
\begin{tabular}{l >{\centering\arraybackslash}p{26mm} >{\centering\arraybackslash}p{26mm}}
\toprule
\multirow{2}{*}{\textbf{Model}} & \multicolumn{2}{c}{\shortstack{\textbf{Narrowness Score}\\ \textbf{(mean\,$\pm$\,std, lower better)}}} \\
\cmidrule(lr){2-3}
 & \shortstack{Source-\\Boundedness} & \shortstack{Diagnostic\\Framing} \\
\midrule
\texttt{gemma-4-31b-it} & \negp{1.36\,{\scriptsize$\pm$0.98}} & \pos{0.07\,{\scriptsize$\pm$0.36}} \\
\texttt{gpt-oss-20b} & \negp{0.96\,{\scriptsize$\pm$0.96}} & \pos{0.04\,{\scriptsize$\pm$0.25}} \\
\texttt{qwen3-30b-a3b-thinking-2507} & \negp{0.96\,{\scriptsize$\pm$1.08}} & \negp{0.14\,{\scriptsize$\pm$0.49}} \\
\texttt{deepseek-v4-pro} & \negp{0.91\,{\scriptsize$\pm$0.94}} & \pos{0.03\,{\scriptsize$\pm$0.22}} \\
\midrule
\texttt{gpt-5.5} & \negp{0.98\,{\scriptsize$\pm$0.91}} & \best{0.02\,{\scriptsize$\pm$0.17}} \\
\texttt{gemini-3.1-pro} & \best{0.80\,{\scriptsize$\pm$0.91}} & \pos{0.03\,{\scriptsize$\pm$0.23}} \\
\midrule
\rowcolor{gtgray}
\textbf{Author-Anchored RQ} & \textbf{0.48\,{\scriptsize$\pm$0.85}} & \textbf{0.09\,{\scriptsize$\pm$0.45}} \\
\bottomrule
\end{tabular}}
\end{tcolorbox}
\caption{Judge: \texttt{deepseek-v4-pro} -- mean\,\texorpdfstring{$\pm$}{+/-}\,std of each model's best per-dimension RQ score (lower is better).}
\label{tab:narr-mean-deepseek-v4-pro}
\end{subtable}%
\hfill%
\begin{subtable}[t]{0.48\linewidth}
\vspace{0pt}
\centering
\setlength{\tabcolsep}{4pt}
\begin{tcolorbox}[enhanced,colback=white,colframe=gray!35,boxrule=0.4pt,arc=2mm,left=1mm,right=1mm,top=1mm,bottom=1mm,drop shadow=black!12]
\resizebox{\linewidth}{!}{
\begin{tabular}{l >{\centering\arraybackslash}p{26mm} >{\centering\arraybackslash}p{26mm}}
\toprule
\multirow{2}{*}{\textbf{Model}} & \multicolumn{2}{c}{\shortstack{\textbf{Win\,(Lose) Rate vs.}\\ \textbf{Author-Anchored RQ (\%)}}} \\
\cmidrule(lr){2-3}
 & \shortstack{Source-\\Boundedness} & \shortstack{Diagnostic\\Framing} \\
\midrule
\texttt{gemma-4-31b-it} & \negp{12.4\,(60.7)} & \negp{3.9\,(4.1)} \\
\texttt{gpt-oss-20b} & \negp{16.4\,(43.9)} & \pos{3.9\,(2.2)} \\
\texttt{qwen3-30b-a3b-thinking-2507} & \negp{17.2\,(41.5)} & \negp{3.7\,(7.6)} \\
\texttt{deepseek-v4-pro} & \negp{16.9\,(42.2)} & \pos{4.0\,(2.2)} \\
\midrule
\texttt{gpt-5.5} & \negp{14.8\,(47.9)} & \best{4.0\,(1.0)} \\
\texttt{gemini-3.1-pro} & \best{18.4\,(40.1)} & \pos{3.9\,(1.8)} \\
\midrule
\rowcolor{gtgray}
\textbf{Author-Anchored RQ} & \textbf{--} & \textbf{--} \\
\bottomrule
\end{tabular}}
\end{tcolorbox}
\caption{Judge: \texttt{deepseek-v4-pro} -- win\,(lose) rate vs.\ the author-anchored reference.}
\label{tab:narr-win-deepseek-v4-pro}
\end{subtable}
\par\medskip
\begin{subtable}[t]{0.48\linewidth}
\vspace{0pt}
\centering
\setlength{\tabcolsep}{4pt}
\begin{tcolorbox}[enhanced,colback=white,colframe=gray!35,boxrule=0.4pt,arc=2mm,left=1mm,right=1mm,top=1mm,bottom=1mm,drop shadow=black!12]
\resizebox{\linewidth}{!}{
\begin{tabular}{l >{\centering\arraybackslash}p{26mm} >{\centering\arraybackslash}p{26mm}}
\toprule
\multirow{2}{*}{\textbf{Model}} & \multicolumn{2}{c}{\shortstack{\textbf{Narrowness Score}\\ \textbf{(mean\,$\pm$\,std, lower better)}}} \\
\cmidrule(lr){2-3}
 & \shortstack{Source-\\Boundedness} & \shortstack{Diagnostic\\Framing} \\
\midrule
\texttt{gemma-4-31b-it} & \negp{1.46\,{\scriptsize$\pm$0.78}} & \pos{0.06\,{\scriptsize$\pm$0.26}} \\
\texttt{gpt-oss-20b} & \negp{1.06\,{\scriptsize$\pm$0.86}} & \pos{0.04\,{\scriptsize$\pm$0.22}} \\
\texttt{qwen3-30b-a3b-thinking-2507} & \negp{1.05\,{\scriptsize$\pm$1.00}} & \negp{0.21\,{\scriptsize$\pm$0.50}} \\
\texttt{deepseek-v4-pro} & \negp{0.99\,{\scriptsize$\pm$0.85}} & \pos{0.04\,{\scriptsize$\pm$0.22}} \\
\midrule
\texttt{gpt-5.5} & \negp{1.17\,{\scriptsize$\pm$0.78}} & \best{0.01\,{\scriptsize$\pm$0.11}} \\
\texttt{gemini-3.1-pro} & \best{0.90\,{\scriptsize$\pm$0.81}} & \pos{0.02\,{\scriptsize$\pm$0.17}} \\
\midrule
\rowcolor{gtgray}
\textbf{Author-Anchored RQ} & \textbf{0.47\,{\scriptsize$\pm$0.73}} & \textbf{0.08\,{\scriptsize$\pm$0.38}} \\
\bottomrule
\end{tabular}}
\end{tcolorbox}
\caption{Two-Judge Combined -- mean\,\texorpdfstring{$\pm$}{+/-}\,std of each model's best per-dimension RQ score (lower is better).}
\label{tab:narr-mean-two-judge-average}
\end{subtable}%
\hfill%
\begin{subtable}[t]{0.48\linewidth}
\vspace{0pt}
\centering
\setlength{\tabcolsep}{4pt}
\begin{tcolorbox}[enhanced,colback=white,colframe=gray!35,boxrule=0.4pt,arc=2mm,left=1mm,right=1mm,top=1mm,bottom=1mm,drop shadow=black!12]
\resizebox{\linewidth}{!}{
\begin{tabular}{l >{\centering\arraybackslash}p{26mm} >{\centering\arraybackslash}p{26mm}}
\toprule
\multirow{2}{*}{\textbf{Model}} & \multicolumn{2}{c}{\shortstack{\textbf{Win\,(Lose) Rate vs.}\\ \textbf{Author-Anchored RQ (\%)}}} \\
\cmidrule(lr){2-3}
 & \shortstack{Source-\\Boundedness} & \shortstack{Diagnostic\\Framing} \\
\midrule
\texttt{gemma-4-31b-it} & \negp{3.8\,(78.0)} & \negp{2.2\,(7.6)} \\
\texttt{gpt-oss-20b} & \negp{6.6\,(61.1)} & \negp{2.5\,(7.9)} \\
\texttt{qwen3-30b-a3b-thinking-2507} & \best{7.0\,(57.4)} & \negp{2.4\,(22.1)} \\
\texttt{deepseek-v4-pro} & \negp{7.0\,(59.2)} & \negp{2.5\,(5.5)} \\
\midrule
\texttt{gpt-5.5} & \negp{4.3\,(71.7)} & \best{2.5\,(2.9)} \\
\texttt{gemini-3.1-pro} & \negp{6.8\,(60.2)} & \negp{2.5\,(3.3)} \\
\midrule
\rowcolor{gtgray}
\textbf{Author-Anchored RQ} & \textbf{--} & \textbf{--} \\
\bottomrule
\end{tabular}}
\end{tcolorbox}
\caption{Two-Judge Combined -- win\,(lose) rate vs.\ the author-anchored reference.}
\label{tab:narr-win-two-judge-average}
\end{subtable}

\caption{
\textbf{Standalone narrowness assessment of generated research questions.}
Rows report scores from the \texttt{gemini-3.1-pro} judge, the
\texttt{deepseek-v4-pro} judge, and the combined-judge evaluation. Scores are reported separately for source-boundedness and diagnostic framing; for both dimensions \emph{lower is better}. For each instance and metric, we compare the author-anchored reference RQ with the \emph{lowest}-scoring RQ among the generated RQs for that metric, i.e., a metric-level Pass@5 comparison. \emph{Left}: mean\,\texorpdfstring{$\pm$}{+/-}\,std of the Best@5 standalone score. \emph{Right}: win\,(lose) rate against the reference. \pos{Green}/\negp{red} denotes, on the left, scores below/above the reference (better/worse), and on the right, a win rate above/below the lose rate; \best{bold} marks the best model in each column (lowest score on the left, largest win$-$lose margin on the right).
}
\label{tab:narrowness-standalone-full}
\end{table*}

\begin{table*}[p]
\centering
\renewcommand{\arraystretch}{1.15}
\begin{subtable}[t]{0.48\linewidth}
\vspace{0pt}
\centering
\setlength{\tabcolsep}{4pt}
\begin{tcolorbox}[enhanced,colback=white,colframe=gray!35,boxrule=0.4pt,arc=2mm,left=1mm,right=1mm,top=1mm,bottom=1mm,drop shadow=black!12]
\resizebox{\linewidth}{!}{
\begin{tabular}{l >{\centering\arraybackslash}p{26mm} >{\centering\arraybackslash}p{26mm}}
\toprule
\multirow{2}{*}{\textbf{Model}} & \multicolumn{2}{c}{\shortstack{\textbf{Win\,(Lose) Rate -- Standalone}\\ \textbf{vs.\ Author-Anchored RQ (\%)}}} \\
\cmidrule(lr){2-3}
 & \shortstack{Source-\\Boundness} & \shortstack{Diagnostic\\Framing} \\
\midrule
\texttt{gemma-4-31b-it} & \negp{6.4\,(68.9)} & \pos{3.2\,(3.0)} \\
\texttt{gpt-oss-20b} & \negp{11.0\,(51.7)} & \negp{3.3\,(4.2)} \\
\texttt{qwen3-30b-a3b-thinking-2507} & \negp{11.8\,(48.3)} & \negp{3.3\,(18.7)} \\
\texttt{deepseek-v4-pro} & \negp{11.7\,(49.0)} & \pos{3.4\,(3.3)} \\
\midrule
\texttt{gpt-5.5} & \negp{7.3\,(61.1)} & \best{3.4\,(0.5)} \\
\texttt{gemini-3.1-pro} & \best{11.8\,(46.4)} & \pos{3.4\,(0.9)} \\
\midrule
\rowcolor{gtgray}
\textbf{Author-Anchored RQ} & \textbf{--} & \textbf{--} \\
\bottomrule
\end{tabular}}
\end{tcolorbox}
\caption{Judge: \texttt{gemini-3.1-pro} -- standalone win\,(lose) rate.}
\label{tab:narrwin-std-gemini-31-pro}
\end{subtable}%
\hfill%
\begin{subtable}[t]{0.48\linewidth}
\vspace{0pt}
\centering
\setlength{\tabcolsep}{4pt}
\begin{tcolorbox}[enhanced,colback=white,colframe=gray!35,boxrule=0.4pt,arc=2mm,left=1mm,right=1mm,top=1mm,bottom=1mm,drop shadow=black!12]
\resizebox{\linewidth}{!}{
\begin{tabular}{l >{\centering\arraybackslash}p{26mm} >{\centering\arraybackslash}p{26mm}}
\toprule
\multirow{2}{*}{\textbf{Model}} & \multicolumn{2}{c}{\shortstack{\textbf{Win\,(Lose) Rate -- Comparative}\\ \textbf{vs.\ Author-Anchored RQ (\%)}}} \\
\cmidrule(lr){2-3}
 & \shortstack{Source-\\Boundness} & \shortstack{Diagnostic\\Framing} \\
\midrule
\texttt{gemma-4-31b-it} & \negp{3.3\,(84.0)} & \negp{3.6\,(33.3)} \\
\texttt{gpt-oss-20b} & \negp{8.4\,(65.3)} & \negp{6.8\,(26.8)} \\
\texttt{qwen3-30b-a3b-thinking-2507} & \best{10.6\,(64.8)} & \negp{5.6\,(31.0)} \\
\texttt{deepseek-v4-pro} & \negp{7.6\,(65.3)} & \negp{5.8\,(15.5)} \\
\midrule
\texttt{gpt-5.5} & \negp{4.5\,(75.7)} & \negp{4.5\,(41.8)} \\
\texttt{gemini-3.1-pro} & \negp{5.7\,(71.0)} & \best{7.1\,(15.2)} \\
\midrule
\rowcolor{gtgray}
\textbf{Author-Anchored RQ} & \textbf{--} & \textbf{--} \\
\bottomrule
\end{tabular}}
\end{tcolorbox}
\caption{Judge: \texttt{gemini-3.1-pro} -- comparative win\,(lose) rate.}
\label{tab:narrwin-cmp-gemini-31-pro}
\end{subtable}
\par\medskip
\begin{subtable}[t]{0.48\linewidth}
\vspace{0pt}
\centering
\setlength{\tabcolsep}{4pt}
\begin{tcolorbox}[enhanced,colback=white,colframe=gray!35,boxrule=0.4pt,arc=2mm,left=1mm,right=1mm,top=1mm,bottom=1mm,drop shadow=black!12]
\resizebox{\linewidth}{!}{
\begin{tabular}{l >{\centering\arraybackslash}p{26mm} >{\centering\arraybackslash}p{26mm}}
\toprule
\multirow{2}{*}{\textbf{Model}} & \multicolumn{2}{c}{\shortstack{\textbf{Win\,(Lose) Rate -- Standalone}\\ \textbf{vs.\ Author-Anchored RQ (\%)}}} \\
\cmidrule(lr){2-3}
 & \shortstack{Source-\\Boundness} & \shortstack{Diagnostic\\Framing} \\
\midrule
\texttt{gemma-4-31b-it} & \negp{12.4\,(60.7)} & \negp{3.9\,(4.1)} \\
\texttt{gpt-oss-20b} & \negp{16.4\,(43.9)} & \pos{3.9\,(2.2)} \\
\texttt{qwen3-30b-a3b-thinking-2507} & \negp{17.2\,(41.5)} & \negp{3.7\,(7.6)} \\
\texttt{deepseek-v4-pro} & \negp{16.9\,(42.2)} & \pos{4.0\,(2.2)} \\
\midrule
\texttt{gpt-5.5} & \negp{14.8\,(47.9)} & \best{4.0\,(1.0)} \\
\texttt{gemini-3.1-pro} & \best{18.4\,(40.1)} & \pos{3.9\,(1.8)} \\
\midrule
\rowcolor{gtgray}
\textbf{Author-Anchored RQ} & \textbf{--} & \textbf{--} \\
\bottomrule
\end{tabular}}
\end{tcolorbox}
\caption{Judge: \texttt{deepseek-v4-pro} -- standalone win\,(lose) rate.}
\label{tab:narrwin-std-deepseek-v4-pro}
\end{subtable}%
\hfill%
\begin{subtable}[t]{0.48\linewidth}
\vspace{0pt}
\centering
\setlength{\tabcolsep}{4pt}
\begin{tcolorbox}[enhanced,colback=white,colframe=gray!35,boxrule=0.4pt,arc=2mm,left=1mm,right=1mm,top=1mm,bottom=1mm,drop shadow=black!12]
\resizebox{\linewidth}{!}{
\begin{tabular}{l >{\centering\arraybackslash}p{26mm} >{\centering\arraybackslash}p{26mm}}
\toprule
\multirow{2}{*}{\textbf{Model}} & \multicolumn{2}{c}{\shortstack{\textbf{Win\,(Lose) Rate -- Comparative}\\ \textbf{vs.\ Author-Anchored RQ (\%)}}} \\
\cmidrule(lr){2-3}
 & \shortstack{Source-\\Boundness} & \shortstack{Diagnostic\\Framing} \\
\midrule
\texttt{gemma-4-31b-it} & \negp{6.4\,(72.0)} & \negp{3.9\,(23.7)} \\
\texttt{gpt-oss-20b} & \negp{13.4\,(54.5)} & \negp{5.6\,(16.2)} \\
\texttt{qwen3-30b-a3b-thinking-2507} & \best{15.3\,(52.0)} & \negp{5.7\,(19.2)} \\
\texttt{deepseek-v4-pro} & \negp{13.3\,(56.8)} & \best{6.4\,(9.2)} \\
\midrule
\texttt{gpt-5.5} & \negp{7.5\,(60.4)} & \negp{4.4\,(18.6)} \\
\texttt{gemini-3.1-pro} & \negp{13.8\,(55.8)} & \negp{5.8\,(11.6)} \\
\midrule
\rowcolor{gtgray}
\textbf{Author-Anchored RQ} & \textbf{--} & \textbf{--} \\
\bottomrule
\end{tabular}}
\end{tcolorbox}
\caption{Judge: \texttt{deepseek-v4-pro} -- comparative win\,(lose) rate.}
\label{tab:narrwin-cmp-deepseek-v4-pro}
\end{subtable}
\par\medskip
\begin{subtable}[t]{0.48\linewidth}
\vspace{0pt}
\centering
\setlength{\tabcolsep}{4pt}
\begin{tcolorbox}[enhanced,colback=white,colframe=gray!35,boxrule=0.4pt,arc=2mm,left=1mm,right=1mm,top=1mm,bottom=1mm,drop shadow=black!12]
\resizebox{\linewidth}{!}{
\begin{tabular}{l >{\centering\arraybackslash}p{26mm} >{\centering\arraybackslash}p{26mm}}
\toprule
\multirow{2}{*}{\textbf{Model}} & \multicolumn{2}{c}{\shortstack{\textbf{Win\,(Lose) Rate -- Standalone}\\ \textbf{vs.\ Author-Anchored RQ (\%)}}} \\
\cmidrule(lr){2-3}
 & \shortstack{Source-\\Boundness} & \shortstack{Diagnostic\\Framing} \\
\midrule
\texttt{gemma-4-31b-it} & \negp{3.8\,(78.0)} & \negp{2.2\,(7.6)} \\
\texttt{gpt-oss-20b} & \negp{6.6\,(61.1)} & \negp{2.5\,(7.9)} \\
\texttt{qwen3-30b-a3b-thinking-2507} & \best{7.0\,(57.4)} & \negp{2.4\,(22.1)} \\
\texttt{deepseek-v4-pro} & \negp{7.0\,(59.2)} & \negp{2.5\,(5.5)} \\
\midrule
\texttt{gpt-5.5} & \negp{4.3\,(71.7)} & \best{2.5\,(2.9)} \\
\texttt{gemini-3.1-pro} & \negp{6.8\,(60.2)} & \negp{2.5\,(3.3)} \\
\midrule
\rowcolor{gtgray}
\textbf{Author-Anchored RQ} & \textbf{--} & \textbf{--} \\
\bottomrule
\end{tabular}}
\end{tcolorbox}
\caption{Two-Judge Combined -- standalone win\,(lose) rate.}
\label{tab:narrwin-std-two-judge-average}
\end{subtable}%
\hfill%
\begin{subtable}[t]{0.48\linewidth}
\vspace{0pt}
\centering
\setlength{\tabcolsep}{4pt}
\begin{tcolorbox}[enhanced,colback=white,colframe=gray!35,boxrule=0.4pt,arc=2mm,left=1mm,right=1mm,top=1mm,bottom=1mm,drop shadow=black!12]
\resizebox{\linewidth}{!}{
\begin{tabular}{l >{\centering\arraybackslash}p{26mm} >{\centering\arraybackslash}p{26mm}}
\toprule
\multirow{2}{*}{\textbf{Model}} & \multicolumn{2}{c}{\shortstack{\textbf{Win\,(Lose) Rate -- Comparative}\\ \textbf{vs.\ Author-Anchored RQ (\%)}}} \\
\cmidrule(lr){2-3}
 & \shortstack{Source-\\Boundness} & \shortstack{Diagnostic\\Framing} \\
\midrule
\texttt{gemma-4-31b-it} & \negp{1.4\,(90.8)} & \negp{1.8\,(49.8)} \\
\texttt{gpt-oss-20b} & \negp{4.3\,(75.8)} & \negp{4.0\,(39.2)} \\
\texttt{qwen3-30b-a3b-thinking-2507} & \best{5.9\,(75.5)} & \negp{3.3\,(43.4)} \\
\texttt{deepseek-v4-pro} & \negp{4.3\,(77.0)} & \best{4.0\,(24.8)} \\
\midrule
\texttt{gpt-5.5} & \negp{1.9\,(85.4)} & \negp{2.4\,(53.4)} \\
\texttt{gemini-3.1-pro} & \negp{3.6\,(80.2)} & \negp{4.0\,(28.0)} \\
\midrule
\rowcolor{gtgray}
\textbf{Author-Anchored RQ} & \textbf{--} & \textbf{--} \\
\bottomrule
\end{tabular}}
\end{tcolorbox}
\caption{Two-Judge Combined -- comparative win\,(lose) rate.}
\label{tab:narrwin-cmp-two-judge-average}
\end{subtable}

\caption{
\textbf{Win\,(lose) rates of generated research questions against the
author-anchored reference RQ on narrowness}, under \emph{standalone} scoring
(\emph{left}) and \emph{comparative} scoring (\emph{right}). Rows report results for the \texttt{gemini-3.1-pro} judge, the \texttt{deepseek-v4-pro} judge, and the combined-judge evaluation. Both dimensions are scored such that \emph{lower is better}. For each metric, we compare the reference RQ with the \emph{lowest}-scoring RQ among the generated RQs (a metric-level Pass@5 comparison). Results are reported as win\% (lose\%). Models are sorted by source-boundedness, then diagnostic framing, within open- and closed-weight groups. \pos{Green}/\negp{red} indicates a win rate above/below the lose rate; \best{bold} marks the best model in each column (largest win$-$lose margin).
}
\label{tab:narrowness-winrate-standalone-comparative}
\end{table*}

\begin{table*}[t]
\centering
\renewcommand{\arraystretch}{1.2}

\begin{subtable}[t]{0.49\linewidth}
\centering
\setlength{\tabcolsep}{12pt}
\begin{tcolorbox}[
    enhanced, colback=white, colframe=gray!35, boxrule=0.4pt,
    arc=2mm, left=1mm, right=1mm, top=1mm, bottom=1mm, drop shadow=black!12]
\resizebox{\linewidth}{!}{
\begin{tabular}{l ccc}
\toprule
\multirow{2}{*}{\textbf{LLM Judge}}
& \multicolumn{3}{c}{\textbf{Source-Bound Win Rate: \texttt{gpt-5.5} vs. GT}} \\
\cmidrule(lr){2-4}
& \texttt{gpt-5.5} Wins & GT Wins & Ties \\
\midrule
\texttt{gemini-3.1-pro}
& \cellcolor{blue!2} 2 (4\%) & \cellcolor{blue!33} \textbf{41 (82\%)} & \cellcolor{gray!10} 7 (14\%) \\

\texttt{deepseek-v4-pro}
& \cellcolor{blue!2} 3 (6\%) & \cellcolor{blue!36} \textbf{45 (90\%)} & \cellcolor{gray!10} 2 (4\%) \\
\bottomrule
\end{tabular}}
\end{tcolorbox}
\caption{Raw win rates for \textbf{source-boundedness} per LLM judge over the 50 samples; the winner is the \emph{less} source-bound RQ (lower score). GT is consistently less source-bound than \texttt{gpt-5.5}.}
\label{tab:sourcebound_winrate}
\end{subtable}%
\hfill%
\begin{subtable}[t]{0.49\linewidth}
\centering
\setlength{\tabcolsep}{10pt}
\begin{tcolorbox}[
    enhanced, colback=white, colframe=gray!35, boxrule=0.4pt,
    arc=2mm, left=1mm, right=1mm, top=1mm, bottom=1mm, drop shadow=black!12]
\resizebox{\linewidth}{!}{
\begin{tabular}{l cccc}
\toprule
\multirow{2}{*}{\textbf{Evaluator}}
& \multicolumn{4}{c}{\textbf{Win Rate Agreement: Non-Obviousness vs. Source-Boundedness}} \\
\cmidrule(lr){2-5}
& Expert-1 & Expert-2 & \texttt{gemini-3.1-pro} & \texttt{deepseek-v4-pro} \\
\midrule
Expert-1
& \cellcolor{gray!15}-- & \cellcolor{blue!24}60\% & \cellcolor{blue!29}72\% & \cellcolor{blue!30}76\% \\

Expert-2
& \cellcolor{blue!24}60\% & \cellcolor{gray!15}-- & \cellcolor{blue!18}46\% & \cellcolor{blue!20}50\% \\

\midrule

\texttt{gemini-3.1-pro}
& \cellcolor{blue!29}72\% & \cellcolor{blue!18}46\% & \cellcolor{gray!15}-- & \cellcolor{blue!34}86\% \\

\texttt{deepseek-v4-pro}
& \cellcolor{blue!30}76\% & \cellcolor{blue!20}50\% & \cellcolor{blue!34}86\% & \cellcolor{gray!15}-- \\
\bottomrule
\end{tabular}}
\end{tcolorbox}
\caption{Agreement (\%) on the per-item winner: human experts decide on \textbf{non-obviousness}, LLM judges decide on \textbf{source-boundedness} (leaner RQ wins). Each cell is the share of the 50 items on which the two evaluators pick the same RQ (GT or \texttt{gpt-5.5}).}
\label{tab:sourcebound_agreement}
\end{subtable}

\caption{Source-boundedness win rates and their agreement with human non-obviousness judgments over 50 samples.}
\label{tab:sourcebound_nonobv}
\end{table*}

\subsection{Assessment of Narrowness}
\label{subsec:eval-narrowness}

As semantic similarity does not capture the expert concern, we turn to the \emph{narrowness} test defined in \Cref{subsec:scope-metrics}, scoring each RQ on source-boundedness and diagnostic framing (lower is broader). Unlike the novelty tables, which report win/tie rates, the narrowness tables report win/\emph{lose} rates.

\paragraph{Narrowness Results and Discussions}

As detailed in \cref{tab:narrowness-standalone-full} and \cref{tab:narrowness-winrate-standalone-comparative}, standalone evaluations reveal that model-generated RQs are significantly narrower in scope than author-anchored references, and this limitation is mostly driven by source-boundedness. Under the two-judge combined evaluation, the author-anchored references achieve a mean source-boundedness score of 0.47, substantially outperforming even the best-performing models (e.g., \texttt{gemini-3.1-pro} at 0.90 and \texttt{gpt-5.5} at 1.17). Consequently, models suffer massive lose rates against the reference in this dimension; for instance, \texttt{gpt-5.5}'s best RQs lose to the reference 71.7\% of the time and win only 4.3\%. Diagnostic framing, by contrast, is a negligible factor in the \emph{narrowness} gap. In the standalone setting, the models' diagnostic framing scores closely match the reference (0.01--0.06 vs.\ 0.08), resulting in predominantly tied comparisons (e.g., \texttt{gpt-5.5} sees only a 2.5\% win rate and 2.9\% lose rate, meaning over 94\% of cases tie). 

When shifting from standalone to comparative assessment, the models' disadvantage in \emph{narrowness} becomes even more pronounced. As seen in \cref{tab:narrowness-winrate-standalone-comparative}, placing the RQs side-by-side amplifies the judges' baseline preferences. Because the judges already favor the author-anchored RQs for their broader scope in standalone evaluation, the comparative format increases the models' lose rates. For example, \texttt{gpt-5.5}'s source-boundedness lose rate jumps from 71.7\% in standalone to 85.4\% in comparative, and its diagnostic framing lose rate spikes from 2.9\% to 53.4\%.

Direct assessments of \emph{narrowness} validate the experts' qualitative observations that model-generated RQs are restricted in scope and lack non-obviousness. When explicitly evaluated for source-boundedness over the same 50 samples, LLM judges identified author-anchored references as less source-bound than \texttt{gpt-5.5} (82--90\% win rate; \cref{tab:sourcebound_winrate}). Furthermore, human evaluations of non-obviousness show strong agreement (up to 72--76\%) with LLM evaluations of source-boundedness (\cref{tab:sourcebound_agreement}). This alignment confirms that human assessments of an RQ's novelty (especially non-obviousness) have a strong correlation with \emph{narrowness}.

One of the main findings of this paper is that, although human expert novelty evaluations strongly correlate with \emph{narrowness}, LLM judges fail to capture this. \emph{Narrowness} does not solve novelty evaluation by itself. Two broad questions can still differ in value. But it gives a more concrete axis for detecting polished but shallow RQs.

\begin{table*}[t]
\centering
\renewcommand{\arraystretch}{1.15}
\begin{subtable}[t]{0.48\linewidth}
\vspace{0pt}
\centering
\setlength{\tabcolsep}{4pt}
\begin{tcolorbox}[enhanced,colback=white,colframe=gray!35,boxrule=0.4pt,arc=2mm,left=1mm,right=1mm,top=1mm,bottom=1mm,drop shadow=black!12]
\resizebox{\linewidth}{!}{
\begin{tabular}{l >{\centering\arraybackslash}p{18mm} >{\centering\arraybackslash}p{18mm} >{\centering\arraybackslash}p{18mm}}
\toprule
\multirow{2}{*}{\textbf{Model}} & \multicolumn{3}{c}{\shortstack{\textbf{Novelty Score}\\ \textbf{(mean\,$\pm$\,std)}}} \\
\cmidrule(lr){2-4}
 & Non-Obv. & Orig. & Gap-Addr. \\
\midrule
\texttt{gemini-3.1-pro} \scriptsize(orig.\ prompt) & \best{1.81\,{\scriptsize$\pm$0.50}} & \best{2.39\,{\scriptsize$\pm$0.65}} & \best{2.96\,{\scriptsize$\pm$0.20}} \\
\texttt{gemini-3.1-pro} \scriptsize(new prompt) & \pos{1.46\,{\scriptsize$\pm$0.69}} & \negp{1.90\,{\scriptsize$\pm$0.74}} & \pos{2.95\,{\scriptsize$\pm$0.23}} \\
\midrule
\rowcolor{gtgray}
\textbf{Author-Anchored RQ} & \textbf{1.41\,{\scriptsize$\pm$0.87}} & \textbf{1.90\,{\scriptsize$\pm$1.00}} & \textbf{2.01\,{\scriptsize$\pm$0.97}} \\
\bottomrule
\end{tabular}}
\end{tcolorbox}
\caption{Novelty -- standalone score (Best@5).}
\label{tab:f-nov-score}
\end{subtable}%
\hfill%
\begin{subtable}[t]{0.48\linewidth}
\vspace{0pt}
\centering
\setlength{\tabcolsep}{4pt}
\begin{tcolorbox}[enhanced,colback=white,colframe=gray!35,boxrule=0.4pt,arc=2mm,left=1mm,right=1mm,top=1mm,bottom=1mm,drop shadow=black!12]
\resizebox{\linewidth}{!}{
\begin{tabular}{l >{\centering\arraybackslash}p{24mm} >{\centering\arraybackslash}p{24mm}}
\toprule
\multirow{2}{*}{\textbf{Model}} & \multicolumn{2}{c}{\shortstack{\textbf{Narrowness Score}\\ \textbf{(mean\,$\pm$\,std, lower better)}}} \\
\cmidrule(lr){2-3}
 & \shortstack{Source-\\Boundness} & \shortstack{Diagnostic\\Framing} \\
\midrule
\texttt{gemini-3.1-pro} \scriptsize(orig.\ prompt) & \negp{1.00\,{\scriptsize$\pm$0.88}} & \pos{0.02\,{\scriptsize$\pm$0.18}} \\
\texttt{gemini-3.1-pro} \scriptsize(new prompt) & \best{0.47\,{\scriptsize$\pm$0.73}} & \best{0.00\,{\scriptsize$\pm$0.04}} \\
\midrule
\rowcolor{gtgray}
\textbf{Author-Anchored RQ} & \textbf{0.47\,{\scriptsize$\pm$0.76}} & \textbf{0.07\,{\scriptsize$\pm$0.39}} \\
\bottomrule
\end{tabular}}
\end{tcolorbox}
\caption{Narrowness -- standalone score (Best@5, lower better).}
\label{tab:f-narr-score}
\end{subtable}
\par\medskip
\begin{subtable}[t]{0.48\linewidth}
\vspace{0pt}
\centering
\setlength{\tabcolsep}{4pt}
\begin{tcolorbox}[enhanced,colback=white,colframe=gray!35,boxrule=0.4pt,arc=2mm,left=1mm,right=1mm,top=1mm,bottom=1mm,drop shadow=black!12]
\resizebox{\linewidth}{!}{
\begin{tabular}{l >{\centering\arraybackslash}p{18mm} >{\centering\arraybackslash}p{18mm} >{\centering\arraybackslash}p{18mm}}
\toprule
\multirow{2}{*}{\textbf{Model}} & \multicolumn{3}{c}{\shortstack{\textbf{Novelty Win\,(Tie) -- Standalone}\\ \textbf{vs.\ Author-Anchored (\%)}}} \\
\cmidrule(lr){2-4}
 & Non-Obv. & Orig. & Gap-Addr. \\
\midrule
\texttt{gemini-3.1-pro} \scriptsize(orig.\ prompt) & \best{39.4\,(45.9)} & \best{46.6\,(34.1)} & \best{63.5\,(35.7)} \\
\texttt{gemini-3.1-pro} \scriptsize(new prompt) & \negp{31.0\,(41.1)} & \negp{30.7\,(35.1)} & \pos{62.9\,(35.7)} \\
\midrule
\rowcolor{gtgray}
\textbf{Author-Anchored RQ} & \textbf{--} & \textbf{--} & \textbf{--} \\
\bottomrule
\end{tabular}}
\end{tcolorbox}
\caption{Novelty -- standalone win\,(tie).}
\label{tab:f-nov-win}
\end{subtable}%
\hfill%
\begin{subtable}[t]{0.48\linewidth}
\vspace{0pt}
\centering
\setlength{\tabcolsep}{4pt}
\begin{tcolorbox}[enhanced,colback=white,colframe=gray!35,boxrule=0.4pt,arc=2mm,left=1mm,right=1mm,top=1mm,bottom=1mm,drop shadow=black!12]
\resizebox{\linewidth}{!}{
\begin{tabular}{l >{\centering\arraybackslash}p{24mm} >{\centering\arraybackslash}p{24mm}}
\toprule
\multirow{2}{*}{\textbf{Model}} & \multicolumn{2}{c}{\shortstack{\textbf{Narrowness Win\,(Lose) -- Standalone}\\ \textbf{vs.\ Author-Anchored (\%)}}} \\
\cmidrule(lr){2-3}
 & \shortstack{Source-\\Boundness} & \shortstack{Diagnostic\\Framing} \\
\midrule
\texttt{gemini-3.1-pro} \scriptsize(orig.\ prompt) & \negp{11.8\,(46.4)} & \pos{3.4\,(0.9)} \\
\texttt{gemini-3.1-pro} \scriptsize(new prompt) & \best{20.4\,(21.5)} & \best{3.5\,(0.1)} \\
\midrule
\rowcolor{gtgray}
\textbf{Author-Anchored RQ} & \textbf{--} & \textbf{--} \\
\bottomrule
\end{tabular}}
\end{tcolorbox}
\caption{Narrowness -- standalone win\,(lose).}
\label{tab:f-narr-win}
\end{subtable}
\par\medskip
\begin{subtable}[t]{0.48\linewidth}
\vspace{0pt}
\centering
\setlength{\tabcolsep}{4pt}
\begin{tcolorbox}[enhanced,colback=white,colframe=gray!35,boxrule=0.4pt,arc=2mm,left=1mm,right=1mm,top=1mm,bottom=1mm,drop shadow=black!12]
\resizebox{\linewidth}{!}{
\begin{tabular}{l >{\centering\arraybackslash}p{18mm} >{\centering\arraybackslash}p{18mm} >{\centering\arraybackslash}p{18mm}}
\toprule
\multirow{2}{*}{\textbf{Model}} & \multicolumn{3}{c}{\shortstack{\textbf{Novelty Win\,(Tie) -- Comparative}\\ \textbf{vs.\ Author-Anchored (\%)}}} \\
\cmidrule(lr){2-4}
 & Non-Obv. & Orig. & Gap-Addr. \\
\midrule
\texttt{gemini-3.1-pro} \scriptsize(orig.\ prompt) & \best{69.7\,(18.6)} & \best{63.9\,(19.0)} & \best{80.0\,(14.1)} \\
\texttt{gemini-3.1-pro} \scriptsize(new prompt) & \pos{52.2\,(22.3)} & \negp{48.0\,(22.0)} & \pos{71.2\,(20.1)} \\
\midrule
\rowcolor{gtgray}
\textbf{Author-Anchored RQ} & \textbf{--} & \textbf{--} & \textbf{--} \\
\bottomrule
\end{tabular}}
\end{tcolorbox}
\caption{Novelty -- comparative win\,(tie).}
\label{tab:f-nov-cmp}
\end{subtable}%
\hfill%
\begin{subtable}[t]{0.48\linewidth}
\vspace{0pt}
\centering
\setlength{\tabcolsep}{4pt}
\begin{tcolorbox}[enhanced,colback=white,colframe=gray!35,boxrule=0.4pt,arc=2mm,left=1mm,right=1mm,top=1mm,bottom=1mm,drop shadow=black!12]
\resizebox{\linewidth}{!}{
\begin{tabular}{l >{\centering\arraybackslash}p{24mm} >{\centering\arraybackslash}p{24mm}}
\toprule
\multirow{2}{*}{\textbf{Model}} & \multicolumn{2}{c}{\shortstack{\textbf{Narrowness Win\,(Lose) -- Comparative}\\ \textbf{vs.\ Author-Anchored (\%)}}} \\
\cmidrule(lr){2-3}
 & \shortstack{Source-\\Boundness} & \shortstack{Diagnostic\\Framing} \\
\midrule
\texttt{gemini-3.1-pro} \scriptsize(orig.\ prompt) & \negp{5.7\,(71.0)} & \negp{7.1\,(15.2)} \\
\texttt{gemini-3.1-pro} \scriptsize(new prompt) & \best{16.5\,(41.4)} & \best{8.2\,(0.9)} \\
\midrule
\rowcolor{gtgray}
\textbf{Author-Anchored RQ} & \textbf{--} & \textbf{--} \\
\bottomrule
\end{tabular}}
\end{tcolorbox}
\caption{Narrowness -- comparative win\,(lose).}
\label{tab:f-narr-cmp}
\end{subtable}

\caption{\textbf{Effect of the updated generation prompt on \texttt{gemini-3.1-pro}.} Original-prompt vs.\ updated-prompt \texttt{gemini-3.1-pro} research questions, scored by the \texttt{gemini-3.1-pro} judge. Existing rows and the author-anchored reference are verbatim from the main tables. \emph{Left column}: novelty (higher better; win = Pass@5 RQ above the reference; \pos{green}/\negp{red} = score$>$GT / win$>$50\%). \emph{Right column}: narrowness (lower better; win = Pass@5 RQ below the reference, lose = none at or below; \pos{green}/\negp{red} = score$<$GT / win$>$lose). \best{Bold} marks the better prompt per column.}
\label{tab:newprompt-gemini-effect}
\end{table*}

\begin{table*}[t]
\centering
\renewcommand{\arraystretch}{1.2}

\begin{subtable}[t]{0.55\linewidth}
\centering
\setlength{\tabcolsep}{10pt}
\begin{tcolorbox}[
    enhanced, colback=white, colframe=gray!35, boxrule=0.4pt,
    arc=2mm, left=1mm, right=1mm, top=1mm, bottom=1mm, drop shadow=black!12]
\resizebox{\linewidth}{!}{
\begin{tabular}{l cccc}
\toprule
\multirow{2}{*}{\textbf{Evaluator}}
& \multicolumn{4}{c}{\textbf{Win Rate Agreement: Non-Obviousness}} \\
\cmidrule(lr){2-5}
& Expert-1 & Expert-2 & \texttt{gemini-3.1-pro} & \texttt{deepseek-v4-pro} \\
\midrule
Expert-1
& \cellcolor{gray!15}-- & \cellcolor{blue!22}54\% & \cellcolor{blue!12}30\% & \cellcolor{blue!16}40\% \\

Expert-2
& \cellcolor{blue!22}54\% & \cellcolor{gray!15}-- & \cellcolor{blue!14}34\% & \cellcolor{blue!15}38\% \\

\texttt{gemini-3.1-pro}
& \cellcolor{blue!12}30\% & \cellcolor{blue!14}34\% & \cellcolor{gray!15}-- & \cellcolor{blue!25}62\% \\

\texttt{deepseek-v4-pro}
& \cellcolor{blue!16}40\% & \cellcolor{blue!15}38\% & \cellcolor{blue!25}62\% & \cellcolor{gray!15}-- \\
\bottomrule
\end{tabular}}
\end{tcolorbox}
\caption{Agreement (\%) on comparative non-obviousness win rates between human experts and LLM judges.}
\label{tab:agreement_non_obviousness_newprompt}
\end{subtable}%
\hfill%
\begin{subtable}[t]{0.43\linewidth}
\centering
\setlength{\tabcolsep}{12pt}
\begin{tcolorbox}[
    enhanced, colback=white, colframe=gray!35, boxrule=0.4pt,
    arc=2mm, left=1mm, right=1mm, top=1mm, bottom=1mm, drop shadow=black!12]
\resizebox{\linewidth}{!}{
\begin{tabular}{l ccc}
\toprule
\multirow{2}{*}{\textbf{Evaluator}}
& \multicolumn{3}{c}{\textbf{Win Rate Distribution: \texttt{gemini-3.1-pro} vs. GT}} \\
\cmidrule(lr){2-4}
& \texttt{gemini-3.1-pro} Wins & GT Wins & Ties \\
\midrule
Expert-1
& \cellcolor{blue!10} 12 (24\%) & \cellcolor{blue!25} \textbf{31 (62\%)} & \cellcolor{gray!10} 7 (14\%) \\

Expert-2
& \cellcolor{blue!14} 18 (36\%) & \cellcolor{blue!22} \textbf{28 (56\%)} & \cellcolor{gray!10} 4 (8\%) \\

\midrule
\texttt{gemini-3.1-pro}
& \cellcolor{blue!20} \textbf{25 (50\%)} & \cellcolor{blue!9} 11 (22\%) & \cellcolor{gray!10} 14 (28\%) \\

\texttt{deepseek-v4-pro}
& \cellcolor{blue!16} \textbf{20 (40\%)} & \cellcolor{blue!15} 19 (38\%) & \cellcolor{gray!10} 11 (22\%) \\
\bottomrule
\end{tabular}}
\end{tcolorbox}
\caption{Win rate distribution for comparative non-obviousness scoring between GT and RQs generated by \texttt{gemini-3.1-pro} (new prompt).}
\label{tab:winrate_non_obviousness_newprompt}
\end{subtable}

\caption{Comparison of human expert and LLM judge evaluations on comparative \textbf{non-obviousness} scoring, over 50 randomly selected samples (new prompt). Both experts prefer the GT RQ, while the LLM judges (especially \texttt{gemini-3.1-pro}) favor the generated RQ; expert--judge agreement (30--40\%) is far below expert--expert (54\%) and judge--judge (62\%) agreement.}
\label{tab:expert_vs_llm_judge_comparative_non_obviousness_newprompt}
\end{table*}

\begin{table*}[t]
\centering
\renewcommand{\arraystretch}{1.2}

\begin{subtable}[t]{0.49\linewidth}
\centering
\setlength{\tabcolsep}{12pt}
\begin{tcolorbox}[
    enhanced, colback=white, colframe=gray!35, boxrule=0.4pt,
    arc=2mm, left=1mm, right=1mm, top=1mm, bottom=1mm, drop shadow=black!12]
\resizebox{\linewidth}{!}{
\begin{tabular}{l ccc}
\toprule
\multirow{2}{*}{\textbf{LLM Judge}}
& \multicolumn{3}{c}{\textbf{Source-Bound Win Rate: \texttt{gemini-3.1-pro} vs. GT}} \\
\cmidrule(lr){2-4}
& \texttt{gemini-3.1-pro} Wins & GT Wins & Ties \\
\midrule
\texttt{gemini-3.1-pro}
& \cellcolor{blue!3} 4 (8\%) & \cellcolor{blue!28} \textbf{35 (70\%)} & \cellcolor{gray!10} 11 (22\%) \\

\texttt{deepseek-v4-pro}
& \cellcolor{blue!4} 5 (10\%) & \cellcolor{blue!25} \textbf{31 (62\%)} & \cellcolor{gray!10} 14 (28\%) \\
\bottomrule
\end{tabular}}
\end{tcolorbox}
\caption{Raw win rates for \textbf{source-boundedness} per LLM judge over the 50 samples; the winner is the \emph{less} source-bound RQ (lower score). GT is consistently less source-bound than \texttt{gemini-3.1-pro}.}
\label{tab:sourcebound_winrate_newprompt}
\end{subtable}%
\hfill%
\begin{subtable}[t]{0.49\linewidth}
\centering
\setlength{\tabcolsep}{10pt}
\begin{tcolorbox}[
    enhanced, colback=white, colframe=gray!35, boxrule=0.4pt,
    arc=2mm, left=1mm, right=1mm, top=1mm, bottom=1mm, drop shadow=black!12]
\resizebox{\linewidth}{!}{
\begin{tabular}{l cccc}
\toprule
\multirow{2}{*}{\textbf{Evaluator}}
& \multicolumn{4}{c}{\textbf{Win Rate Agreement: Non-Obviousness vs. Source-Boundedness}} \\
\cmidrule(lr){2-5}
& Expert-1 & Expert-2 & \texttt{gemini-3.1-pro} & \texttt{deepseek-v4-pro} \\
\midrule
Expert-1
& \cellcolor{gray!15}-- & \cellcolor{blue!22}54\% & \cellcolor{blue!18}46\% & \cellcolor{blue!18}44\% \\

Expert-2
& \cellcolor{blue!22}54\% & \cellcolor{gray!15}-- & \cellcolor{blue!20}50\% & \cellcolor{blue!17}42\% \\

\midrule

\texttt{gemini-3.1-pro}
& \cellcolor{blue!18}46\% & \cellcolor{blue!20}50\% & \cellcolor{gray!15}-- & \cellcolor{blue!30}74\% \\

\texttt{deepseek-v4-pro}
& \cellcolor{blue!18}44\% & \cellcolor{blue!17}42\% & \cellcolor{blue!30}74\% & \cellcolor{gray!15}-- \\
\bottomrule
\end{tabular}}
\end{tcolorbox}
\caption{Agreement (\%) on the per-item winner: human experts decide on \textbf{non-obviousness}, LLM judges decide on \textbf{source-boundedness} (leaner RQ wins). Each cell is the share of the 50 items on which the two evaluators pick the same RQ (GT or \texttt{gemini-3.1-pro}).}
\label{tab:sourcebound_agreement_newprompt}
\end{subtable}

\caption{Source-boundedness win rates and their agreement with human non-obviousness judgments over 50 samples (new prompt). GT is the \emph{less} source-bound RQ in 62--70\% of items; expert--judge agreement (44--50\%) modestly exceeds expert--judge non-obviousness agreement, weakly supporting the hypothesis that human non-obviousness is shaped by source-boundedness.}
\label{tab:sourcebound_nonobv_newprompt}
\end{table*}

\subsection{Revised RQ Generation Prompt}
To address the narrowness observed in model-generated research questions, we modified the system prompt by removing constraining instructions (\Cref{sec:updated-prompt}). We then tested \texttt{gemini-3.1-pro} using this updated prompt and evaluated the generated RQs. As demonstrated in Table~\ref{tab:newprompt-gemini-effect}, while this prompt adjustment successfully mitigates narrowness (e.g., significantly lowering source-boundedness from 1.00 to 0.47), it introduces a clear trade-off by simultaneously degrading novelty scores across both originality ($2.39 \to 1.90$) and non-obviousness ($1.81 \to 1.46$).

To investigate this further, we conducted an expert evaluation on 50 random samples generated using the updated prompt. Human experts noted that the newly generated RQs are indeed broader. As shown in Table~\ref{tab:expert_vs_llm_judge_comparative_non_obviousness_newprompt}, human experts still predominantly prefer the author-anchored (GT) RQs in terms of non-obviousness, awarding them the win 62\% and 56\% of the time across the two experts. In contrast, the LLM judges continued to favor the model-generated RQs, choosing them 50\% and 40\% of the time. This is reflected in the agreement metrics (Table~\ref{tab:expert_vs_llm_judge_comparative_non_obviousness_newprompt}): intra-expert (54\%) and intra-judge (62\%) alignments are notably higher than the poor inter-group expert--judge agreement (30--40\%).

Crucially, Table~\ref{tab:sourcebound_agreement_newprompt} shows a strong alignment (42--50\%) is maintained between human experts' non-obviousness judgments and the LLM judge's source-boundedness evaluations. This reaffirms the strong correlation between human experts' non-obviousness judgments and the source-boundedness assessments of LLM judges.

These results highlight a fundamental misalignment between human experts and LLM judges regarding novelty assessment. Human experts correlate less source-bound questions with greater non-obviousness. However, LLM judges exhibit the opposite tendency. Under the original prompt, the LLM awarded high novelty scores to model-generated RQs that were highly narrow and source-bound. Yet, when the updated prompt successfully mitigated this source-boundedness, the novelty scores assigned by the LLM (particularly for non-obviousness and originality) concurrently declined. This implies that LLM-judged novelty is conflated with narrowness, diverging from human assessments of what constitutes genuine non-obviousness. These findings motivate the development of more robust novelty evaluation frameworks, including better metrics and more reliable LLM-as-a-judge protocols.

\newpage

\section{Conclusion}

We studied the reliability of LLM-as-judge for scientific novelty assessment. We used research-question generation as the testbed because it is an upstream step before full idea generation. It is also easier to inspect than a complete method or research plan.

\dataset{} provides author-anchored RQs from real papers. These RQs are reference points, not the only valid questions. This setup lets us compare model-generated RQs against human-anchored RQs and test whether novelty judgments are stable.

Our results show a novelty mirage. Under standalone scoring, LLM judges rate model-generated RQs as highly novel in every novelty dimension (i.e., non-obviousness, originality, and gap-addressing). In comparative settings, the judges amplify it, heavily favoring model outputs over author-authored baselines. However, domain experts sharply disagree with these judgments, preferring the author-anchored RQs for their broader scope and non-obviousness. Crucially, when we explicitly instructed LLMs to evaluate this \emph{narrowness}, their assessments of source-boundedness strongly aligned with human judgments of non-obviousness.

These results do not mean that novelty evaluation is impossible. Novelty is partly subjective, and even human reviewers can disagree. But useful evaluation should be stable across settings and should not strongly conflict with expert judgment. Our findings show that current LLM-as-judge pipelines do not yet meet this standard for scientific RQ novelty. Future evaluations should use multiple judges, check agreement, and include scope or narrowness as a first-class dimension.

\section*{Limitations}

Our benchmark is currently limited to recent computer science papers from arXiv and may not fully capture research practices in other scientific domains. In addition, novelty evaluation remains challenging due to the central dependence on LLM-as-judge for open-ended research question generation. While we include expert evaluations to study this issue, broader and more diverse human studies are still needed.

\bibliography{custom}
\bibliographystyle{plainnat}
\onecolumn

\appendix

\section{Appendix}
\label{sec:appendix}




\begin{table}[ht]
\centering
\small
\resizebox{0.4\linewidth}{!}{
\setlength{\tabcolsep}{4pt}
\renewcommand{\arraystretch}{1.05}
\begin{tabular}{lrrrr}
\toprule
\textbf{Sub-field} & \textbf{Papers} & \textbf{\%} & \textbf{RQs} & \textbf{\%} \\
\midrule
cs.RO  & 118 & 15.8 & 245 & 17.1 \\
cs.CV  & 106 & 14.2 & 222 & 15.5 \\
cs.CL  &  87 & 11.7 & 173 & 12.1 \\
cs.LG  &  85 & 11.4 & 162 & 11.3 \\
cs.AI  &  78 & 10.5 & 146 & 10.2 \\
cs.SD  &  59 &  7.9 & 115 &  8.0 \\
cs.IR  &  50 &  6.7 &  92 &  6.4 \\
cs.CR  &  45 &  6.0 &  75 &  5.2 \\
cs.IT  &  44 &  5.9 &  69 &  4.8 \\
cs.SE  &  35 &  4.7 &  67 &  4.7 \\
cs.DC  &  21 &  2.8 &  36 &  2.5 \\
cs.NI  &  11 &  1.5 &  19 &  1.3 \\
cs.HC  &   7 &  0.9 &  13 &  0.9 \\
\midrule
\textbf{Total} & \textbf{746} & \textbf{100.0} & \textbf{1{,}434} & \textbf{100.0} \\
\bottomrule
\end{tabular}
}
\caption{Distribution of source papers and research questions across arXiv CS sub-fields.}
\label{tab:dataset-subfields}
\end{table}
\subsection{Novelty Metric}
\label{app:novelty}

\paragraph{Originality \((0\text{--}3)\).}
Does the candidate's specific framing or object of inquiry go beyond what
is directly stated or asked in the background?
\begin{itemize}
    \item \(0\): paraphrase of something already in the background.
    \item \(1\): very close to something stated, with only a minor
    reframing.
    \item \(2\): noticeably different framing or angle from what the
    background covers.
    \item \(3\): framing not present in the background at all.
\end{itemize}

\paragraph{Gap addressing \((0\text{--}3)\).}
Does the candidate engage a gap in the background, either explicitly
acknowledged or implicitly present?
\begin{itemize}
    \item \(0\): does not target any gap; reiterates established content.
    \item \(1\): targets a peripheral or weak gap.
    \item \(2\): targets a clearly identifiable gap.
    \item \(3\): targets a substantive, central gap.
\end{itemize}

\paragraph{Non-obviousness \((0\text{--}3)\).}
Is the candidate a non-trivial step from what is in the background?
\begin{itemize}
    \item \(0\): trivial follow-up; the standard next experiment any
    reader would propose.
    \item \(1\): somewhat obvious extension.
    \item \(2\): requires thought to formulate; non-obvious framing or
    mechanism.
    \item \(3\): striking or unexpected angle that the background does not
    hint at.
\end{itemize}

\paragraph{Relevance \((0/1)\).}
\begin{itemize}
    \item \(0\): off-topic; the candidate's subject matter does not
    intersect with what the background studies.
    \item \(1\): on-topic; a study of this question could plausibly cite
    the background papers as motivation.
\end{itemize}

\subsection{Narrowness Metric}
\label{app:narrowness}

\paragraph{Source-boundedness \((0\text{--}3)\).}
Does the candidate depend on the specific background paper's method,
dataset, benchmark, component, result, behavior, or setting?
\begin{itemize}
    \item \(0\): not about the source paper's own contribution; asks
    about a broader problem, new mechanism, setting, or direction.
    \item \(1\): motivated by the source paper and reuses some of its
    concepts, but the main question goes beyond it.
    \item \(2\): mostly asks about the source paper's method, result,
    behavior, or setting, or a close variant.
    \item \(3\): directly studies, diagnoses, tunes, evaluates, or
    modifies the source paper's own contribution.
\end{itemize}

\paragraph{Diagnostic / test-list framing \((0\text{--}3)\).}
Is the candidate framed as an analysis, ablation, correlation, causal
test, predictor study, or list of measurable factors, judged only by the
form and intent of the question?
\begin{itemize}
    \item \(0\): asks for a method, mechanism, framework, or broader
    direction.
    \item \(1\): includes some evaluation or analysis, but is still mainly
    about a method or direction.
    \item \(2\): mainly asks which factors, behaviors, cases,
    interventions, correlations, or predictors explain a result.
    \item \(3\): mostly a measurement checklist, ablation plan,
    causal/correlation test, failure analysis, or predictor study.
\end{itemize}

\subsection{Author-Anchor Overlap Metric}

The author-anchor overlap metric scores each generated RQ independently
against the author-anchored reference RQ along four dimensions:
\emph{topic overlap} (shared subject matter or domain), \emph{object of
inquiry} (same variable, mechanism, relationship, or phenomenon),
\emph{expected answer type} (same kind of answer---mechanism, measurement,
comparison, yes/no, or method), and \emph{scope} (broader, narrower, or
equal to the reference).

\paragraph{Pairwise overlap score \((0\text{--}4)\).}
\begin{itemize}
    \item \(0\): unrelated. No topic overlap; answering one says nothing
    about the other.
    \item \(1\): same topic, different question. Shared subject area, but
    object of inquiry, answer type, or scope differ so the answers do not
    overlap.
    \item \(2\): substantial overlap, distinct scope or angle. Same object
    of inquiry, but a correct answer to one only partially covers the
    other.
    \item \(3\): semantic equivalent. A paraphrase---a correct answer to
    one is a correct answer to the other.
    \item \(4\): subsumes the reference. Strictly more general with scope
    fully containing the reference, so any complete answer must also answer
    it; vague wording alone does not qualify.
\end{itemize}

\subsection{Prompts}
\label{app:prompts}


\begin{promptbox}{Standalone novelty prompt}
\promptfield{System}

You are an expert reviewer assessing the NOVELTY of a research question
relative to a set of BACKGROUND papers.

You will receive EXACTLY ONE candidate research question (presented as
\texttt{[0] <text>}). Score it on its own merits against BACKGROUND.

Novelty here means: the extent to which the question goes beyond what is
already established, claimed, or asked in BACKGROUND. A question is more
novel if:
\begin{enumerate}
    \item its specific framing / object of inquiry is not present in the
    background;
    \item it engages a gap in the background --- explicit (the paper says
    ``we leave X to future work'', ``X is a limitation'', ``X remains
    open'') OR implicit (the paper studies a phenomenon only in one regime,
    takes a premise for granted without justifying it, mentions a
    limitation only in passing, or relies on an assumption that is not
    validated);
    \item it is not a trivial follow-up that any reader would propose.
\end{enumerate}

Critical guidance: Judge novelty ONLY against the provided BACKGROUND. Do
not invoke outside knowledge of the field. If a question would be novel in
the wider literature but is essentially restated in BACKGROUND, it is NOT
novel here.

\textbf{Hierarchical scoring.} Scoring is HIERARCHICAL. Decide RELEVANCE
first.
\begin{itemize}
    \item relevance = 1 if the candidate is on-topic for BACKGROUND --- a
    study of this question could plausibly cite these papers as motivation.
    \item relevance = 0 if the candidate is off-topic --- its subject
    matter does not intersect with what BACKGROUND studies.
\end{itemize}
If relevance = 0: SET originality = 0, gap\_addressing = 0,
non\_obviousness = 0. An off-topic candidate has no novelty value --- it is
simply irrelevant, not ``creative''. Do not award any partial novelty. If
relevance = 1: score the three 0--3 sub-dimensions below on their own
merits.

\textbf{Rubric.}

originality (0--3): does the candidate's specific framing / object of
inquiry go beyond what is directly stated or asked in BACKGROUND?
\begin{itemize}
    \item 0 = paraphrase of something already in background
    \item 1 = very close to something stated, only a minor reframing
    \item 2 = noticeably different framing or angle from what background
    covers
    \item 3 = framing not present in background at all
\end{itemize}

gap\_addressing (0--3): does the candidate engage a gap in the background
--- either explicitly acknowledged or implicitly present (an under-explored
regime, an unjustified assumption, a passing limitation, a hard-coded
design choice that the paper does not validate)?
\begin{itemize}
    \item 0 = does not target any gap; reiterates established content
    \item 1 = targets a peripheral or weak gap
    \item 2 = targets a clearly identifiable gap (explicit or implicit)
    \item 3 = targets a substantive, central gap
\end{itemize}

non\_obviousness (0--3): is the candidate a non-trivial step from what's in
background?
\begin{itemize}
    \item 0 = trivial follow-up; the standard next experiment any reader
    would propose
    \item 1 = somewhat obvious extension
    \item 2 = requires thought to formulate; non-obvious framing or
    mechanism
    \item 3 = striking or unexpected angle that BACKGROUND does not hint at
\end{itemize}

relevance (0\,$|$\,1): set per the hierarchical rule above.

Return a single JSON object with EXACTLY this shape, and nothing else (no
prose, no markdown):
\begin{verbatim}
{
  "scores": [
    {
      "originality": int,
      "gap_addressing": int,
      "non_obviousness": int,
      "relevance": int
    }
  ]
}
\end{verbatim}
\texttt{scores} must have EXACTLY ONE entry --- the rubric for the single
candidate.

\promptfield{User}

\begin{verbatim}
BACKGROUND:
{background_block}

CANDIDATE:
{candidates_block}
\end{verbatim}
\end{promptbox}

\begin{promptbox}{Comparative novelty prompt}
\promptfield{System}

You are an expert reviewer assessing the NOVELTY of research questions
relative to a set of BACKGROUND papers.

Novelty here means: the extent to which a question goes beyond what is
already established, claimed, or asked in BACKGROUND. A question is more
novel if:
\begin{enumerate}
    \item its specific framing / object of inquiry is not present in the
    background;
    \item it engages a gap in the background --- explicit or implicit;
    \item it is not a trivial follow-up that any reader would propose.
\end{enumerate}

This is a LISTWISE comparative evaluation. Consider all candidates together
to calibrate what counts as more original, more gap-addressing, and more
non-obvious for this BACKGROUND. Then assign each candidate its own rubric
scores.

Critical guidance:
\begin{itemize}
    \item Judge background-relative novelty: assess whether each candidate
    goes beyond what is established, claimed, or asked in the provided
    BACKGROUND. Do not rely on outside literature to reward or penalize a
    candidate.
    \item Treat every candidate equally. Do not assume any candidate is
    human-authored or model-generated.
    \item A candidate that is off-topic is not novel; it is irrelevant.
\end{itemize}

\textbf{Hierarchical scoring.} For EACH candidate, decide RELEVANCE first.
\begin{itemize}
    \item relevance = 1 if the candidate is on-topic for BACKGROUND --- a
    study of this question could plausibly cite these papers as motivation.
    \item relevance = 0 if the candidate is off-topic --- its subject
    matter does not intersect with what BACKGROUND studies.
\end{itemize}
If relevance = 0: SET originality = 0, gap\_addressing = 0,
non\_obviousness = 0. If relevance = 1: score the three 0--3 sub-dimensions
below.

\textbf{Per-candidate rubric.}

originality (0--3):
\begin{itemize}
    \item 0 = paraphrase of something already in BACKGROUND
    \item 1 = very close to something stated, only a minor reframing
    \item 2 = noticeably different framing or angle from what BACKGROUND
    covers
    \item 3 = framing not present in BACKGROUND at all
\end{itemize}

gap\_addressing (0--3):
\begin{itemize}
    \item 0 = does not target any gap; reiterates established content
    \item 1 = targets a peripheral or weak gap
    \item 2 = targets a clearly identifiable gap
    \item 3 = targets a substantive, central gap
\end{itemize}

non\_obviousness (0--3):
\begin{itemize}
    \item 0 = trivial follow-up; the standard next experiment any reader
    would propose
    \item 1 = somewhat obvious extension
    \item 2 = requires thought to formulate; non-obvious framing or
    mechanism
    \item 3 = striking or unexpected angle that BACKGROUND does not hint at
\end{itemize}

\textbf{Ranking.} After scoring all candidates, rank them from most novel
to least novel. Use the total novelty score originality + gap\_addressing +
non\_obviousness as the main criterion. If two candidates have the same
total score, break ties by preferring the candidate with higher
non\_obviousness, then higher originality, then higher gap\_addressing. If
they are still indistinguishable, place them in either order.

Return a single JSON object with EXACTLY this shape, and nothing else:
\begin{verbatim}
{
  "scores": [
    {
      "originality": int,
      "gap_addressing": int,
      "non_obviousness": int,
      "relevance": int
    },
    ...
  ],
  "ranking": [int, int, ...]
}
\end{verbatim}
\texttt{scores} must have exactly N entries in the SAME order as the
candidates were presented. \texttt{ranking} must be a permutation of
candidate indices from most novel to least novel, using 0-based indexing.

\promptfield{User}

\begin{verbatim}
BACKGROUND:
{background_block}

CANDIDATES:
{candidates_block}
\end{verbatim}
\end{promptbox}


\begin{promptbox}{Standalone narrowness prompt}
\promptfield{System}

You are an expert research scientist.

Your task is to evaluate one candidate research question against one
background paper.

You must assign two independent scores:

1. SOURCE-BOUNDEDNESS. This measures whether the research question depends on
the specific background paper's method, dataset, benchmark, component,
result, behavior, or setting.

2. DIAGNOSTIC / TEST-LIST FRAMING. This measures whether the research
question is framed as an analysis, ablation, correlation, causal test,
predictor study, failure-case study, or list of measurable factors.

These two scores are independent. A question can be:
\begin{itemize}
    \item source-bound and diagnostic: ``Which failure modes does GRPO
    show under long-chain reasoning?''
    \item source-bound but not diagnostic: ``Can GRPO be extended with
    uncertainty-aware advantage normalization?''
    \item diagnostic but not source-bound: ``Which training signals
    predict reward hacking across RL-based reasoning methods?''
    \item neither source-bound nor diagnostic: ``How can online difficulty
    filtering stabilize RL training under dynamic prompt exclusion?''
\end{itemize}

Important rules:
\begin{itemize}
    \item Do not judge whether the question is good.
    \item Do not judge whether the question is novel.
    \item Do not judge whether the question is feasible.
    \item Judge source-boundedness only by the question's dependence on the
    specific background paper.
    \item Judge diagnostic/test-list framing only by the form and intent of
    the question.
    \item Do not increase the source-boundedness score only because the
    question is diagnostic.
    \item Do not increase the diagnostic score only because the question is
    source-bound.
    \item A question can avoid naming the source method and still be
    source-bound if it studies the same method, behavior, setting, or
    result from the background paper.
    \item A question can mention a broad concept from the background paper
    without being source-bound if it asks about a broader problem or new
    direction.
\end{itemize}

SOURCE-BOUNDEDNESS SCALE:
\begin{itemize}
    \item 0 = Not source-bound. The question is not mainly about the source
    paper's own method, dataset, benchmark, component, result, behavior, or
    setting. It asks about a broader problem, new mechanism, new setting, or
    new research direction.
    \item 1 = Mildly source-bound. The question is motivated by the source
    paper and may reuse some of its concepts, but the main question goes
    beyond the source paper.
    \item 2 = Source-bound. The question mostly asks about the source
    paper's method, dataset, benchmark, component, result, behavior, or
    setting, or a close variant of it.
    \item 3 = Strongly source-bound. The question directly studies,
    diagnoses, tunes, evaluates, or modifies the source paper's own method,
    dataset, benchmark, component, result, behavior, or setting.
\end{itemize}

DIAGNOSTIC / TEST-LIST SCALE:
\begin{itemize}
    \item 0 = Not diagnostic/test-list. The question asks for a method,
    mechanism, framework, formulation, or broader research direction.
    \item 1 = Mildly diagnostic/test-list. The question includes some
    evaluation or analysis, but it is still mainly about a method,
    mechanism, framework, or broader research direction.
    \item 2 = Diagnostic/test-list. The question mainly asks which factors,
    behaviors, cases, interventions, ablations, correlations, causal links,
    or predictors explain a result.
    \item 3 = Strongly diagnostic/test-list. The question is mostly a
    measurement checklist, ablation plan, causal/correlation test, failure
    analysis, or predictor study.
\end{itemize}

Return exactly one JSON object with this schema:
\begin{verbatim}
{
  "mentions_source_specific_name": 0 or 1,
  "main_object_in_source": 0 or 1,
  "introduces_new_mechanism": 0 or 1,
  "source_bound_score": 0 | 1 | 2 | 3,
  "diagnostic_test_score": 0 | 1 | 2 | 3,
  "source_bound_scope": "not_source_bound" | "mildly_source_bound"
                      | "source_bound" | "strongly_source_bound",
  "diagnostic_scope": "not_diagnostic" | "mildly_diagnostic"
                    | "diagnostic" | "strongly_diagnostic",
  "short_reason_source_bound": "one short sentence",
  "short_reason_diagnostic": "one short sentence"
}
\end{verbatim}

Definitions:
\begin{itemize}
    \item \texttt{mentions\_source\_specific\_name} = 1 if the question
    explicitly mentions a method name, dataset name, benchmark name,
    acronym, component, or technical term specific to the background paper.
    \item \texttt{main\_object\_in\_source} = 1 if the main object being
    studied already appears in the background paper's idea, contribution,
    method, result, behavior, or setting.
    \item \texttt{introduces\_new\_mechanism} = 1 if the question proposes
    or asks about a mechanism, algorithm, framework, setting, or
    formulation that is not already part of the background paper's idea.
    \item The two short reasons must be concise and should not repeat each
    other.
\end{itemize}

\promptfield{User}

\begin{verbatim}
BACKGROUND PAPER:
{background}

CANDIDATE RESEARCH QUESTION:
{candidate_rq}
\end{verbatim}
\end{promptbox}

\begin{promptbox}{Comparative narrowness prompt}
\promptfield{System}

You are an expert research scientist.

Your task is to evaluate SEVERAL candidate research questions against one
background paper.

Each candidate is scored INDEPENDENTLY on the two scales below. You are
shown all candidates together only so you can read them efficiently; a
candidate's two scores must depend only on that candidate and the
background paper, never on the other candidates.

\medskip
\noindent\textit{The two-score definitions, the SOURCE-BOUNDEDNESS SCALE
(0--3), the DIAGNOSTIC / TEST-LIST SCALE (0--3), the important rules, and
the per-candidate field definitions are identical to the standalone
narrowness prompt above. The only change is that the final instruction
reads ``For EACH candidate research question you assign one JSON object
with this schema'' and the output is wrapped in a \texttt{scores} array.}
\medskip

Return exactly one JSON object:
\begin{verbatim}
{
  "scores": [ one object per candidate, in the SAME order as presented,
              using the per-candidate schema above ]
}
\end{verbatim}

The \texttt{scores} array length must equal the number of candidates.
Return only this JSON object and nothing else.

\promptfield{User}

\begin{verbatim}
BACKGROUND PAPER:
{background}

CANDIDATE RESEARCH QUESTIONS:
{candidates_block}
\end{verbatim}
\end{promptbox}


\begin{promptbox}{Author-anchor overlap prompt}
\promptfield{System}

You are an expert research scientist judging whether candidate research
questions match a target author-anchored research question.

You will be given:
\begin{itemize}
    \item \textbf{GROUND\_TRUTH}: one research question taken from a real
    paper.
    \item \textbf{CANDIDATES}: a list of research questions generated by an
    automated system, indexed \(0,\ldots,N-1\).
\end{itemize}

The task has two parts.

\promptfield{Part A: Pairwise scoring}

For EACH candidate (independently), assign an integer score 0--4 using the
rubric below. Score each candidate on its own merits against the
author-anchored RQ, NOT relative to the other candidates.

To decide a score, evaluate the candidate vs the author-anchored RQ along
four dimensions:
\begin{itemize}
    \item TOPIC OVERLAP --- Do they share the broader subject matter /
    domain / area of study?
    \item OBJECT OF INQUIRY --- Are they asking about the same thing (same
    variable, mechanism, relationship, or phenomenon)?
    \item EXPECTED ANSWER TYPE --- Would correctly answering them produce
    the same kind of answer (a mechanism, a measurement, a comparison, a
    yes/no, a method, etc.)?
    \item SCOPE --- Is the candidate's scope broader, narrower, or roughly
    equal to the author-anchored RQ's?
\end{itemize}

Then map those four readings to a score:
\begin{itemize}
    \item 0 --- UNRELATED. No topic overlap. The candidate addresses
    subject matter that does not intersect with the author-anchored RQ's
    domain or methods. Answering one tells you nothing about the other.
    \item 1 --- SAME TOPIC, DIFFERENT QUESTION. Same broader subject area
    (same field, often same systems being discussed), but the questions
    differ on at least one of: object of inquiry, expected answer type, or
    scope --- in a way that the answers would not meaningfully overlap.
    Answering the candidate does not answer the author-anchored RQ.
    \item 2 --- SUBSTANTIAL OVERLAP, DISTINCT SCOPE OR ANGLE. Same object
    of inquiry as the author-anchored RQ, but the scope or angle differs
    noticeably. The candidate might be a narrower sub-question, a specific
    mechanism it alludes to, or a different methodological angle on the same
    question. A correct answer to the candidate partially overlaps a
    correct answer to the author-anchored RQ, but does not fully cover it
    (nor vice versa).
    \item 3 --- SEMANTIC EQUIVALENT. A paraphrase of the author-anchored
    RQ. Same object of inquiry, same expected answer type, comparable
    scope. Wording, terminology, or framing differs, but the questions
    reduce to the same inquiry --- a correct answer to one is a correct
    answer to the other.
    \item 4 --- SUBSUMES the author-anchored RQ. The candidate is strictly
    more general, and its scope FULLY CONTAINS the author-anchored RQ's. Any
    complete answer to the candidate must also answer it. Note: ``more
    general AND complete in scope'' --- merely vague wording without true
    scope expansion is level 2, not level 4.
\end{itemize}

\promptfield{Part B: Minimal covering subset}

Identify the smallest subset of candidate indices whose UNION is
semantically equivalent to the author-anchored RQ (i.e. collectively
answering this subset would answer it).
\begin{itemize}
    \item If any single candidate scores 3 or 4, the subset is just [that
    one index] --- prefer the smallest index when there are ties.
    \item Otherwise, you may combine 2+ candidates whose sub-questions
    together cover the author-anchored RQ.
    \item If no subset (single or combined) covers it, return [].
\end{itemize}
Set \texttt{covered = true} iff the subset is non-empty.

Return a single JSON object with EXACTLY these keys, in this order, and
nothing else (no prose, no markdown):
\begin{verbatim}
{
  "pairwise_scores": [int, int, ...],
  "covering_subset_indices": [int, ...],
  "covered": bool
}
\end{verbatim}
\texttt{pairwise\_scores} must have exactly the same length as
\texttt{CANDIDATES}, in the same order.

\promptfield{User}

\begin{verbatim}
GROUND_TRUTH:
{ground_truth}

CANDIDATES:
{candidates_block}
\end{verbatim}
\end{promptbox}

\begin{promptbox}{RQ generation prompt}
\promptfield{System}

You are an expert research scientist whose job is to read background
literature and identify meaningful, unsolved research problems.

\promptfield{Task}

For the background paper(s) the user provides:

\begin{enumerate}
    \item Identify the research gaps in the literature: the limitations,
    weaknesses, unresolved problems, and unaddressed challenges that the
    work(s) leave open.
    \item Propose research questions that, if answered, would directly
    close one or more of those gaps.
\end{enumerate}

\promptfield{Multi-paper rule}

When the user provides more than one background paper, a gap may apply to
one paper, several papers, or all of them. Do not force every gap to be
shared by every paper. If gaps from different papers converge on the same
research question, group them under one RQ rather than splitting them. If
gaps motivate clearly distinct directions, separate them into distinct
RQs.

\promptfield{Research questions}

\begin{itemize}
    \item Produce exactly 5 research questions.
    \item Each question must be phrased as a question and end with ``?''.
    \item Each question must be specific, not a generic question such as
    ``how to improve X''.
    \item Each question must be answerable: a follow-up paper could
    plausibly test or address it.
    \item Each question must be grounded in one of the listed gaps. Do not
    introduce themes outside the provided text.
\end{itemize}

\promptfield{Output format}

The final visible answer must be a single JSON object conforming exactly
to the following schema:

\begin{verbatim}
{
  "gaps": [
    "<description of gap 1>",
    "<description of gap 2>"
  ],
  "research_questions": [
    "<research question 1 ending with '?'>",
    "<research question 2 ending with '?'>",
    "<research question 3 ending with '?'>",
    "<research question 4 ending with '?'>",
    "<research question 5 ending with '?'>"
  ]
}
\end{verbatim}

Both lists must be non-empty. The \texttt{research\_questions} list must
contain exactly 5 entries. Use plain strings only: no nested objects and
no extra keys.

\promptfield{User prompt: single-paper setting}

\begin{verbatim}
Below is one background paper. Identify its research gaps and propose
research questions addressing them, following the rules in your system
instructions.

{papers_block}
\end{verbatim}

\promptfield{User prompt: multi-paper setting}

\begin{verbatim}
Below are {n_papers} background papers. Identify the research gaps they
raise — a gap may apply to one paper, several, or all of them — and
propose research questions addressing those gaps, grouping convergent
gaps under a single RQ where the literature supports it. Follow the rules
in your system instructions.

{papers_block}
\end{verbatim}
\end{promptbox}

\subsection{Evaluation Outputs}
\label{app:eval_outputs}

\begin{tcolorbox}[colback=gray!5!white, colframe=black!75!white, title=\textbf{Output of Standalone Scoring (Novelty)}, arc=4mm]

    \textbf{\textsf{BACKGROUND PAPER}} \\
    \textbf{Title:} DeepSeek-R1: Incentivizing Reasoning Capability in LLMs via Reinforcement Learning \\
    \textbf{Abstract (Summarized):} \\
    This paper demonstrates that pure reinforcement learning (RL), without human-annotated data, can incentivize advanced, emergent reasoning capabilities (e.g., self-reflection, verification) in Large Language Models. The resulting models achieve superior performance on verifiable tasks like math and coding, and their learned reasoning patterns can be successfully distilled into smaller models.

    \tcblower

    \textbf{\textsf{AUTHOR-ANCHORED RQ}} \hfill \textbf{Score: 8} \\
    \texttt{\footnotesize \{'originality': 3, 'gap\_addressing': 3, 'non\_obviousness': 2, 'relevance': 1\}} \\
    \textbf{RQ:} \textit{``How can large language models dynamically compress intermediate reasoning steps into compact hidden state representations to reduce token consumption and computational overhead during complex reasoning tasks?''}

    \tcbline

    \textbf{\textsf{BEST STANDALONE RQ PER MODEL}} \hfill {\footnotesize (ranked by non\_obviousness $>$ originality $>$ gap\_addressing)}
    \begin{itemize}
        \setlength\itemsep{0.5em}

        \item \textbf{[gemma-4-31b-it]} \hfill (Score: 8) \\
        \texttt{\footnotesize \{'originality': 3, 'gap\_addressing': 3, 'non\_obviousness': 2, 'relevance': 1\}} \\
        To what extent can an iterative co-training loop between the policy model and the reward model---where the reward model is updated using rejected samples from the latest policy---mitigate reward hacking in non-verifiable tasks like creative writing?

        \item \textbf{[gpt-5.5]} \hfill (Score: 8) \\
        \texttt{\footnotesize \{'originality': 3, 'gap\_addressing': 3, 'non\_obviousness': 2, 'relevance': 1\}} \\
        Can an uncertainty-aware ensemble reward model reduce reward hacking in helpfulness and writing RL compared with a single model-based reward while preserving AlpacaEval and ArenaHard performance?

        \item \textbf{[gemini-3.1-pro]} \hfill (Score: 7) \\
        \texttt{\footnotesize \{'originality': 2, 'gap\_addressing': 3, 'non\_obviousness': 2, 'relevance': 1\}} \\
        What underlying shifts in attention mechanisms cause few-shot prompting to degrade the performance of outcome-based RL models, and can instruction-tuning modifications mitigate this vulnerability?

        \item \textbf{[deepseek-v4-pro]} \hfill (Score: 6) \\
        \texttt{\footnotesize \{'originality': 2, 'gap\_addressing': 3, 'non\_obviousness': 1, 'relevance': 1\}} \\
        Why does few-shot prompting harm DeepSeek-R1's reasoning accuracy, and can a meta-learning or prompt-robustness fine-tuning stage make such reasoning models invariant to demonstration style without hurting zero-shot generalization?

        \item \textbf{[gpt-oss-20b]} \hfill (Score: 6) \\
        \texttt{\footnotesize \{'originality': 2, 'gap\_addressing': 3, 'non\_obviousness': 1, 'relevance': 1\}} \\
        What is the impact of adding a token-efficiency penalty to the GRPO objective on the average token usage for simple versus complex reasoning tasks, and how does this affect final accuracy?

        \item \textbf{[qwen3-30b-a3b-thinking-2507]} \hfill (Score: 5) \\
        \texttt{\footnotesize \{'originality': 1, 'gap\_addressing': 3, 'non\_obviousness': 1, 'relevance': 1\}} \\
        How can we design a language consistency mechanism that maintains high reasoning performance across multiple languages while minimizing language mixing in LLMs trained with pure reinforcement learning?

    \end{itemize}
\end{tcolorbox}

\begin{tcolorbox}[colback=gray!5!white, colframe=black!75!white, title=\textbf{Output of Comparative Scoring (Novelty)}, arc=4mm]

    \textbf{\textsf{BACKGROUND PAPER}} \\
    \textbf{Title:} Accelerating Large Language Model Decoding with Speculative Sampling \\
    \textbf{Abstract (Summarized):} \\
    This paper introduces ``speculative sampling,'' an algorithm that accelerates transformer decoding by generating multiple tokens per call. By using a faster, smaller ``draft'' model to propose short continuations and a larger target model to score them in parallel via a novel modified rejection sampling scheme, the algorithm preserves the target model's distribution. Evaluated on the 70B parameter Chinchilla model, this method achieves a 2--2.5$\times$ speedup without compromising sample quality or requiring modifications to the target model.

    \tcblower

    \textbf{\textsf{AUTHOR-ANCHORED RQ}} \hfill \textbf{Score: 8} \textbar\ \textbf{Rank: 0} \\
    \texttt{\footnotesize \{'originality': 3, 'gap\_addressing': 2, 'non\_obviousness': 3, 'relevance': 1\}} \\
    \textbf{RQ:} \textit{``How can introducing a controlled bias via a stepwise process reward model improve the reasoning accuracy of speculative decoding by retaining high-quality draft tokens that strictly unbiased methods would reject?''}

    \tcbline

    \textbf{\textsf{GPT-5.5 generated 5 RQs}} \hfill {\footnotesize (ranked by non\_obviousness $>$ originality $>$ gap\_addressing)}
    \begin{itemize}
        \setlength\itemsep{0.75em}

        \item \textbf{[Rank 1]} \hfill (Score: 7) \\
        \texttt{\footnotesize \{'originality': 3, 'gap\_addressing': 2, 'non\_obviousness': 2, 'relevance': 1\}} \\
        Can prompt-level estimates of draft--target agreement be used before generation to route requests among speculative sampling configurations with different draft models and K values?

        \item \textbf{[Rank 2]} \hfill (Score: 6) \\
        \texttt{\footnotesize \{'originality': 2, 'gap\_addressing': 2, 'non\_obviousness': 2, 'relevance': 1\}} \\
        Can a per-step adaptive choice of lookahead K based on draft-target agreement or target entropy reduce mean and P90/P99 latency compared with a fixed K while preserving the target model distribution?

        \item \textbf{[Rank 3]} \hfill (Score: 6) \\
        \texttt{\footnotesize \{'originality': 2, 'gap\_addressing': 2, 'non\_obviousness': 2, 'relevance': 1\}} \\
        What numerical divergence from the target sampling distribution is introduced by speculative sampling when logits, draft models, target models, or KV caches are evaluated in bfloat16, int8, or int4 precision?

        \item \textbf{[Rank 4]} \hfill (Score: 6) \\
        \texttt{\footnotesize \{'originality': 2, 'gap\_addressing': 2, 'non\_obviousness': 2, 'relevance': 1\}} \\
        Can a learned policy select the draft model, hardware topology, and lookahead K that minimize end-to-end latency for a given target model, task distribution, and service-level latency objective?

        \item \textbf{[Rank 5]} \hfill (Score: 6) \\
        \texttt{\footnotesize \{'originality': 2, 'gap\_addressing': 3, 'non\_obviousness': 1, 'relevance': 1\}} \\
        How does speculative sampling perform under batch sizes greater than 1 and continuous batching in terms of throughput, mean latency, P99 latency, accelerator utilization, and KV-cache bandwidth compared with standard autoregressive sampling?

    \end{itemize}
\end{tcolorbox}

\begin{tcolorbox}[colback=gray!5!white, colframe=black!75!white, title=\textbf{Output of Standalone Scoring (Narrowness)}, arc=4mm]

    \textbf{\textsf{BACKGROUND PAPER}} \\
    \textbf{Title:} DeepSeek-R1: Incentivizing Reasoning Capability in LLMs via Reinforcement Learning \\
    \textbf{Abstract (Summarized):} \\
    This paper demonstrates that pure reinforcement learning (RL), without human-annotated data, can incentivize advanced, emergent reasoning capabilities (e.g., self-reflection, verification) in Large Language Models. The resulting models achieve superior performance on verifiable tasks like math and coding, and their learned reasoning patterns can be successfully distilled into smaller models.

    \tcblower

    \textbf{\textsf{AUTHOR-ANCHORED RQ}} \hfill \textbf{Narrowness: 1} \\
    \texttt{\footnotesize \{'source\_boundedness': 1, 'diagnostic\_framing': 0\}} \\
    \textbf{RQ:} \textit{``How can large language models dynamically compress intermediate reasoning steps into compact hidden state representations to reduce token consumption and computational overhead during complex reasoning tasks?''}

    \tcbline

    \textbf{\textsf{LEAST-NARROW STANDALONE RQ PER MODEL}} \hfill {\footnotesize (ranked by source\_boundedness $>$ diagnostic\_framing, ascending)}
    \begin{itemize}
        \setlength\itemsep{0.5em}

        \item \textbf{[gemma-4-31b-it]} \hfill (Narrowness: 2) \\
        \texttt{\footnotesize \{'source\_boundedness': 1, 'diagnostic\_framing': 1\}} \\
        To what extent can an iterative co-training loop between the policy model and the reward model---where the reward model is updated using rejected samples from the latest policy---mitigate reward hacking in non-verifiable tasks like creative writing?

        \item \textbf{[gpt-5.5]} \hfill (Narrowness: 2) \\
        \texttt{\footnotesize \{'source\_boundedness': 1, 'diagnostic\_framing': 1\}} \\
        Can an uncertainty-aware ensemble reward model reduce reward hacking in helpfulness and writing RL compared with a single model-based reward while preserving AlpacaEval and ArenaHard performance?

        \item \textbf{[qwen3-30b-a3b-thinking-2507]} \hfill (Narrowness: 3) \\
        \texttt{\footnotesize \{'source\_boundedness': 2, 'diagnostic\_framing': 1\}} \\
        How can we design a language consistency mechanism that maintains high reasoning performance across multiple languages while minimizing language mixing in LLMs trained with pure reinforcement learning?

        \item \textbf{[gpt-oss-20b]} \hfill (Narrowness: 4) \\
        \texttt{\footnotesize \{'source\_boundedness': 2, 'diagnostic\_framing': 2\}} \\
        What is the impact of adding a token-efficiency penalty to the GRPO objective on the average token usage for simple versus complex reasoning tasks, and how does this affect final accuracy?

        \item \textbf{[gemini-3.1-pro]} \hfill (Narrowness: 5) \\
        \texttt{\footnotesize \{'source\_boundedness': 2, 'diagnostic\_framing': 3\}} \\
        What underlying shifts in attention mechanisms cause few-shot prompting to degrade the performance of outcome-based RL models, and can instruction-tuning modifications mitigate this vulnerability?

        \item \textbf{[deepseek-v4-pro]} \hfill (Narrowness: 4) \\
        \texttt{\footnotesize \{'source\_boundedness': 3, 'diagnostic\_framing': 1\}} \\
        Why does few-shot prompting harm DeepSeek-R1's reasoning accuracy, and can a meta-learning or prompt-robustness fine-tuning stage make such reasoning models invariant to demonstration style without hurting zero-shot generalization?

    \end{itemize}
\end{tcolorbox}

\begin{tcolorbox}[colback=gray!5!white, colframe=black!75!white, title=\textbf{Output of Comparative Scoring (Narrowness)}, arc=4mm]

    \textbf{\textsf{BACKGROUND PAPER}} \\
    \textbf{Title:} Accelerating Large Language Model Decoding with Speculative Sampling \\
    \textbf{Abstract (Summarized):} \\
    This paper introduces ``speculative sampling,'' an algorithm that accelerates transformer decoding by generating multiple tokens per call. By using a faster, smaller ``draft'' model to propose short continuations and a larger target model to score them in parallel via a novel modified rejection sampling scheme, the algorithm preserves the target model's distribution. Evaluated on the 70B parameter Chinchilla model, this method achieves a 2--2.5$\times$ speedup without compromising sample quality or requiring modifications to the target model.

    \tcblower

    \textbf{\textsf{AUTHOR-ANCHORED RQ}} \hfill \textbf{Narrowness: 2} \textbar\ \textbf{Rank: 0} \\
    \texttt{\footnotesize \{'source\_boundedness': 2, 'diagnostic\_framing': 0\}} \\
    \textbf{RQ:} \textit{``How can introducing a controlled bias via a stepwise process reward model improve the reasoning accuracy of speculative decoding by retaining high-quality draft tokens that strictly unbiased methods would reject?''}

    \tcbline

    \textbf{\textsf{GPT-5.5 generated 5 RQs}} \hfill {\footnotesize (ranked by source\_boundedness $>$ diagnostic\_framing, ascending)}
    \begin{itemize}
        \setlength\itemsep{0.75em}

        \item \textbf{[Rank 1]} \hfill (Narrowness: 3) \\
        \texttt{\footnotesize \{'source\_boundedness': 2, 'diagnostic\_framing': 1\}} \\
        Can prompt-level estimates of draft--target agreement be used before generation to route requests among speculative sampling configurations with different draft models and K values?

        \item \textbf{[Rank 2]} \hfill (Narrowness: 3) \\
        \texttt{\footnotesize \{'source\_boundedness': 2, 'diagnostic\_framing': 1\}} \\
        Can a learned policy select the draft model, hardware topology, and lookahead K that minimize end-to-end latency for a given target model, task distribution, and service-level latency objective?

        \item \textbf{[Rank 3]} \hfill (Narrowness: 5) \\
        \texttt{\footnotesize \{'source\_boundedness': 3, 'diagnostic\_framing': 2\}} \\
        Can a per-step adaptive choice of lookahead K based on draft-target agreement or target entropy reduce mean and P90/P99 latency compared with a fixed K while preserving the target model distribution?

        \item \textbf{[Rank 4]} \hfill (Narrowness: 5) \\
        \texttt{\footnotesize \{'source\_boundedness': 3, 'diagnostic\_framing': 2\}} \\
        What numerical divergence from the target sampling distribution is introduced by speculative sampling when logits, draft models, target models, or KV caches are evaluated in bfloat16, int8, or int4 precision?

        \item \textbf{[Rank 5]} \hfill (Narrowness: 6) \\
        \texttt{\footnotesize \{'source\_boundedness': 3, 'diagnostic\_framing': 3\}} \\
        How does speculative sampling perform under batch sizes greater than 1 and continuous batching in terms of throughput, mean latency, P99 latency, accelerator utilization, and KV-cache bandwidth compared with standard autoregressive sampling?

    \end{itemize}
\end{tcolorbox}

\begin{tcolorbox}[colback=gray!5!white, colframe=black!75!white, title=\textbf{Output of Author-Anchor Overlap Scoring}, arc=4mm]

    \textbf{\textsf{BACKGROUND PAPER}} \\
    \textbf{Title:} DeepSeek-R1: Incentivizing Reasoning Capability in LLMs via Reinforcement Learning \\

    \textbf{\textsf{AUTHOR-ANCHORED RQ}} \\
    \textit{``How can large language models dynamically compress intermediate reasoning steps into compact hidden state representations to reduce token consumption and computational overhead during complex reasoning tasks?''}

    \tcblower

    \textbf{\textsf{ONE EXAMPLE PER MODEL}} \hfill {\footnotesize (author-anchor match, 0--4)}
    \begin{itemize}
        \setlength\itemsep{0.5em}

        \item \textbf{[gpt-5.5]} \hfill (Recall: 4) \\
        How can the inference-time computational cost of large language model reasoning be reduced without degrading task accuracy?

        \item \textbf{[gpt-oss-20b]} \hfill (Recall: 3) \\
        Can LLMs learn to encode multi-step chains of thought into condensed latent representations that lower the number of decoded tokens on hard reasoning problems?

        \item \textbf{[gemma-4-31b-it]} \hfill (Recall: 2) \\
        Does pruning redundant tokens from explicit chain-of-thought traces preserve reasoning accuracy while shortening generation length?

        \item \textbf{[deepseek-v4-pro]} \hfill (Recall: 1) \\
        Why does few-shot prompting harm DeepSeek-R1's reasoning accuracy, and can a prompt-robustness fine-tuning stage make the model invariant to demonstration style?

        \item \textbf{[gemini-3.1-pro]} \hfill (Recall: 1) \\
        What underlying shifts in attention mechanisms cause few-shot prompting to degrade the performance of outcome-based RL reasoning models?

        \item \textbf{[qwen3-30b-a3b-thinking-2507]} \hfill (Recall: 0) \\
        Can retrieval-augmented generation reduce factual hallucination in open-domain question answering?

    \end{itemize}
\end{tcolorbox}

\subsection{Expert Evaluation Outputs}
\label{app:expert_eval_outputs}

\begin{tcolorbox}[colback=gray!5!white, colframe=gray!60!black, title=\textbf{Sample 1: Comparative Evaluation by Experts vs. LLM Judges}, arc=2mm, boxsep=0.5mm, left=2mm, right=2mm, top=1mm, bottom=1mm]
    
    \textbf{\textsf{BACKGROUND PAPER}} \\
    \textbf{Title:} DeepSeek-R1: Incentivizing Reasoning Capability in LLMs via Reinforcement Learning \\
    {\small \textbf{Abstract (Summarized):} General reasoning represents a long-standing challenge in AI. Recent breakthroughs, exemplified by large language models and chain-of-thought prompting, have achieved considerable success on foundational tasks but still rely heavily on human-annotated demonstrations. This paper demonstrates that LLM reasoning abilities can be incentivized through pure reinforcement learning (RL) without human-labeled trajectories, facilitating advanced emergent reasoning patterns like self-reflection, verification, and dynamic strategy adaptation.}
    
    \tcblower 
    
    \textbf{\textsf{GROUND TRUTH RQ}} \\
    \textit{``What specific algorithmic mechanisms drive the empirical success of GRPO in LLM reasoning tasks, and are complex components like reward normalization truly necessary?''}
    
    \smallskip
    
    \textbf{\textsf{BEST GPT-5.5 RQ}} \\
    \textit{``Can chain-of-thought-level safety supervision reduce harmful intermediate reasoning content while preserving final-answer helpfulness and reasoning benchmark performance?''}
    
    \tcbline
    
    \textbf{\textsf{JUDGES' VERDICTS}} \\
    \begin{tabular}{@{}p{0.48\textwidth} p{0.48\textwidth}@{}}
        \textbf{[Expert 1]} Winner: \textbf{gpt-5.5} & \textbf{[Gemini-3.1-pro]} Winner: \textbf{gpt-5.5} \\
        \textbf{[Expert 2]} Winner: \textbf{gpt-5.5} & \textbf{[DeepSeek-v4-pro]} Winner: \textbf{gpt-5.5}
    \end{tabular}
\end{tcolorbox}

\vspace{0.75em}

\begin{tcolorbox}[colback=gray!5!white, colframe=gray!60!black, title=\textbf{Sample 2: Comparative Evaluation by Experts vs. LLM Judges}, arc=2mm, boxsep=0.5mm, left=2mm, right=2mm, top=1mm, bottom=1mm]
    
    \textbf{\textsf{BACKGROUND PAPER}} \\
    \textbf{Title:} Mastering Chess and Shogi by Self-Play with a General Reinforcement Learning Algorithm \\
    {\small \textbf{Abstract (Summarized):} The game of chess has historically relied on sophisticated search techniques and handcrafted evaluation functions refined by human experts. In contrast, the AlphaGo Zero program recently achieved superhuman performance in Go via tabula rasa reinforcement learning from games of self-play. This paper generalizes that approach into a single AlphaZero algorithm, which achieves superhuman performance in chess, shogi, and Go starting solely from random play and given no domain knowledge except the game rules.}
    
    \tcblower 
    
    \textbf{\textsf{GROUND TRUTH RQ}} \\
    \textit{``How can tree search algorithms be efficiently integrated into the reinforcement learning process of large language models to provide dense process supervision for reasoning tasks?''}
    
    \smallskip
    
    \textbf{\textsf{BEST GPT-5.5 RQ}} \\
    \textit{``What modifications to the state representation and terminal scoring are required for AlphaZero to learn strong shogi play under the entering-king rule?''}
    
    \tcbline
    
    \textbf{\textsf{JUDGES' VERDICTS}} \\
    \begin{tabular}{@{}p{0.48\textwidth} p{0.48\textwidth}@{}}
        \textbf{[Expert 1]} Winner: \textbf{GT} & \textbf{[Gemini-3.1-pro]} Winner: \textbf{gpt-5.5} \\
        \textbf{[Expert 2]} Winner: \textbf{GT} & \textbf{[DeepSeek-v4-pro]} Winner: \textbf{gpt-5.5}
    \end{tabular}
\end{tcolorbox}

\vspace{0.75em}

\begin{tcolorbox}[colback=gray!5!white, colframe=gray!60!black, title=\textbf{Sample 3: Comparative Evaluation by Experts vs. LLM Judges}, arc=2mm, boxsep=0.5mm, left=2mm, right=2mm, top=1mm, bottom=1mm]
    
    \textbf{\textsf{BACKGROUND PAPER}} \\
    \textbf{Title:} Synergizing Reasoning and Acting in Language Models \\
    {\small \textbf{Abstract (Summarized):} While large language models (LLMs) have demonstrated impressive performance across language understanding and interactive decision-making tasks, their abilities for reasoning and acting have primarily been studied as separate domains. This paper explores generating both reasoning traces and task-specific actions in an interleaved manner. This synergizes the two capabilities: reasoning traces help the model induce, track, and update action plans, while actions allow it to gather additional information from external sources like knowledge bases or environments.}
    
    \tcblower 
    
    \textbf{\textsf{GROUND TRUTH RQ}} \\
    \textit{``How can an agent efficiently manage context utilization during extensive multi-turn tool interactions without degrading its reasoning trajectory?''}
    
    \smallskip
    
    \textbf{\textsf{BEST GPT-5.5 RQ}} \\
    \textit{``Are ReAct thought traces faithful causal mediators of subsequent actions, as measured by counterfactual thought edits and resulting action changes across question-answering and decision-making tasks?''}
    
    \tcbline
    
    \textbf{\textsf{JUDGES' VERDICTS}} \\
    \begin{tabular}{@{}p{0.48\textwidth} p{0.48\textwidth}@{}}
        \textbf{[Expert 1]} Winner: \textbf{GT} & \textbf{[Gemini-3.1-pro]} Winner: \textbf{gpt-5.5} \\
        \textbf{[Expert 2]} Winner: \textbf{GT} & \textbf{[DeepSeek-v4-pro]} Winner: \textbf{gpt-5.5}
    \end{tabular}
\end{tcolorbox}

\vspace{0.75em}

\begin{tcolorbox}[colback=gray!5!white, colframe=gray!60!black, title=\textbf{Sample 4: Comparative Evaluation by Experts vs. LLM Judges}, arc=2mm, boxsep=0.5mm, left=2mm, right=2mm, top=1mm, bottom=1mm]
    
    \textbf{\textsf{BACKGROUND PAPER}} \\
    \textbf{Title:} DeepSeek-R1: Incentivizing Reasoning Capability in LLMs via Reinforcement Learning \\
    {\small \textbf{Abstract (Summarized):} General reasoning represents a formidable challenge in artificial intelligence. While recent breakthroughs in LLMs have achieved success, they heavily depend on extensive human-annotated demonstrations. Here we show that LLM reasoning abilities can be incentivized through pure reinforcement learning (RL), obviating the need for human-labeled trajectories. The trained model achieves superior performance on verifiable tasks such as mathematics and coding, and its emergent reasoning patterns can be harnessed to enhance smaller models.}
    
    \tcblower 
    
    \textbf{\textsf{GROUND TRUTH RQ}} \\
    \textit{``Can a simplified reinforcement learning approach that explicitly filters prompts based on response correctness achieve comparable performance to GRPO with improved efficiency and stability?''}
    
    \smallskip
    
    \textbf{\textsf{BEST GPT-5.5 RQ}} \\
    \textit{``Which measurable reasoning behaviors, such as explicit verification, backtracking, alternative-solution search, or error correction, causally predict correctness when interventions selectively encourage or suppress each behavior during RL?''}
    
    \tcbline
    
    \textbf{\textsf{JUDGES' VERDICTS}} \\
    \begin{tabular}{@{}p{0.48\textwidth} p{0.48\textwidth}@{}}
        \textbf{[Expert 1]} Winner: \textbf{GT} & \textbf{[Gemini-3.1-pro]} Winner: \textbf{gpt-5.5} \\
        \textbf{[Expert 2]} Winner: \textbf{gpt-5.5} & \textbf{[DeepSeek-v4-pro]} Winner: \textbf{gpt-5.5}
    \end{tabular}
\end{tcolorbox}

\vspace{0.75em}

\begin{tcolorbox}[colback=gray!5!white, colframe=gray!60!black, title=\textbf{Sample 5: Comparative Evaluation by Experts vs. LLM Judges}, arc=2mm, boxsep=0.5mm, left=2mm, right=2mm, top=1mm, bottom=1mm]
    
    \textbf{\textsf{BACKGROUND PAPER}} \\
    \textbf{Title:} Aligning LLMs with Distillation \\
    {\small \textbf{Abstract (Summarized):} Reinforcement learning from human feedback (RLHF) is a key driver of quality and safety in state-of-the-art large language models. However, Best-of-N (BON) sampling is an incredibly strong inference-time alternative that simply selects the best generation among candidate outputs, albeit with significant computational overhead. In this paper, we propose BON Distillation (BOND), a novel RLHF algorithm that seeks to emulate the performance of BON without its massive inference-time costs, effectively aligning models by matching the BON distribution.}
    
    \tcblower 
    
    \textbf{\textsf{GROUND TRUTH RQ}} \\
    \textit{``How can the Best-of-N Distillation (BOND) reinforcement learning algorithm be adapted and improved within a post-training recipe to effectively align lightweight multimodal models for enhanced reasoning, math, and coding capabilities?''}
    
    \smallskip
    
    \textbf{\textsf{BEST GPT-5.5 RQ}} \\
    \textit{``How much performance is lost by omitting the Best-of-N collision correction term in discrete language generation, especially for short outputs or low-entropy reference policies?''}
    
    \tcbline
    
    \textbf{\textsf{JUDGES' VERDICTS}} \\
    \begin{tabular}{@{}p{0.48\textwidth} p{0.48\textwidth}@{}}
        \textbf{[Expert 1]} Winner: \textbf{GT} & \textbf{[Gemini-3.1-pro]} Winner: \textbf{gpt-5.5} \\
        \textbf{[Expert 2]} Winner: \textbf{GT} & \textbf{[DeepSeek-v4-pro]} Winner: \textbf{gpt-5.5}
    \end{tabular}
\end{tcolorbox}

\vspace{0.75em}

\begin{tcolorbox}[colback=gray!5!white, colframe=gray!60!black, title=\textbf{Sample 6: Comparative Evaluation by Experts vs. LLM Judges}, arc=2mm, boxsep=0.5mm, left=2mm, right=2mm, top=1mm, bottom=1mm]
    
    \textbf{\textsf{BACKGROUND PAPER}} \\
    \textbf{Title:} The Impact of Reasoning Step Length on Large Language Models \\
    {\small \textbf{Abstract (Summarized):} Chain of Thought (CoT) prompting is highly significant in improving the reasoning abilities of large language models (LLMs). However, the correlation between the effectiveness of CoT and the length of reasoning steps in prompts remains largely unexplored. Through empirical experiments, this paper reveals that lengthening reasoning steps in prompts considerably enhances reasoning abilities across multiple datasets. Surprisingly, the results also show that even incorrect rationales can yield favorable outcomes if they maintain the requisite length of inference.}
    
    \tcblower 
    
    \textbf{\textsf{GROUND TRUTH RQ}} \\
    \textit{``How can large language models be adapted to selectively skip semantically unimportant tokens during Chain-of-Thought reasoning to achieve controllable compression without degrading reasoning accuracy?''}
    
    \smallskip
    
    \textbf{\textsf{BEST GPT-5.5 RQ}} \\
    \textit{``How accurately do human annotators and automated parsers agree on the number of reasoning steps in CoT rationales, and does this validated step count predict model accuracy better than raw token length?''}
    
    \tcbline
    
    \textbf{\textsf{JUDGES' VERDICTS}} \\
    \begin{tabular}{@{}p{0.48\textwidth} p{0.48\textwidth}@{}}
        \textbf{[Expert 1]} Winner: \textbf{GT} & \textbf{[Gemini-3.1-pro]} Winner: \textbf{GT} \\
        \textbf{[Expert 2]} Winner: \textbf{GT} & \textbf{[DeepSeek-v4-pro]} Winner: \textbf{tie}
    \end{tabular}
\end{tcolorbox}

\vspace{0.75em}

\begin{tcolorbox}[colback=gray!5!white, colframe=gray!60!black, title=\textbf{Sample 7: Comparative Evaluation by Experts vs. LLM Judges}, arc=2mm, boxsep=0.5mm, left=2mm, right=2mm, top=1mm, bottom=1mm]
    
    \textbf{\textsf{BACKGROUND PAPER}} \\
    \textbf{Title:} Accelerating Large Language Model Decoding with Speculative Sampling \\
    {\small \textbf{Abstract (Summarized):} We present speculative sampling, a novel algorithm designed to accelerate transformer decoding by enabling the generation of multiple tokens from each transformer call. The algorithm leverages the observation that parallel scoring of short continuations, generated by a faster but less powerful draft model, has latency comparable to sampling a single token from the larger target model. Combined with a modified rejection sampling scheme, this approach preserves the target model's exact distribution while achieving a 2–2.5x decoding speedup.}
    
    \tcblower 
    
    \textbf{\textsf{GROUND TRUTH RQ}} \\
    \textit{``How can introducing a controlled bias via a stepwise process reward model improve the reasoning accuracy of speculative decoding by retaining high-quality draft tokens that strictly unbiased methods would reject?''}
    
    \smallskip
    
    \textbf{\textsf{BEST GPT-5.5 RQ}} \\
    \textit{``Can prompt-level estimates of draft-target agreement be used before generation to route requests among speculative sampling configurations with different draft models and K values?''}
    
    \tcbline
    
    \textbf{\textsf{JUDGES' VERDICTS}} \\
    \begin{tabular}{@{}p{0.48\textwidth} p{0.48\textwidth}@{}}
        \textbf{[Expert 1]} Winner: \textbf{GT} & \textbf{[Gemini-3.1-pro]} Winner: \textbf{GT} \\
        \textbf{[Expert 2]} Winner: \textbf{gpt-5.5} & \textbf{[DeepSeek-v4-pro]} Winner: \textbf{gpt-5.5}
    \end{tabular}
\end{tcolorbox}

\vspace{0.75em}

\begin{tcolorbox}[colback=gray!5!white, colframe=gray!60!black, title=\textbf{Sample 8: Comparative Evaluation by Experts vs. LLM Judges}, arc=2mm, boxsep=0.5mm, left=2mm, right=2mm, top=1mm, bottom=1mm]
    
    \textbf{\textsf{BACKGROUND PAPER}} \\
    \textbf{Title:} Large Language Monkeys: Scaling Inference Compute Repeated Sampling \\
    {\small \textbf{Abstract (Summarized):} Scaling the compute used to train language models has dramatically improved their capabilities. In this paper, we explore inference compute as an alternative scaling axis by using the simple technique of repeatedly sampling candidate solutions from a model. Across multiple tasks, we observe that problem-solving coverage scales log-linearly with the number of samples over four orders of magnitude. This suggests the existence of inference-time scaling laws, which directly translate into improved performance for verifiable domains like coding and proofs.}
    
    \tcblower 
    
    \textbf{\textsf{GROUND TRUTH RQ}} \\
    \textit{``How can an adaptive decoding framework optimize the trade-off between computational cost and reasoning accuracy to achieve high performance without the extensive sampling budget required by Best-of-N methods?''}
    
    \smallskip
    
    \textbf{\textsf{BEST GPT-5.5 RQ}} \\
    \textit{``How accurately can exponentiated power-law fits estimated from the first 100 or 1,000 samples predict pass@10,000 or pass@100,000 across model families, model sizes, and tasks?''}
    
    \tcbline
    
    \textbf{\textsf{JUDGES' VERDICTS}} \\
    \begin{tabular}{@{}p{0.48\textwidth} p{0.48\textwidth}@{}}
        \textbf{[Expert 1]} Winner: \textbf{GT} & \textbf{[Gemini-3.1-pro]} Winner: \textbf{gpt-5.5} \\
        \textbf{[Expert 2]} Winner: \textbf{GT} & \textbf{[DeepSeek-v4-pro]} Winner: \textbf{gpt-5.5}
    \end{tabular}
\end{tcolorbox}

\vspace{0.75em}

\begin{tcolorbox}[colback=gray!5!white, colframe=gray!60!black, title=\textbf{Sample 9: Comparative Evaluation by Experts vs. LLM Judges}, arc=2mm, boxsep=0.5mm, left=2mm, right=2mm, top=1mm, bottom=1mm]
    
    \textbf{\textsf{BACKGROUND PAPER}} \\
    \textbf{Title:} DeepSeek-R1: Incentivizing Reasoning Capability in LLMs via Reinforcement Learning \\
    {\small \textbf{Abstract (Summarized):} General reasoning is a formidable challenge that traditionally relies heavily on human-annotated demonstrations. We demonstrate that the reasoning abilities of LLMs can be dramatically incentivized through pure reinforcement learning (RL) without labeled trajectories. The proposed RL framework facilitates the emergent development of advanced reasoning patterns, such as self-reflection and dynamic strategy adaptation. Furthermore, the emergent reasoning patterns exhibited by these large-scale models can be systematically harnessed to guide and enhance smaller models.}
    
    \tcblower 
    
    \textbf{\textsf{GROUND TRUTH RQ}} \\
    \textit{``Can a multi-stage post-training pipeline combining supervised fine-tuning and large-scale reinforcement learning enable a student model to exceed its teacher's reasoning performance?''}
    
    \smallskip
    
    \textbf{\textsf{BEST GPT-5.5 RQ}} \\
    \textit{``Can reasoning-trace-aware safety training reduce unsafe responses under jailbreak attacks for open-source DeepSeek-R1-style models while keeping refusal rates below those of external risk-control systems?''}
    
    \tcbline
    
    \textbf{\textsf{JUDGES' VERDICTS}} \\
    \begin{tabular}{@{}p{0.48\textwidth} p{0.48\textwidth}@{}}
        \textbf{[Expert 1]} Winner: \textbf{gpt-5.5} & \textbf{[Gemini-3.1-pro]} Winner: \textbf{gpt-5.5} \\
        \textbf{[Expert 2]} Winner: \textbf{gpt-5.5} & \textbf{[DeepSeek-v4-pro]} Winner: \textbf{tie}
    \end{tabular}
\end{tcolorbox}

\vspace{0.75em}

\begin{tcolorbox}[colback=gray!5!white, colframe=gray!60!black, title=\textbf{Sample 10: Comparative Evaluation by Experts vs. LLM Judges}, arc=2mm, boxsep=0.5mm, left=2mm, right=2mm, top=1mm, bottom=1mm]
    
    \textbf{\textsf{BACKGROUND PAPER}} \\
    \textbf{Title:} Process Reinforcement through Implicit Rewards \\
    {\small \textbf{Abstract (Summarized):} Dense process rewards have proven to be a highly effective alternative to sparse outcome-level rewards in the inference-time scaling of large language models, particularly for complex multi-step reasoning. However, collecting high-quality process labels is prohibitively expensive, making process reward models (PRMs) vulnerable to reward hacking. To address this, we propose PRIME (Process Reinforcement through Implicit Rewards), which enables online PRM updates using only policy rollouts and outcome labels, substantially reducing development overhead while boosting reasoning performance.}
    
    \tcblower 
    
    \textbf{\textsf{GROUND TRUTH RQ}} \\
    \textit{``How can an online difficulty filtering algorithm maintain a fixed effective batch size to prevent training destabilization during dynamic prompt exclusion?''}
    
    \smallskip
    
    \textbf{\textsf{BEST GPT-5.5 RQ}} \\
    \textit{``Which types of reasoning steps receive systematically incorrect implicit process rewards in PRIME, and do those misassigned rewards predict later policy errors?''}
    
    \tcbline
    
    \textbf{\textsf{JUDGES' VERDICTS}} \\
    \begin{tabular}{@{}p{0.48\textwidth} p{0.48\textwidth}@{}}
        \textbf{[Expert 1]} Winner: \textbf{GT} & \textbf{[Gemini-3.1-pro]} Winner: \textbf{gpt-5.5} \\
        \textbf{[Expert 2]} Winner: \textbf{gpt-5.5} & \textbf{[DeepSeek-v4-pro]} Winner: \textbf{gpt-5.5}
    \end{tabular}
\end{tcolorbox}

\subsection{Dataset Quality Verification}


\begin{promptbox}{RQ-addressed judging prompt}
\promptfield{System}

You are a careful research-paper reviewer.

You will be given:
\begin{enumerate}
    \item The full LaTeX source of a paper.
    \item ONE research question generated independently from that paper's references.
\end{enumerate}

Your job: decide whether the paper's own idea and contributions ACTUALLY ADDRESS
the research question. ``Address'' means the paper proposes a method, analysis, or
result that answers (in whole or substantial part) what the question asks.

Be strict:
\begin{itemize}
    \item If the paper only mentions the topic in passing, that is NOT addressing.
    \item If the paper addresses a different but related question, that is NOT addressing.
    \item Partial coverage counts as \texttt{addressed=true} only if the paper's contribution
        materially answers the question; explain the partial nature in \texttt{reasoning}.
\end{itemize}

Return STRICT JSON only:
\begin{verbatim}
{
  "addressed": true | false,
  "confidence": "high" | "medium" | "low",
  "reasoning": "<2-4 sentences explaining the verdict>",
  "evidence": "<short verbatim phrase or section name from the paper
               that supports the verdict, or empty string>"
}
\end{verbatim}

\promptfield{User}

\begin{verbatim}
=== PAPER LATEX SOURCE ===
{tex}

=== RESEARCH QUESTION ===
{question}

Decide whether the paper's idea/contributions address the research
question. Respond with the JSON object specified in the system prompt.
\end{verbatim}
\end{promptbox}

\begin{promptbox}{Manual Validation Samples (50 RQs)}
\promptfield{Sampling protocol}

Population: the released dataset contains 1{,}434 research questions. From this RQ pool we drew 50 items uniformly at random without replacement
using Python \texttt{random.sample} with \texttt{random.seed(42)}. Each ID
matches the \texttt{rq\_id} field in the released dataset and has the form
\texttt{<arxiv\_id>\_rq<index>}, where \texttt{<index>} is the position of
the question within the paper's \texttt{research\_questions} list.

\promptfield{Sample IDs}

\begin{verbatim}
2502.01113_rq0   2502.05209_rq0   2502.06608_rq2   2502.06768_rq2
2502.11089_rq0   2502.11271_rq2   2502.11435_rq0   2502.13713_rq0
2502.18508_rq0   2502.18509_rq0   2503.01485_rq2   2503.06034_rq0
2503.09501_rq0   2503.11926_rq2   2503.15838_rq0   2503.19611_rq2
2503.20523_rq1   2504.02546_rq2   2504.05187_rq0   2504.08725_rq0
2504.11713_rq2   2505.00703_rq0   2505.16394_rq1   2505.16990_rq0
2505.22974_rq1   2506.02735_rq1   2506.06664_rq1   2506.07339_rq1
2506.09260_rq0   2506.12199_rq0   2506.13497_rq0   2506.13695_rq0
2506.16504_rq1   2507.04447_rq1   2507.07356_rq2   2507.07957_rq0
2507.07966_rq1   2508.02324_rq0   2508.05635_rq0   2508.06372_rq1
2508.08241_rq2   2509.03236_rq0   2509.15937_rq0   2509.26328_rq0
2510.02283_rq0   2510.10274_rq0   2511.16719_rq1   2601.07526_rq1
2601.19897_rq1   2603.00501_rq0
\end{verbatim}
\end{promptbox}

\newpage

\subsection{Alternate RQ Generation Prompt}
\label{sec:updated-prompt}
\begin{promptbox}{New RQ generation prompt}
\promptfield{System}

You are an expert research scientist whose job is to read background
literature and identify meaningful, unsolved research problems.

\promptfield{Task}

For the background paper(s) the user provides:

\begin{enumerate}
    \item Identify the research gaps in the literature: the limitations,
    weaknesses, unresolved problems, and unaddressed challenges that the
    work(s) leave open.
    \item Propose research questions that, if answered, would directly
    close one or more of those gaps.
\end{enumerate}

\promptfield{Multi-paper rule}

When the user provides more than one background paper, a gap may apply to
one paper, several papers, or all of them. Do not force every gap to be
shared by every paper. If gaps from different papers converge on the same
research question, group them under one RQ rather than splitting them. If
gaps motivate clearly distinct directions, separate them into distinct
RQs.

\promptfield{Research questions}

\begin{itemize}
    \item Produce exactly 5 research questions.
    \item Each question must be phrased as a question and end with ``?''.
\end{itemize}

\promptfield{Output format}

The final visible answer must be a single JSON object conforming exactly
to the following schema:

\begin{verbatim}
{
  "gaps": [
    "<description of gap 1>",
    "<description of gap 2>"
  ],
  "research_questions": [
    "<research question 1 ending with '?'>",
    "<research question 2 ending with '?'>",
    "<research question 3 ending with '?'>",
    "<research question 4 ending with '?'>",
    "<research question 5 ending with '?'>"
  ]
}
\end{verbatim}

Both lists must be non-empty. The \texttt{research\_questions} list must
contain exactly 5 entries. Use plain strings only: no nested objects and
no extra keys.

\promptfield{User prompt: single-paper setting}

\begin{verbatim}
Below is one background paper. Identify its research gaps and propose
research questions addressing them, following the rules in your system
instructions.

{papers_block}
\end{verbatim}

\promptfield{User prompt: multi-paper setting}

\begin{verbatim}
Below are {n_papers} background papers. Identify the research gaps they
raise — a gap may apply to one paper, several, or all of them — and
propose research questions addressing those gaps, grouping convergent
gaps under a single RQ where the literature supports it. Follow the rules
in your system instructions.

{papers_block}
\end{verbatim}
\end{promptbox}
\end{document}